\documentclass{aa}
\usepackage[varg]{txfonts}
\usepackage{graphicx}
\usepackage{subcaption}
\usepackage{color}

\bibpunct{(}{)}{;}{a}{}{,} 

\begin{document}

\title{A multiplicity study of transiting exoplanet host stars. I.}
\subtitle{High-contrast imaging with VLT/SPHERE\thanks{Based on observations collected at the European Organisation for Astronomical Research in the Southern Hemisphere under ESO programmes 098.C-0589(A) and 099.C-0155(A).}}

\author{
A.~J.~Bohn\inst{1}
\and J.~Southworth\inst{2}
\and C.~Ginski\inst{3}
\and M.~A.~Kenworthy\inst{1}
\and P.~F.~L.~Maxted\inst{2}
\and D.~F.~Evans\inst{2}
}

\institute{Leiden Observatory, Leiden University, PO Box 9513, 2300 RA Leiden, The Netherlands\\
              \email{bohn@strw.leidenuniv.nl}
              \and Astrophysics Group, Keele University, Staffordshire ST5 5BG, UK
              \and Sterrenkundig Instituut Anton Pannekoek, Science Park 904, 1098 XH Amsterdam, The Netherlands
       }

\date{Received November 17, 2019/ Accepted January 17, 2020}

\abstract 
{
Many main sequence stars are part of multiple systems.
The impact of stellar multiplicity on planet formation and migration, however, is poorly understood.
} 
{
We study the multiplicity of host stars to known transiting extra-solar planets to test competing theories on the formation mechanisms of hot Jupiters.
} 
{
We observed 45 exoplanet host stars using VLT/SPHERE/IRDIS to search for potential companions.
For each identified candidate companion we determined the probability that it is gravitationally bound to its host by performing common proper motion checks and modelling of synthetic stellar populations around the host.
In addition, we derived contrast limits as a function of angular separation to set upper limits on further companions in these systems.
We converted the derived contrast to mass thresholds using AMES-Cond, AMES-Dusty, and BT-Settl models.
} 
{
We detected new candidate companions around K2-38, WASP-72, WASP-80, WASP-87, WASP-88, WASP-108, WASP-118, WASP-120, WASP-122, WASP123, WASP-130, WASP-131 and WASP-137.
The closest candidates were detected at separations of $0\farcs124\pm0\farcs007$ and $0\farcs189\pm0\farcs003$ around WASP-108 and WASP-131;
the measured $K$ band contrasts indicate that these are stellar companions of $0.35\pm0.02\,M_{\sun}$ and $0.62^{+0.05}_{-0.04}\,M_{\sun}$, respectively.
Including the re-detection and confirmation of previously known companions in 13 other systems we derived a multiplicity fraction of $55.4^{+5.9}_{-9.4}\,\%$.
For the representative sub-sample of 40 hot Jupiter host stars among our targets, the derived multiplicity rate is $54.8^{+6.3}_{-9.9}\,\%$.
Our data do not confirm any trend that systems with eccentric planetary companions are preferably part of multiple systems.
On average, we reached a magnitude contrast of $8.5\pm0.9$\,mag at an angular separation of 0\farcs5.
This allows to exclude additional stellar companions with masses larger than $0.08$\,M$_\sun$ for almost all observed systems;
around the closest and youngest systems this sensitivity is achieved at physical separations as small as 10\,au.
} 
{
The presented study shows that SPHERE is an ideal instrument to detect and characterize close companions to exoplanetary host stars.
Although the second data release of the Gaia mission also provides useful constraints for some of the systems, the achieved sensitivity provided by the current data release of this mission is not good enough to measure parallaxes and proper motions for all detected candidates.
For 14 identified companion candidates further astrometric epochs are required to confirm their common proper motion at 5$\sigma$ significance.
}


\maketitle

\section{Introduction}
\label{sec:introduction}

The detection and characterization of extrasolar planets has evolved rapidly during the past decades.
Many large-scale radial velocity surveys \citep[RV; e.g.][]{Baranne1996,Mayor2003,Cosentino2012} and transit surveys \citep[e.g.][]{Bakos2004,Pollacco2006,Auvergne2009,Borucki2010} have provided a statistically highly significant sample consisting of several thousands of exoplanets with various physical properties that mostly differ from what we had known from the solar system so far.
Already the first exoplanet detected around a main sequence star, \object{51~Peg~b} \citep{Mayor1995}, showed drastically deviating attributes compared to all solar system planets.
With the detection of several similarly behaved Jovian planets on very close-in orbits with periods of a few days \citep{Butler1997,Fischer1999}, a new class of so called hot Jupiters was established.
These gas giants typically have masses larger than 0.3\,M$_\mathrm{Jup}$ and separations to their host stars that are smaller than 0.1\,au.

Although hundreds of hot Jupiter systems are known today, there is no consensus on a consistent formation pathway of these environments.
Shortly after the discovery of \object{51~Peg~b}, \citet{Lin1996} argued that in-situ formation of hot Jupiters via core accretion is disfavoured, as the typical temperatures in protoplanetary discs at their characteristic separations are too high to facilitate the condensation of solids, hence preventing rocky cores from forming in these regions \citep{Pollack1996}.
Simulations of \citet{Bodenheimer2000} and more recent results of \citet{Boley2016} and \citet{Batygin2016}, however, challenge this hypothesis:
previous assumptions on the amount of condensable solids in the circumstellar disc were based on abundances in the solar nebula,
which might be too simplistic to cope with the huge variety observed in exoplanetary systems.

Alternatively to the in-situ formation scenario, hot Jupiters might form at wider separations of several astronomical units and migrate inwards towards their detected position \citep{Lin1996}.
Theories that describe this migration process, however, are still a highly controversial topic.
Potential scenarios of this inward migration are required not only to reproduce the small orbital separations but also to provide useful explanations for other properties of known hot Jupiters, as for instance highly eccentric orbits \citep{Udry2007} or orbital misalignments with respect to the stellar rotation axis \citep{Winn2010}.
Recent research shows that the observed spin-orbit misalignments may have a primordial origin caused by either magnetic fields of the star interacting with the protoplanetary disc \citep{Lai2011} or gravitational interaction with massive stellar binaries \citep{Batygin2012}.
The high eccentricities, however, are not reproduced by an inward migration as first proposed by \citet{Lin1996} due to damping of excited modes caused by gravitational interaction with material of the circumstellar disc \citep{Kley2012}.
Other theories hypothesize a high-eccentricity migration of the companion after its formation \citep{Socrates2012}:
after the planet has formed in a circular orbit of several astronomical units, it gets excited to high eccentricities, and tidal dissipation at subsequent periastron passages reduces the orbital semi-major axis as well as the eccentricity gained.
The excitation of high eccentricities may be caused by planet-planet scattering \citep{Rasio1996,Chatterjee2008,Wu2011}, through Kozai-Lidov (KL) oscillations due to a stellar binary \citep{Eggleton2001, Wu2003, Fabrycky2007}, or a combination of these mechanisms \citep{Nagasawa2008}.

To test these theories, additional data of exoplanet host systems is required.
Especially stellar binaries may play an important role in the evolution of exoplanetary systems, as they are essential ingredients for explaining primordial spin-orbit misalignments or high-eccentricity migration due to KL mechanisms.
Current estimates on the multiplicity fractions among transiting exoplanet host stars are not very conclusive and range from 7.6$\pm$2.3\% \citep{Ngo2017} to 13.5\% \citep{Law2014} for RV planet hosts, but are usually higher for transiting planetary systems as the sample selection criteria for RV surveys impose an intrinsic bias against multiple stellar systems.
\citet{Ngo2015} recently estimated a much higher multiplicity rate of 49$\pm$9\% for systems with transiting hot Jupiters compared to their RV analogues.
To reduce the uncertainties on these ratios it is necessary to expand the samples to achieve statistically more significant results.

For transiting planet hosts stars, observations at high spatial resolution are also an important tool to reject other scenarios that might cause the periodic dip in the light curve, in particular background eclipsing binaries. Furthermore, the derived properties of the exoplanet and its host star are normally measured under the assumption that all the light from the system comes from the host star, i.e.\ there is no contamination from unresolved sources at very small projected separations. If this assumption is violated and the data are not corrected for the contaminating light, its presence may cause both the mass and radius of the planet to be systematically underestimated. In the worst-case scenario, a not-much-fainter nearby star could even be the planet host star, and measurements of the planet's mass and radius under the assumption that the brightest star is the host would lead to planetary properties that are severely biased away from their true values \citep[e.g.][]{Evans2016b}. In a companion paper \citep{paper2} we reanalyze the most strongly affected of the planetary systems included in the current work, in order to correct measurements of their physical properties for the light arising from the nearby companion stars we have found.

A state-of-the-art method for the detection of stellar companions at small angular separations is adaptive optics (AO)-assisted, coronagraphic, high-contrast imaging.
We therefore launched a direct imaging survey targeting host stars of transiting exoplanets.
Starting with the TEPCat catalogue \citep{Southworth2011}, we selected all targets that are observable from the Very Large Telescope (VLT) and that have an $R$ band magnitude brighter than 11\,mag to enable the AO system to lock on the source as a natural guide star.
A detailed list of the 45 studied objects and their properties is presented in Table~\ref{tbl:star_properties}.
\begin{table*}
\caption{
Stellar and planetary properties of the targets that were observed within the scope of our survey.
}
\label{tbl:star_properties}
\tiny
\def\arraystretch{1.2}
\setlength{\tabcolsep}{8pt}
\centering
\begin{tabular}{@{}llllllrlllrl@{}}
\hline\hline
Star & $M_\star$  & $R_\star$ & $T_\mathrm{eff}$ & Distance\tablefootmark{a} & Age & Period & Eccentricity & $M_{\rm p}$ & $R_{\rm p}$ & $T_{\rm eq}$ & References \\  
     & (M$_\odot$) & (R$_\odot$) & (K) & (pc) & (Gyr) & (d) &  & (M$_{\rm jup}$) & (R$_{\rm jup}$) & (K) &  \\
\hline
\object{HAT-P-41} & 1.418 & 1.683 & 6390 & $337.7^{+3.7}_{-3.8}$ & $ 2.32 \pm  0.42 $ & 2.694  & 0      & 0.800 & 1.685 & 1941 & 1 \\
\object{HAT-P-57} & 1.47  & 1.500 & 7500 & $279.9^{+3.2}_{-3.2}$ & $ 1.04 \pm  0.47 $ & 2.465  & 0      &       & 1.413 & 2200 & 2 \\
\object{K2-2}     & 0.775 & 0.716 & 5089 & $62.4^{+0.2}_{-0.2}$  & $ 5.65 \pm  3.63 $ & 9.121  & 0.205  & 0.037 & 0.226 &  690 & 3 \\
\object{K2-24b}   & 1.07  & 1.16  & 5625 & $170.6^{+1.3}_{-1.4}$ & $ 6.49 \pm  1.81 $ & 20.890 & 0.06   & 0.057 & 0.482 &  767 & 4, 5 \\
\object{K2-24c}   & 1.07  & 1.16  & 5625 & $170.6^{+1.3}_{-1.4}$ & $ 6.49 \pm  1.81 $ & 42.339 & 0      & 0.048 & 0.669 &  606 & 4, 5 \\
\object{K2-38b}   & 1.07  & 1.10  & 5757 & $192.7^{+2.6}_{-2.7}$ & $ 2.51 \pm  1.40 $ & 4.016  & 0      & 0.038 & 0.138 & 1184 & 6 \\
\object{K2-38c}   & 1.07  & 1.10  & 5757 & $192.7^{+2.6}_{-2.7}$ & $ 2.51 \pm  1.40 $ & 10.561 & 0      & 0.031 & 0.216 &  858 & 6 \\
\object{K2-39}    & 1.192 & 2.93  & 4912 & $307.5^{+4.6}_{-4.7}$ & $ 4.71 \pm  0.92 $ & 4.605  & 0.152  & 0.125 & 0.509 & 1670 & 7, 8 \\
\object{K2-99}    & 1.60  & 3.1   & 5990 & $519.2^{+12.4}_{-13.0}$ & $ 2.12 \pm 0.09$ & 18.249 & 0.19   & 0.97  & 1.29  &      & 9 \\
\object{KELT-10}  & 1.112 & 1.209 & 5948 & $188.4^{+2.1}_{-2.2}$ & $ 2.82 \pm  1.45 $ & 4.166  & 0      & 0.679 & 1.399 & 1377 & 10 \\
\object{WASP-2}   & 0.851 & 0.823 & 5170 & $153.2^{+1.6}_{-1.6}$ & $ 7.40 \pm  2.83 $ & 2.152  & 0      & 0.880 & 1.063 & 1286 & 11, 12 \\
\object{WASP-7}   & 1.317 & 1.478 & 6520 & $162.3^{+1.3}_{-1.3}$ & $ 2.05 \pm  0.47 $ & 4.955  & 0      & 0.98  & 1.374 & 1530 & 13, 12 \\
\object{WASP-8}   & 1.030 & 0.945 & 5600 & $90.0^{+0.4}_{-0.4}$  & $ 3.27 \pm  2.05 $ & 8.159  & 0.3100 & 2.25  & 1.038 &      & 14 \\
\object{WASP-16}  & 0.980 & 1.087 & 5630 & $194.1^{+1.9}_{-1.9}$ & $ 8.93 \pm  2.17 $ & 3.119  & 0      & 0.832 & 1.218 & 1389 & 15, 16 \\
\object{WASP-20}  & 1.089 & 1.142 & 6000 & $235^{+20}_{-20}$     & $ 4.34 \pm  1.76 $  & 4.900  & 0      & 0.378 & 1.28  & 1282 & 43 \\                
\object{WASP-21}  & 0.890 & 1.136 & 5924 & $258.4^{+2.8}_{-2.9}$ & $ 8.47 \pm  1.63 $ & 4.323  & 0      & 0.276 & 1.162 & 1333 & 17, 18 \\
\object{WASP-29}  & 0.825 & 0.808 & 4875 & $87.6^{+0.3}_{-0.3}$  & $10.10 \pm  4.05 $ & 3.923  & 0.03   & 0.244 & 0.776 &  970 & 19, 20 \\
\object{WASP-30}  & 1.249 & 1.389 & 6190 & $353.5^{+8.8}_{-9.3}$ & $ 3.42 \pm  0.70 $ & 4.157  & 0      & 62.5  & 0.951 & 1474 & 21, 22 \\
\object{WASP-54}  & 1.213 & 1.828 & 6296 & $251.3^{+4.3}_{-4.5}$ & $ 3.02 \pm  0.57 $ & 3.694  & 0.067  & 0.636 & 1.653 & 1759 & 23 \\
\object{WASP-68}  & 1.24  & 1.69  & 5910 & $226.4^{+1.6}_{-1.6}$ & $ 3.02 \pm  0.57 $ & 5.084  & 0      & 0.95  & 1.24  & 1490 & 24 \\
\object{WASP-69}  & 0.826 & 0.813 & 4700 & $50.0^{+0.1}_{-0.1}$  & $13.52 \pm  2.80 $ & 3.868  & 0      & 0.260 & 1.057 &  963 & 25 \\
\object{WASP-70}  & 1.106 & 1.215 & 5700 & $222.4^{+2.8}_{-2.9}$ & $ 9.35 \pm  2.01 $ & 3.713  & 0      & 0.590 & 1.164 & 1387 & 25 \\
\object{WASP-71}  & 1.559 & 2.26  & 6180 & $362.7^{+6.7}_{-7.0}$ & $ 2.22 \pm  0.45 $ & 2.904  & 0      & 2.242 & 1.46  & 2049 & 26 \\
\object{WASP-72}  & 1.386 & 1.98  & 6250 & $434.8^{+8.2}_{-8.5}$ & $ 3.55 \pm  0.82 $ & 2.217  & 0      & 1.461 & 1.27  & 2210 & 27 \\
\object{WASP-73}  & 1.34  & 2.07  & 6030 & $316.7^{+2.9}_{-3.0}$ & $ 3.59 \pm  0.94 $ & 4.087  & 0      & 1.88  & 1.16  & 1790 & 24 \\
\object{WASP-74}  & 1.191 & 1.536 & 5984 & $149.2^{+1.1}_{-1.1}$ & $ 3.67 \pm  0.48 $ & 2.138  & 0      & 0.826 & 1.404 & 1926 & 28, 29 \\
\object{WASP-76}  & 1.46  & 1.70  & 6250 & $194.5^{+5.8}_{-6.2}$ & $ 2.72 \pm  0.46 $  & 1.810    & 0      & 0.87  & 1.73  & 2154 & 44 \\
\object{WASP-80}  & 0.596 & 0.593 & 4145 & $49.8^{+0.1}_{-0.1}$  & $10.51 \pm  4.45 $ & 3.068  & 0      & 0.562 & 0.986 &  825 & 30, 31 \\
\object{WASP-87}  & 1.204 & 1.627 & 6450 & $298.4^{+3.5}_{-3.6}$ & $ 4.04 \pm  1.00 $ & 1.683  & 0      & 2.18  & 1.385 & 2322 & 32 \\
\object{WASP-88} & 1.45  & 2.08  & 6430 & $523.8^{+8.5}_{-8.8}$ & $ 2.60 \pm  0.65 $ & 4.954  & 0      & 0.56  & 1.70  & 1772 & 24 \\
\object{WASP-94}  & 1.45  & 1.62  & 6170 & $211.2^{+2.5}_{-2.5}$ & $ 3.07 \pm  0.61 $ & 3.950  & 0      & 0.452 & 1.72  & 1604 & 33 \\
\object{WASP-95}  & 1.11  & 1.13  & 5830 & $137.5^{+0.8}_{-0.8}$ & $ 5.62 \pm  2.59 $ & 2.185  & 0      & 1.13  & 1.21  & 1570 & 34 \\
\object{WASP-97}  & 1.12  & 1.06  & 5670 & $151.1^{+0.5}_{-0.5}$ & $ 4.65 \pm  2.33 $ & 2.073  & 0      & 1.32  & 1.13  & 1555 & 34 \\
\object{WASP-99}  & 1.48  & 1.76  & 6150 & $158.7^{+0.8}_{-0.8}$ & $ 3.26 \pm  0.80 $ & 5.753  & 0      & 2.78  & 1.10  & 1480 & 34 \\
\object{WASP-108} & 1.167 & 1.215 & 6000 & $258.8^{+3.2}_{-3.3}$ & $ 4.64 \pm  1.94 $ & 2.676  & 0      & 0.892 & 1.284 & 1590 & 32 \\
\object{WASP-109} & 1.203 & 1.346 & 6520 & $356.1^{+4.8}_{-5.0}$ & $ 2.68 \pm  0.92 $ & 3.319  & 0      & 0.91  & 1.443 & 1685 & 32 \\
\object{WASP-111} & 1.50  & 1.85  & 6400 & $293.1^{+6.2}_{-6.4}$ & $ 2.59 \pm  0.59 $ & 2.311  & 0      & 1.83  & 1.443 & 2140 & 32 \\
\object{WASP-117} & 1.126 & 1.170 & 6040 & $158.0^{+0.6}_{-0.6}$ & $ 4.98 \pm  1.89 $ & 10.022 & 0.302  & 0.275 & 1.021 & 1024 & 35 \\
\object{WASP-118} & 1.319 & 1.754 & 6410 & $376.7^{+10.6}_{-11.2}$ & $2.34 \pm  0.44$ & 4.046  & 0      & 0.52  & 1.394 & 1753 & 36, 37 \\
\object{WASP-120} & 1.393 & 1.87  & 6450 & $381.2^{+3.2}_{-3.2}$ & $ 2.66 \pm  0.51 $ & 3.611  & 0.057  & 4.85  & 1.473 & 1880 & 38 \\
\object{WASP-121} & 1.353 & 1.458 & 6460 & $269.9^{+1.6}_{-1.6}$ & $ 1.90 \pm  0.60 $ & 1.275  & 0      & 1.183 & 1.865 & 2358 & 39 \\
\object{WASP-122} & 1.239 & 1.52  & 5720 & $250.1^{+1.5}_{-1.5}$ & $ 6.24 \pm  1.93 $ & 1.710  & 0      & 1.284 & 1.743 & 1970 & 38 \\
\object{WASP-123} & 1.166 & 1.285 & 5740 & $198.0^{+3.0}_{-3.1}$ & $ 7.17 \pm  2.11 $ & 2.978  & 0      & 0.899 & 1.318 & 1520 & 38 \\
\object{WASP-130} & 1.04  & 0.96  & 5600 & $172.3^{+1.4}_{-1.4}$ & $ 2.82 \pm  1.87 $ & 11.551 & 0      & 1.23  & 0.89  &  833 & 40 \\
\object{WASP-131} & 1.06  & 1.53  & 5950 & $200.1^{+2.6}_{-2.7}$ & $ 7.25 \pm  1.55 $ & 5.322  & 0      & 0.27  & 1.22  & 1460 & 40 \\
\object{WASP-136} & 1.41  & 2.21  & 6250 & $275.6^{+4.5}_{-4.6}$ & $ 3.71 \pm  0.67 $ & 5.215  & 0      & 1.51  & 1.38  & 1742 & 41 \\
\object{WASP-137} & 1.216 & 1.52  & 6100 & $286.5^{+3.6}_{-3.7}$ & $ 4.29 \pm  1.24 $ & 3.908  & 0      & 0.681 & 1.27  & 1601 & 42 \\
\hline
\end{tabular}
\tablefoot{
\tablefoottext{a}{Distances are based on Gaia DR2 parallaxes \citep{GAIA2018} and calculations by \citet{BailerJones18}. 
The distance estimate for WASP-20 presented in \citet{BailerJones18} is $1383.1^{+526.1}_{-813.6}$, which does not agree with previous literature on this system.
This disagreement might be caused by confusion due the binary nature of this target. 
For that reason, we adopt the distance derived by \citet{Evans2016b} for WASP-20.}
}
\tablebib{
(1)~\citet{2012AJ....144..139H};
(2)~\citet{2015AJ....150..197H};
(3)~\citet{2015ApJ...800...59V};
(4)~\citet{2016ApJ...818...36P};
(5)~\citet{2018AJ....156...89P};
(6)~\citet{2016ApJ...827...78S};
(7)~\citet{2016AJ....152..143V};
(8)~\citet{2017AJ....153..142P};
(9)~\citet{2017MNRAS.464.2708S};
(10)~\citet{2016MNRAS.459.4281K};
(11)~\citet{2007MNRAS.375..951C};
(12)~\citet{2012MNRAS.426.1291S};
(13)~\citet{2009ApJ...690L..89H};
(14)~\citet{2010A+A...517L...1Q};
(15)~\citet{2009ApJ...703..752L};
(16)~\citet{2013MNRAS.434.1300S};
(17)~\citet{2010A+A...519A..98B};
(18)~\citet{2013A+A...557A..30C};
(19)~\citet{2010ApJ...723L..60H};
(20)~\citet{2013MNRAS.428.3680G};
(21)~\citet{2011ApJ...726L..19A};
(22)~\citet{2013A+A...549A..18T};
(23)~\citet{2013A+A...551A..73F};
(24)~\citet{2014A+A...563A.143D};
(25)~\citet{2014MNRAS.445.1114A};
(26)~\citet{2013A+A...552A.120S};
(27)~\citet{2013A+A...552A..82G};
(28)~\citet{2015AJ....150...18H};
(29)~\citet{Mancini2019};
(30)~\citet{2013A+A...551A..80T};
(31)~\citet{2014A+A...562A.126M};
(32)~\citet{Anderson2014a};
(33)~\citet{2014A+A...572A..49N};
(34)~\citet{2014MNRAS.440.1982H};
(35)~\citet{2014A+A...568A..81L};
(36)~\citet{2016MNRAS.463.3276H};
(37)~\citet{2017MNRAS.469.1622M};
(38)~\citet{2016PASP..128f4401T};
(39)~\citet{2016MNRAS.458.4025D};
(40)~\citet{2017MNRAS.465.3693H};
(41)~\citet{2017A+A...599A...3L};
(42)~\citet{arXiv181209264};
(43)~\citet{Evans2016b};
(44)~\citet{2017MNRAS.464..810B}.
}
\end{table*}

In Sect.~\ref{sec:observations} of this article we describe the observations we have carried out and in Sect.~\ref{sec:data_reduction} we explain the applied data reduction techniques.
We present the detected candidate companions (CCs), analyze the likelihood of each to be a gravitationally bound component within a multiple stellar system, and present detection limits for all targets of our sample within Sect.~\ref{sec:analysis}.
Finally, we discuss our results within the scope of previous literature in Sect.~\ref{sec:discussion} and we conclude in Sect.~\ref{sec:conclusions}.

\section{Observations}
\label{sec:observations}

Our observations (PI: D.~F.~Evans) were carried out with the Spectro-Polarimetric High-contrast Exoplanet REsearch \citep[SPHERE;][]{Beuzit2019} instrument that is mounted on the Nasmyth platform of Unit 3 telescope (UT3) at the ESO Very Large Telescope (VLT).
SPHERE is assisted by the SAXO extreme AO system \citep{Fusco2006} to obtain diffraction-limited data.
The targets were observed using the instrument's integral field spectrograph \citep[IFS,][]{Claudi2008} and the infrared dual imaging spectrograph \citep[IRDIS,][]{Dohlen2008} simultaneously.
Within the scope of this article we focus on the analysis of the IRDIS data, which provide similar inner working angle (IWA) capabilities down to 100\,mas (Wilby et al. in prep.), yet a much larger field of view up to 5\farcs5 in radial separation compared to the IFS.
IRDIS was operated in classical imaging \citep[CI,][]{Vigan2010} mode applying a broadband \textit{K}$_s$-band filter (Filter ID: \texttt{BB\_Ks}).
The filter has a bandwidth of $\Delta\lambda^{{K}_{s}}=313.5$\,nm centred around $\lambda_\text{c}^{{K}_{s}}=2181.3$\,nm.
To suppress the stellar flux, an apodized pupil Lyot coronagraph  \citep{Soummer2005,Carbillet2011,Guerri2011} was used (Coronagraph ID: \texttt{N\_ALC\_YJH\_S}).
To locate the star's position behind the coronagraphic mask, centre frames were taken alongside the science observations.
For these frames, a sinusoidal pattern was applied to the deformable mirror to create four reference spots around the star.
To perform precise photometry of potential companions, we obtained additional unsaturated, non-coronagraphic flux images of each target with a neutral density filter in place.
Furthermore, the observations in ESO period 98 were conducted in pupil stabilized imaging mode, whereas the data in period 99 were collected in field stabilized mode.
A detailed description of the observational setup and the atmospheric conditions for all observations are presented in Table \ref{tbl:observational_setup}.
\begin{table*}
\caption{
Observational setup and weather conditions.
}
\label{tbl:observational_setup}
\small
\def\arraystretch{1.3}
\setlength{\tabcolsep}{12pt}
\centering
\begin{tabular}{@{}lrrllllll@{}}
\hline\hline
Star & $V$\tablefootmark{a} & $K$\tablefootmark{b} & Observation date & Mode\tablefootmark{c} & NDITH$\times$NDIT$\times$DIT\tablefootmark{d} & $\langle\omega\rangle$\tablefootmark{e} & $\langle X\rangle$\tablefootmark{f} & $\langle\tau_0\rangle$\tablefootmark{g} \\  
& (mag) & (mag) & (yyyy-mm-dd) & & (1$\times$1$\times$s) & (\arcsec) & & (ms)\\
\hline
\object{HAT-P-41} & 11.36 &  9.73 & 2016-10-24 & P & 26$\times$4$\times$4 & 1.53 & 1.24 & 5.70 \\
\object{HAT-P-57} & 10.47 &  9.43 & 2016-10-09 & P & 16$\times$4$\times$4 & 0.61 & 1.52 & 7.61 \\
\object{HAT-P-57} & 10.47 &  9.43 & 2017-05-15 & F & 16$\times$4$\times$4 & 0.92 & 1.22 & 2.94 \\
\object{K2-02}    & 10.19 &  8.03 & 2017-05-15 & F & 16$\times$4$\times$4 & 0.99 & 1.50 & 2.77 \\
\object{K2-24}    & 11.28 &  9.18 & 2017-06-23 & F & 16$\times$4$\times$4 & 2.13 & 1.58 & 1.60 \\
\object{K2-38}    & 11.39 &  9.47 & 2017-03-06 & P & 16$\times$4$\times$4 & 0.56 & 1.01 & 7.38 \\
\object{K2-39}    & 10.83 &  8.52 & 2017-05-15 & F & 16$\times$4$\times$4 & 1.13 & 1.21 & 2.47 \\
\object{K2-99}    & 11.15 &  9.72 & 2017-08-28 & F & 16$\times$4$\times$4 & 0.66 & 1.83 & 3.17 \\
\object{KELT-10}  & 10.70 &  9.34 & 2017-05-15 & F & 16$\times$4$\times$4 & 0.96 & 1.09 & 3.14 \\
\object{WASP-2}  & 11.98 &  9.63 & 2017-05-15 & F & 16$\times$4$\times$4 & 1.04 & 1.27 & 2.38 \\
\object{WASP-7}  &  9.48 &  8.40 & 2016-10-06 & P & 16$\times$4$\times$4 & 0.69 & 1.17 & 4.90 \\
\object{WASP-8}  &  9.77 &  8.09 & 2016-10-06 & P & 16$\times$4$\times$4 & 0.69 & 1.03 & 4.82 \\
\object{WASP-16}  & 11.31 &  9.59 & 2017-03-06 & P & 16$\times$4$\times$4 & 0.52 & 1.07 & 9.41 \\
\object{WASP-20}  & 10.79  &  9.39 & 2016-10-06 & P & 16$\times$4$\times$4 & 0.94 & 1.02 & 3.20 \\
\object{WASP-21}  & 11.59 &  9.98 & 2016-10-24 & P & 16$\times$4$\times$4 & 0.94 & 1.42 & 2.84 \\
\object{WASP-29}  & 11.21 &  8.78 & 2016-10-09 & P & 16$\times$4$\times$4 & 0.46 & 1.04 & 11.84 \\
\object{WASP-30}  & 11.46 & 10.20 & 2017-05-15 & F & 16$\times$4$\times$4 & 1.05 & 1.37 & 3.12 \\
\object{WASP-54}  & 10.42 &  9.04 & 2017-03-05 & P & 16$\times$4$\times$4 & 0.57 & 1.23 & 5.81 \\
\object{WASP-68}  & 10.68 &  8.95 & 2017-06-29 & F & 16$\times$4$\times$4 & 1.41 & 1.01 & 1.78 \\
\object{WASP-69}  &  9.87 &  7.46 & 2016-10-06 & P & 12$\times$4$\times$4 & 0.69 & 1.08 & 4.90 \\
\object{WASP-70}  & 10.79 &  9.58 & 2017-05-15 & F & 16$\times$4$\times$4 & 1.28 & 1.07 & 2.49 \\
\object{WASP-71}  & 10.56 &  9.32 & 2016-11-08 & P & 16$\times$4$\times$4 & 0.76 & 1.63 & 9.40 \\
\object{WASP-72}  & 10.87 &  9.62 & 2017-07-06 & F & 16$\times$4$\times$4 & 0.95 & 1.22 & 3.41 \\
\object{WASP-73}  & 10.48 &  9.03 & 2016-10-09 & P & 26$\times$4$\times$4 & 0.56 & 1.20 & 7.79 \\
\object{WASP-74}  &  9.76 &  8.22 & 2017-06-22 & F & 16$\times$4$\times$4 & 1.07 & 1.10 & 2.23 \\
\object{WASP-76}  & 9.53  &  8.24 & 2016-11-07 & P & 16$\times$4$\times$4 & 0.81 & 1.76 & 9.40 \\
\object{WASP-80}  & 11.87 &  8.35 & 2017-06-22 & F & 16$\times$4$\times$4 & 1.25 & 1.08 & 2.56 \\
\object{WASP-87}  & 10.74 &  9.60 & 2017-04-02 & F & 16$\times$4$\times$4 & 1.74 & 1.19 & 1.54 \\
\object{WASP-88}  & 11.39 & 10.32 & 2017-05-15 & F & 16$\times$4$\times$4 & 0.93 & 1.14 & 2.90 \\
\object{WASP-94}  & 10.06 &  8.87 & 2016-10-09 & P & 16$\times$4$\times$4 & 0.54 & 1.01 & 9.59 \\
\object{WASP-95}  & 10.09 &  8.56 & 2016-10-21 & P & 16$\times$4$\times$4 & 0.84 & 1.25 & 2.78 \\
\object{WASP-97}  & 10.58 &  9.03 & 2016-10-09 & P & 16$\times$4$\times$4 & 0.47 & 1.17 & 11.72 \\
\object{WASP-99}  &  9.48 &  8.09 & 2017-07-06 & F & 16$\times$4$\times$4 & 0.78 & 1.23 & 3.32 \\
\object{WASP-108} & 11.22 &  9.80 & 2017-03-05 & P & 16$\times$4$\times$4 & 0.81 & 1.10 & 6.20 \\
\object{WASP-109} & 11.44 & 10.20 & 2017-07-23 & F & 12$\times$4$\times$4 & 1.41 & 1.62 & 2.65 \\
\object{WASP-111} & 10.26 &  9.00 & 2017-05-15 & F & 16$\times$4$\times$4 & 1.26 & 1.11 & 2.24 \\
\object{WASP-117} & 10.15 &  8.78 & 2016-10-21 & P & 16$\times$4$\times$4 & 0.78 & 1.14 & 3.37 \\
\object{WASP-118} & 11.02 &  9.79 & 2017-07-06 & F & 16$\times$4$\times$4 & 1.12 & 1.23 & 3.41 \\
\object{WASP-120} & 10.96 &  9.88 & 2016-12-20 & P & 9$\times$4$\times$4  & 0.86 & 1.07 & 7.61 \\
\object{WASP-121} & 10.52 &  9.37 & 2016-12-25 & P & 16$\times$4$\times$4 & 1.39 & 1.04 & 2.51 \\
\object{WASP-122} & 11.00 &  9.42 & 2016-12-25 & P & 16$\times$4$\times$4 & 1.41 & 1.07 & 2.26 \\
\object{WASP-123} & 11.03 &  9.36 & 2016-10-22 & P & 16$\times$4$\times$4 & 0.89 & 1.17 & 2.20 \\
\object{WASP-130} & 11.11 &  9.46 & 2017-03-11 & P & 16$\times$4$\times$4 & 0.42 & 1.20 & 11.18 \\
\object{WASP-131} & 10.08 &  8.57 & 2017-07-05 & F & 16$\times$4$\times$4 & 1.01 & 1.66 & 2.77 \\
\object{WASP-136} &  9.98 &  8.81 & 2016-10-25 & P & 16$\times$4$\times$4 & 1.45 & 1.26 & 5.70 \\
\object{WASP-137} & 10.89 &  9.46 & 2016-10-26 & P & 6$\times$4$\times$4  & 0.58 & 1.49 & 8.47 \\
\hline
\end{tabular}
\tablefoot{
\tablefoottext{a}{$V$-band apparent magnitudes are from a range of sources and are those reported in TEPCat \citep{Southworth2011}.}
\tablefoottext{b}{$K$-band system magnitudes from 2MASS \citep{Cutri2012a}.}
\tablefoottext{c}{Observation mode is either pupil (P) or field (F) stabilized}
\tablefoottext{d}{NDITH denotes the number of dithering positions, NDIT describes the number of integrations per dithering position and DIT is the detector integration time for each individual exposure.}
\tablefoottext{e}{$\langle\omega\rangle$ denotes the average seeing conditions during the observation.}
\tablefoottext{f}{$\langle X\rangle$ denotes the average airmass during the observation.}
\tablefoottext{g}{$\langle\tau_0\rangle$ denotes the average coherence time during the observation.}
}
\end{table*}

\section{Data reduction}
\label{sec:data_reduction}

The data reduction was performed using a custom processing pipeline based on the latest release of PynPoint \citep[version 0.8.1;][]{Stolker2019} that includes standard dark and flatfield calibrations.
Bad pixels were replaced by the average inside a 5$\times$5 box around the corresponding pixel.
Furthermore, we corrected for the instrumental anamorphic distortion according to the description in the SPHERE manual.
To achieve photon-noise-limited sensitivities, an accurate model of the thermal background is essential for $K_s$ band imaging.
Unfortunately, no sky images without any source in the field of view were taken alongside the science observations of the program.
We thus searched the ESO archive to find useful calibration files that were obtained with the same instrumental setup (i.e.\ exposure time, coronagraph and filter choice).
Within these constraints, we found exactly one suitable sky image taken as part of another program (PI: M.~Kenworthy, ESO ID: 0101.C-0153).
For an optimal background subtraction, we performed the sky subtraction for both sides of the detector individually.
We cropped all images around the rough position of the star in the science frames and aligned the sky images to prominent features induced by the substrate of the inserted coronagraph.
The alignment was performed using a cross-correlation in Fourier space according to \citet{Guizar2008} and \citet{Fienup1997}.
While masking a region of 0\farcs86 around the star, the aligned sky image was fitted to each individual science frame by a simple linear least squares approach.
This yielded one optimized scaling coefficient per science frame that the sky image had to be multiplied with, before the subtraction.
The sky subtraction afterwards was applied to the full frame to ensure a precise background subtraction even for the location of the star.
After sky subtraction, the science images were shifted to correct for their corresponding dither positions and centred by using the centre frames as described in the SPHERE manual.
At this stage we averaged both detector sides for each exposure to dampen noise introduced by bad pixels.
Finally, we de-rotated the data that were obtained in pupil stabilized mode according to the difference in parallactic angle.
An additional constant pupil offset of -135\fdg99 was taken into account as well.
The rotation was skipped for data that were taken in field stabilized imaging mode.
For both pupil and field stabilized data, we finally performed a correction for the true north position given by a rotation of -1\fdg75 according to \citet{Maire2016}.
No further PSF removal was performed and our final image was obtained as the median of the processed stack.

\section{Results and analysis}
\label{sec:analysis}

\subsection{Determining consistent ages for the exoplanet host stars}
\label{subsec:age_host_stars}
We used version 1.2 of the program {\tt bagemass}\footnote{\url{https://sourceforge.net/projects/bagemass/}} \citep{Maxted2015} to estimate the age of each star based on the observed values of $T_{\rm eff}$, [Fe/H] and the mean stellar density $\rho_{\star}$. These values were obtained from the references listed in Table\,\ref{tbl:star_properties}. The methods and assumptions used for the calculation of the stellar model grid using the GARSTEC stellar evolution code are described in \citet{Serenelli2013} and  \citet{Maxted2015}. We set lower limits of 80\,K on the standard error for T$_{\rm eff}$ and 0.07 dex for the standard error on [Fe/H] and assumed flat prior distributions for the stellar mass and age. The ages derived are shown in Table~\ref{tbl:star_properties}. The values and errors quoted are the median and standard deviation of the sampled posterior age distributions provided by {\tt bagemass}.

\subsection{Characterization of candidate companions}
\label{subsec:characterization_of_ccs}

\begin{figure*}
\resizebox{\hsize}{!}{\includegraphics{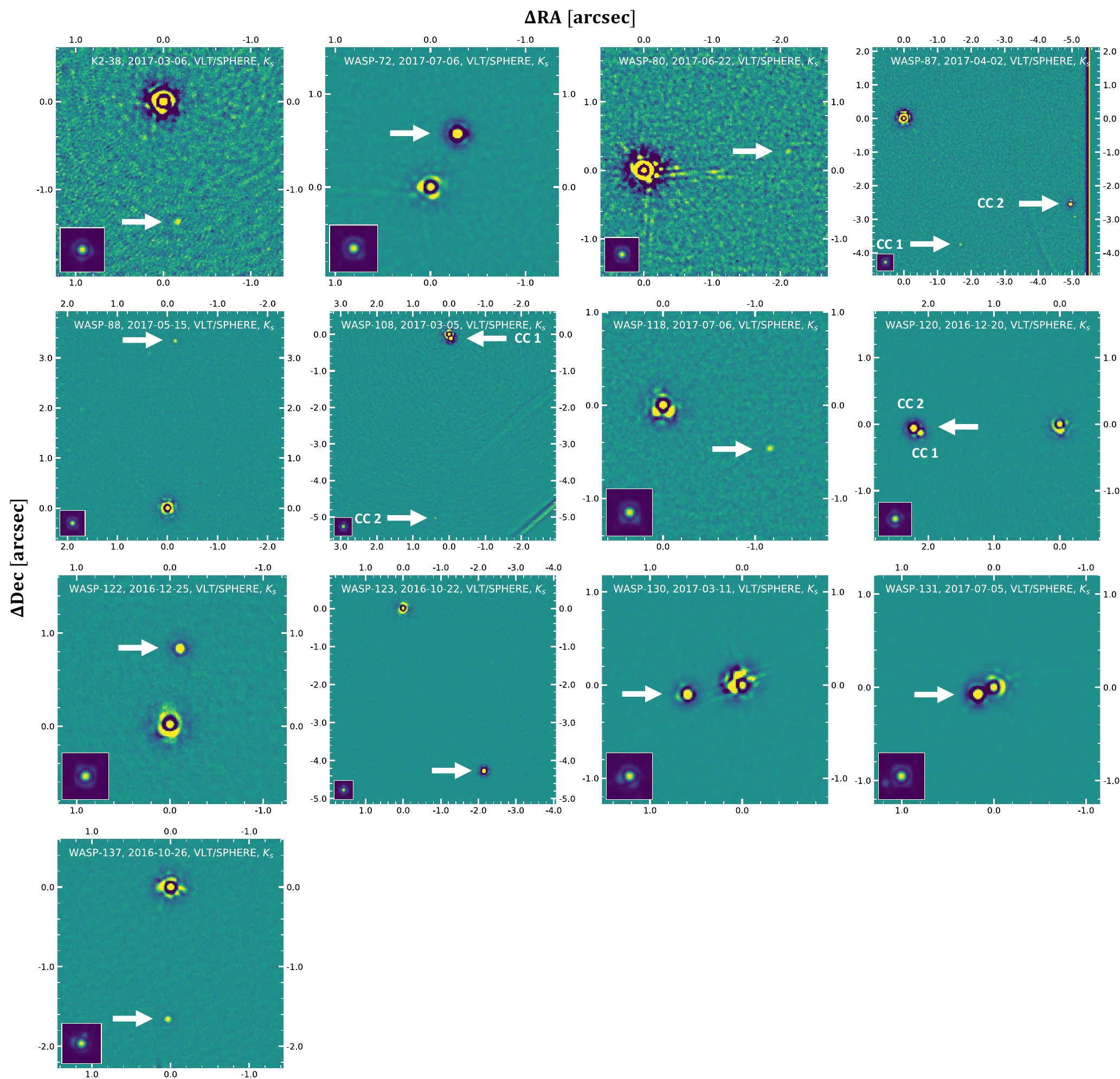}}
\caption{
Newly detected candidate companions around transiting exoplanet host stars from the SPHERE/IRDIS data.
An unsharp mask was applied to highlight point sources.
The origin of the axes is located at the position of the host star.
The images are displayed using a logarithmic scale with arbitrary offsets and stretches to highlight the candidate companions.
In all images north points up and east towards the left.
The lower left corner of each image shows the reduced non-coronagraphic flux image with the same spatial scale and field orientation.
}
\label{fig:observational_results_1}
\end{figure*}

\begin{figure*}
\resizebox{\hsize}{!}{\includegraphics{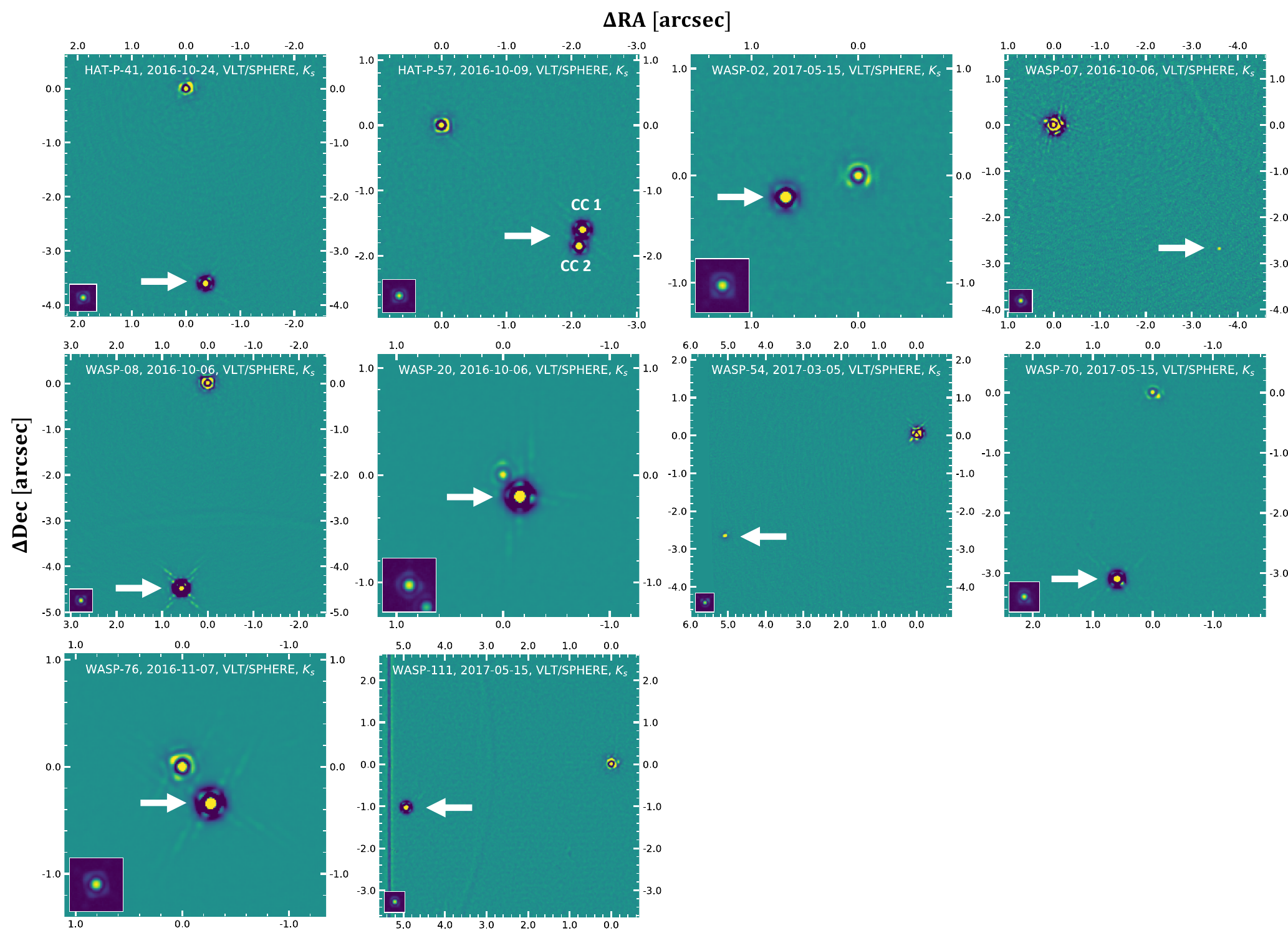}}
\caption{
Previously detected candidate companions around transiting exoplanet host stars from the SPHERE/IRDIS data.
An unsharp mask was applied to highlight point sources.
The origin of the axes is located at the position of the host star.
The images are displayed using a logarithmic scale with arbitrary offsets and stretches to highlight the candidate companions.
In all images north points up and east towards the left.
The lower left corner of each image shows the reduced non-coronagraphic flux image with the same spatial scale and field orientation.
}
\label{fig:observational_results_2}
\end{figure*}

In the IRDIS data we detected 27 off-axis point sources around 23 stars of our sample.
Compilations of these detections are presented in Fig.~\ref{fig:observational_results_1} and Fig.~\ref{fig:observational_results_2}, which show new detections by our survey and previously known sources, respectively.
Sixteen of the 27 candidate companions have not been detected by similar surveys of the multiplicity of these exoplanet host stars.
This impressively demonstrates the ability of high-contrast imaging with SPHERE.
Only 256\,s of on-target integration are sufficient to reach better sensitivities than previous surveys that have been carried out either with different AO-assisted instruments or with other observing strategies such as lucky imaging.

As we did not perform any PSF subtraction, we characterized the companions directly in the median-combined images, applying the standard astrometric solution of IRDIS with a plate scale of $12.265$\,mas in $K_s$ band.
For the astrometric characterization, we fitted a two dimensional Gaussian function to the PSF of the companion.
The magnitude contrast was estimated with aperture photometry that we applied on both flux and science image around the previously determined centroid.
We used an aperture size that is equivalent to the FWHM of the SPHERE PSF in $K_s$ band of 55\,mas and scaled the counts from the flux image to account for the difference in exposure time and applied neutral density filter.
A detailed list of all detected candidate companions including their separations, position angles (PAs), and magnitude contrasts is presented in Table~\ref{tbl:cc_astrometry_photometry}.
\begin{table*}
\caption{
Astrometry and photometry of CCs within the IRDIS field of view.
Furthermore, we present the primaries' $K$ band magnitudes corrected for the contribution of the CCs (see equation~\ref{eqn:magnitude_correction}).
}
\label{tbl:cc_astrometry_photometry}
\tiny
\def\arraystretch{1.2}
\setlength{\tabcolsep}{8pt}
\centering
\begin{tabular}{@{}lllllllllll@{}}
\hline\hline
Star & CC ID & Epoch & Separation & PA & $K_\star$ & $\Delta K$ & Status\tablefootmark{a} & $p^\mathrm{B}$ & $M$\tablefootmark{b} & $T_\mathrm{eff}$\tablefootmark{b} \\  
& &  (yyyy-mm-dd) & (\arcsec) & (\degr) & (mag) & (mag) & & (\%) & ($M_\sun$) & ($K$) \\
\hline
HAT-P-41 & 1 & 2016-10-24 & $3.621\pm0.004$ & $183.9\pm0.1$ & 9.83 & $2.50\pm0.21$ & C & - & $0.69^{+0.06}_{-0.05}$ & $4336^{+250}_{-199}$ \\
HAT-P-57 & 1 & 2016-10-09 & $2.688\pm0.004$ & $231.8\pm0.1$ & 9.55 & $2.91\pm0.05$ & C & - & $0.59^{+0.01}_{-0.01}$ & $3942^{+50}_{-37}$ \\
HAT-P-57 & 2 & 2016-10-09 & $2.807\pm0.004$ & $226.9\pm0.1$ & 9.55 & $3.47\pm0.05$ & C & - & $0.50^{+0.01}_{-0.01}$ & $3684^{+40}_{-23}$ \\
HAT-P-57 & 1 & 2017-05-15 & $2.689\pm0.004$ & $231.8\pm0.1$ & 9.55 & $2.90\pm0.12$ & C & - & $0.59^{+0.03}_{-0.03}$ & $3944^{+114}_{-75}$ \\
HAT-P-57 & 2 & 2017-05-15 & $2.809\pm0.004$ & $227.0\pm0.1$ & 9.55 & $3.45\pm0.12$ & C & - & $0.50^{+0.03}_{-0.03}$ & $3691^{+77}_{-48}$ \\
K2-38 & 1 & 2017-03-06 & $1.378\pm0.014$ & $185.2\pm0.6$ & 9.47 & $8.72\pm0.31$ & A & 1.59 & $0.07^{+0.01}_{-0.01}$ & $1699^{+150}_{-106}$ \\
WASP-2 & 1 & 2017-05-15 & $0.710\pm0.003$ & $104.9\pm0.2$ & 9.73 & $2.55\pm0.07$ & C & - & $0.40^{+0.02}_{-0.02}$ & $3523^{+28}_{-19}$ \\
WASP-7 & 1 & 2016-10-06 & $4.474\pm0.007$ & $231.5\pm0.1$ & 8.40 & $8.70\pm0.27$ & B & - & - & - \\
WASP-8 & 1 & 2016-10-06 & $4.520\pm0.005$ & $170.9\pm0.1$ & 8.09 & $2.29\pm0.08$ & C & - & $0.53^{+0.02}_{-0.02}$ & $3758^{+47}_{-43}$ \\
WASP-20 & 1 & 2016-10-06 & $0.259\pm0.003$ & $216.0\pm0.6$ & 9.79 & $0.86\pm0.06$ & A & 0.004 & $0.88^{+0.08}_{-0.07}$ & $5235^{+270}_{-275}$ \\
WASP-54 & 1 & 2017-03-05 & $5.728\pm0.006$ & $115.9\pm0.1$ & 9.04 & $5.94\pm0.06$ & C & - & $0.19^{+0.01}_{-0.01}$ & $3216^{+26}_{-25}$ \\
WASP-70 & 1 & 2017-05-15 & $3.160\pm0.004$ & $167.4\pm0.1$ & 9.85 & $1.38\pm0.18$ & C & - & $0.70^{+0.06}_{-0.05}$ & $4504^{+263}_{-213}$ \\
WASP-72 & 1 & 2017-07-06 & $0.639\pm0.003$ & $331.9\pm0.3$ & 9.67 & $3.34\pm0.06$ & A & 0.02 & $0.66^{+0.02}_{-0.02}$ & $4234^{+80}_{-81}$ \\
WASP-76 & 1 & 2016-11-07 & $0.436\pm0.003$ & $215.9\pm0.4$ & 8.37 & $2.30\pm0.05$ & C & - & $0.79^{+0.03}_{-0.03}$ & $4824^{+128}_{-132}$ \\
WASP-80 & 1 & 2017-06-22 & $2.132\pm0.010$ & $275.5\pm0.3$ & 8.35 & $9.25\pm0.28$ & A & 3.29 & $0.07^{+0.01}_{-0.01}$ & $1306^{+84}_{-53}$ \\
WASP-87 & 1 & 2017-04-02 & $4.109\pm0.016$ & $202.3\pm0.2$ & 9.56 & $8.48\pm1.19$ & A & 19.83 & $0.08^{+0.02}_{-0.01}$ & $2289^{+540}_{-621}$ \\
WASP-87 & 2 & 2017-04-02 & $5.569\pm0.007$ & $241.0\pm0.1$ & 9.56 & $5.57\pm0.70$ & B & - & - & - \\
WASP-88 & 1 & 2017-05-15 & $3.350\pm0.015$ & $355.5\pm0.5$ & 10.32 & $7.60\pm0.53$ & A & 1.65 & $0.11^{+0.03}_{-0.02}$ & $2844^{+155}_{-209}$ \\
WASP-108 & 1 & 2017-03-05 & $0.124\pm0.007$ & $203.0\pm3.3$ & 9.83 & $3.90\pm0.06$ & A & 32.82 & $0.35^{+0.02}_{-0.02}$ & $3471^{+18}_{-18}$ \\
WASP-108 & 2 & 2017-03-05 & $5.039\pm0.019$ & $174.2\pm0.2$ & 9.83 & $7.48\pm0.43$ & B & - & - & - \\
WASP-111 & 1 & 2017-05-15 & $5.039\pm0.005$ & $100.1\pm0.1$ & 9.08 & $3.01\pm0.17$ & C & - & $0.67^{+0.05}_{-0.04}$ & $4285^{+195}_{-172}$ \\
WASP-118 & 1 & 2017-07-06 & $1.251\pm0.004$ & $246.5\pm0.2$ & 9.79 & $6.73\pm0.13$ & A & 0.09 & $0.15^{+0.01}_{-0.01}$ & $3034^{+52}_{-52}$ \\
WASP-120 & 1 & 2016-12-20 & $2.124\pm0.004$ & $91.7\pm0.1$ & 9.95 & $4.44\pm0.23$ & A & 0.47 & $0.39^{+0.04}_{-0.04}$ & $3504^{+60}_{-44}$ \\
WASP-120 & 2 & 2016-12-20 & $2.221\pm0.005$ & $89.8\pm0.1$ & 9.95 & $3.27\pm0.32$ & A & 0.51 & $0.57^{+0.06}_{-0.06}$ & $3897^{+227}_{-167}$ \\
WASP-122 & 1 & 2016-12-25 & $0.837\pm0.003$ & $350.7\pm0.2$ & 9.43 & $5.09\pm0.30$ & A & 0.50 & $0.23^{+0.04}_{-0.04}$ & $3311^{+60}_{-63}$ \\
WASP-123 & 1 & 2016-10-22 & $4.786\pm0.005$ & $205.0\pm0.1$ & 9.36 & $3.47\pm0.11$ & C & - & $0.40^{+0.02}_{-0.02}$ & $3524^{+37}_{-26}$ \\
WASP-130 & 1 & 2017-03-11 & $0.600\pm0.003$ & $98.0\pm0.3$ & 9.50 & $3.73\pm0.12$ & A & 0.22 & $0.30^{+0.03}_{-0.02}$ & $3410^{+29}_{-32}$ \\
WASP-131 & 1 & 2017-07-05 & $0.189\pm0.003$ & $111.5\pm0.9$ & 8.65 & $2.82\pm0.20$ & A & 0.01 & $0.62^{+0.05}_{-0.04}$ & $4109^{+200}_{-163}$ \\
WASP-137 & 1 & 2016-10-26 & $1.660\pm0.003$ & $177.0\pm0.1$ & 9.46 & $6.20\pm0.28$ & A & 0.14 & $0.17^{+0.02}_{-0.02}$ & $3106^{+85}_{-85}$ \\
\hline
\end{tabular}
\tablefoot{
\tablefoottext{a}{
Status is either companion (C), background (B), or ambiguous (A).
The latter classification indicates that neither the background nor the companion hypothesis are confirmed by proper motion analysis at the 5$\sigma$ level.
For the ambiguous cases we also present the background probability $p^\mathrm{B}$ based on our TRILEGAL analysis (equation~\ref{eqn:background_probability}) in the next column.
}
\tablefoottext{b}{
For confirmed background objects, we do not provide masses and effective temperatures, as these parameters depend on the distance to the object, which is not known in these cases.
For all dubious cases the distances and temperatures are calculated for the case that the object is at the same distance as the primary.
}
}
\end{table*}
Furthermore, we calculated mass and temperature estimates based on the derived photometry using evolutionary models of (sub-)stellar objects \citep[e.g.][]{Allard2001,Baraffe2003}.
As various physical processes play major roles for objects of different temperatures, we used AMES-Cond, AMES-Dusty, and BT-Settl models for the characterization of candidate companions with $T_\mathrm{eff}<1400\,K$, $1400\,K<T_\mathrm{eff}<2700\,K$, and $T_\mathrm{eff}>2700\,K$, respectively.

There are three potential scenarios -- depending on the available data -- to assess the likelihood that a CC is gravitationally bound to its host:
\begin{enumerate}
    \item Gaia DR2 provides parallax and proper motion of the CC.
    \item Previous studies have detected the CC and provide astrometric measurements of it. This includes the case that Gaia DR2 only provides the position of the CC at reference epoch J2015.5, but no parallax or proper motion estimates.
    \item None of the information above is accessible.
\end{enumerate}
In the first case, the hypothesis whether the CC is bound or not could be easily tested by the provided parallaxes and proper motions of primary and CC.
For the second scenario, we tested the proper motion of the object instead and determined whether its astrometry over several epochs agrees with a co-moving companion.
In case that no other data on the CC was available, we estimated the likelihood of its companionship by a synthetic model of the stellar population around the stellar coordinates.
This analysis was performed in a similar way to that described by \citet{Dietrich2018}.
First we used TRILEGAL \citep{Girardi2005} to simulate a stellar population for 1 square degree centred around the exoplanet host star.
We chose the 2MASS $K$-band filter which is in good agreement with the actual SPHERE filter used for the observations.
The limiting magnitude provided for the simulation was based on the maximum contrast we reached around the particular target (see Sect.~\ref{subsec:detection_limits}).
Other than this, we used the default parameters of TRILEGAL v1.6.
Following the description of \citet{Lillo-Box2014}, we measured the likelihood of a CC to be a background object as
\begin{align}
\label{eqn:background_probability}
    p^\mathrm{B}=\pi r^2\rho_\mathrm{sim}\,,
\end{align}
where $\rho_\mathrm{sim}$ denotes the number of simulated stars per square degree around the exoplanet host and $r$ is the radial separation of the corresponding CC.
As this analysis is purely based on statistical arguments, we do not classify the CCs within this category as background or bound, but rather flag these as ambiguous objects, whose common proper motion needs to be confirmed by future studies.
Since we base the further analysis of these ambiguous candidates only on the derived background probabilities (see Sect.~\ref{subsec:analysis_multiplicity_rate}), this classification does not affect the derived multiplicity fractions in any way.
A detailed analysis for each detected CC is presented in the following subsections.

Most of the CCs that we detected with IRDIS are unresolved in the 2MASS catalogue \citep{Cutri2012a} that we used for calibrating the $K$ band magnitude of the host star.
In fact, only for WASP-8, WASP-111 and WASP-123 does the 2MASS catalogue provides spatially resolved flux measurements for the primary and CC.
For the remaining cases, we had to assume that the flux of potential CCs is contributing to the listed 2MASS $K$ band magnitude of the primary, but of course this contribution is negligible for large contrasts between both components.
The corrected $K$ band magnitude for primary $j$ from our sample that is hosting $n_j$ CCs with corresponding magnitude contrasts of $\Delta K_{j,\ell}$ for $\ell=1,\dots,n_j$, is
\begin{align}
    K_j = K_{\mathrm{2MASS}, j} + 2.5\log_{10}\left(1+\sum_{\ell=1}^{n_j}\left(10^{-\frac{\Delta K_{j,\ell}}{2.5}}\right)\right)\,.
    \label{eqn:magnitude_correction}
\end{align}
We applied this correction directly to the 2MASS system magnitudes that are presented in Table~\ref{tbl:observational_setup}.
The updated $K$ band magnitudes of primaries with companions that are unresolved in 2MASS photometric data, are listed in Table~\ref{tbl:cc_astrometry_photometry} instead. 

\subsubsection{HAT-P-41}
\label{subsubsec:analysis_HAT-P-41}

In the discovery paper of a transiting hot Jupiter around HAT-P-41, \citet{2012AJ....144..139H} detected a potential stellar companion south of the star.
The candidate was also detected by the lucky imaging surveys of \citet{Wollert2015a} and \citet{Wollert2015b}.
Based on stellar population synthesis models these studies concluded that the object is probably bound.
\citet{Ngo2016} also detected the candidate companion in Keck/NIRC2 $K_s$ data and their colour analysis supported the theory that HAT-P-41 is a candidate multiple stellar system.
\citet{Evans2016a} carried out an additional high-resolution imaging campaign and they determined a common proper motion with 2$\sigma$ significance.
An additional companion to the system that was also detected by \citet{Evans2016a} was ruled out at a later stage and identified as an instrumental artifact \citep{Evans2018}.
Therefore, previous studies present a lot of evidence that HAT-P-41 is actually a binary system.
A conclusive common proper motion analysis and an accurate distance determination, however, has not been published so far.

These previous results were confirmed by our SPHERE survey.
We detected exactly one off-axis point source within the IRDIS field of view at the position of the previously detected candidate companion with a separation of $3\farcs621\pm0\farcs004$ and a position angle of $183\fdg9\pm0\fdg1$.
Furthermore, this companion was also detected by the second data release of the Gaia mission \citep[Gaia DR2;][]{GAIA2018}.
\citet{BailerJones18} provided distance estimates based on the Gaia parallaxes of $348\pm4$\,pc and $338\pm4$\,pc for HAT-P-41 and the candidate companion, respectively.
Considering the reported proper motions of $(\mu_{\alpha}^A,\mu_{\delta}^A)=(-3.28\pm0.06,-6.39\pm0.04)$\,mas per year for the primary and $(\mu_{\alpha}^B,\mu_{\delta}^B)=(-3.71\pm0.05,-6.78\pm0.04)$\,mas per year for the secondary, we could conclude that both sources are co-distant and co-moving.
Thus, the former candidate companion is proven to be a stellar binary to HAT-P-41 and should be named  HAT-P-41~B accordingly.
From our comparison to BT-Settl models we derived a mass of $0.71^{+0.06}_{-0.05}$\,M$_\sun$ for the secondary component of the system.

\subsubsection{HAT-P-57}
\label{subsubsec:analysis_HAT-P-57}

We re-detected the binary pair southwest of HAT-P-57 that was already found in the discovery paper of the transiting exoplanet HAT-P-57~b \citep{2015AJ....150..197H}.
\citet{2015AJ....150..197H} already concluded that HAT-P-57~b must orbit the primary star, as the detected binary is too faint in the optical to be responsible for the measured transit depth.
Additional RV data of the system confirmed this hypothesis.
From photometric $H$ and $L$ band analysis in a colour magnitude diagram, \citet{2015AJ....150..197H} concluded that both binary components are co-evolutionary with the primary.
Consequently, they argued that all three stars form a hierarchical triple system and should be named HAT-P-57~ABC.
The masses of the smaller companions were estimated as $0.61\pm0.10$\,M$_\sun$ and $0.53\pm0.08$\,M$_\sun$.
However, no other test for actual companionship -- such as common proper motion analysis -- was performed.

\begin{figure}
\resizebox{\hsize}{!}{\includegraphics{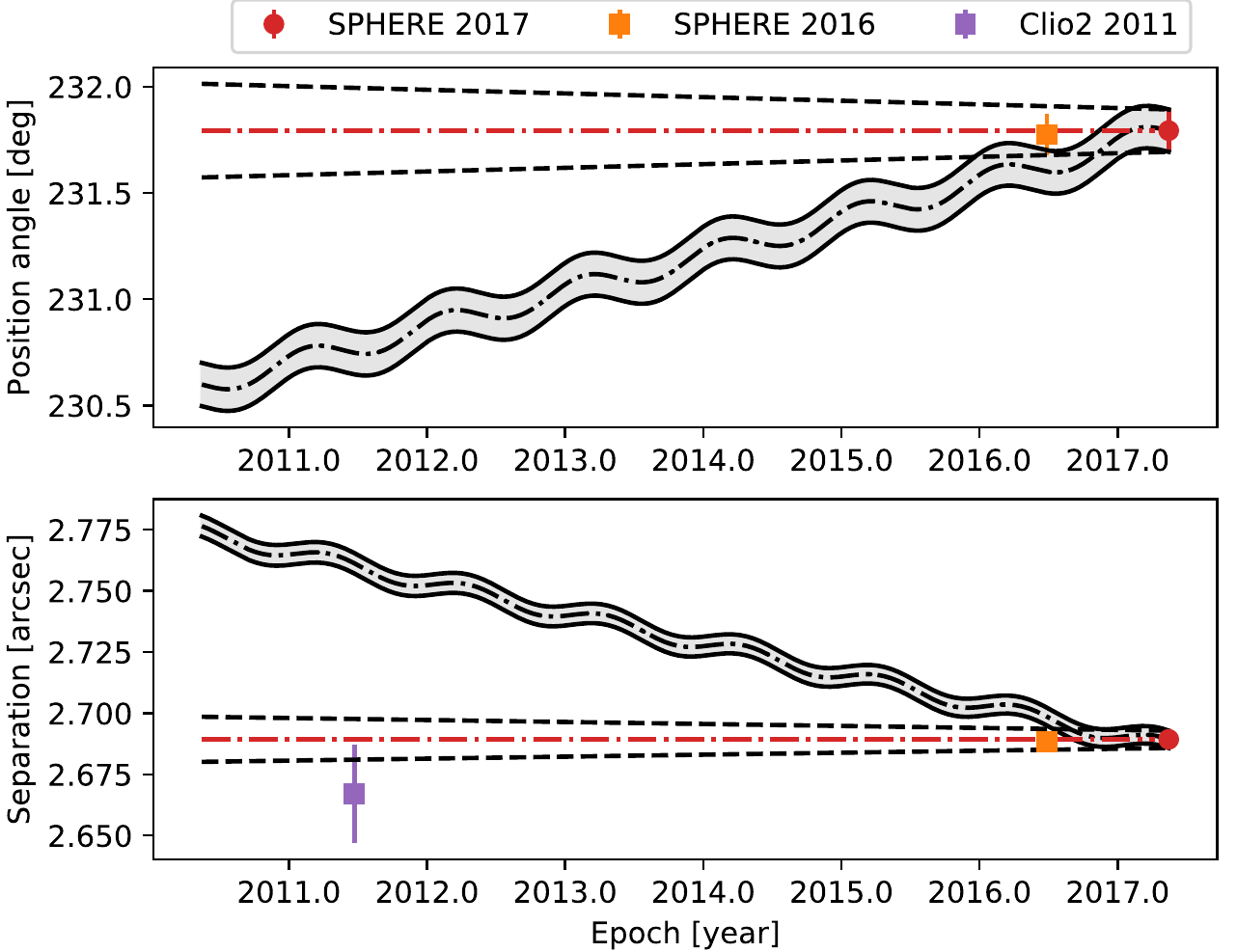}}
\caption{
Proper motion analysis of CC 1 and 2 detected around HAT-P-57.
PA and separation are evaluated individually.
The dashed cone presents the expected position of a gravitationally bound companion considering potential orbital motion of the object.
The grey trajectory represents the expected location of a stationary background object, instead.
For the MMT/Clio2 data we adopted the separation measurement presented \citet{2015AJ....150..197H}; 
no PA of the source at this epoch is provided.
}
\label{fig:HAT-P-57_CPM}
\end{figure}

With the two SPHERE epochs, we aimed to perform such an analysis.
\citet{2015AJ....150..197H} only provided a separation of $2\farcs667\pm0\farcs001$ from the primary to the binary pair and a separation of $0\farcs225\pm0\farcs002$ between both components of the binary itself.
No individual separations from the primary to each component of the binary and no PAs were presented in their article.
For that reason, we considered the binary pair as a single component and we performed the proper motion test by splitting up the evaluation of separation and PA.
The results of this analysis are visualized in Fig~\ref{fig:HAT-P-57_CPM}.
Already the two SPHERE epochs imply that the binary agrees better with the hypothesis of being bound to HAT-P-57 than with being an unrelated background object.
The additional separation measurement adapted from \citet{2015AJ....150..197H} that is based on MMT/Clio2 data from June 22, 2011, confirmed this hypothesis.
As their presented uncertainty in separation, only 1\,mas, seemed to be very optimistic -- as the primary is heavily saturated -- we adjusted this value to 20\,mas to also account for the difference in separation of both CCs.
This analysis proved that the binary pair is clearly incompatible with a stationary background object at more than 5$\sigma$ significance.
Therefore, CC 1 and CC 2 should be named HAT-P-57~B and HAT-P-57~C, respectively.

From the $K_s$-band photometry, we derived masses of $0.60^{+0.02}_{-0.01}$\,M$_\sun$ and $0.51^{+0.01}_{-0.01}$\,M$_\sun$ for components B and C, respectively.
Furthermore, we measured separations of $0\farcs260\pm0\farcs004$ and $0\farcs261\pm0\farcs004$ as well as PAs of $168\fdg3\pm0\fdg1$ and $168\fdg4\pm0\fdg1$ between components B and C for the SPHERE epochs.
This is compatible with the increasing trend in separation when additionally considering the separation of $0\farcs225\pm0\farcs002$ between both components in 2011 \citep{2015AJ....150..197H}.
For a conclusive orbital motion fit of these two objects, a detailed analysis and another epoch at high astrometric precision are required, which is beyond the scope of the current work.

\subsubsection{K2-38}
\label{subsubsec:analysis_K2-38}

\citet{Evans2018} reported a potential companion around K2-38 at a separation of $10\farcs7752\pm0\farcs0950$, which is unfortunately outside the IRDIS field of view.
The potential companion, however, was picked up by Gaia DR2 and, together with two additional sources listed -- but already considered unlikely to be bound -- by \citet{Evans2018}, these three objects were clearly proven to be background based on their parallaxes.

In our SPHERE data we detected a previously-unknown CC south of the star at a separation of $1\farcs378\pm0\farcs014$.
As no other astrometric data of this CC is available, we estimated its likelihood to be a background object using TRILEGAL.
This provided a probability of 1.59\% that the candidate is a background object.

\subsubsection{WASP-2}
\label{subsubsec:analysis_WASP-2}

In addition to the detection of the hot Jupiter WASP-2~b, \citet{2007MNRAS.375..951C} also reported a potential stellar companion to WASP-2~b at a separation of 0\farcs7 and a magnitude contrast of $\Delta H=2.7$\,mag.
This companion was detected by several follow-up surveys \citep{Daemgen2009,Bergfors2013,Adams2013,Ngo2015,Wollert2015a} and photometric analysis suggests a spectral type of late K to early M dwarf.
The most recent astrometric measurements by \citet{Evans2016a} proved a common proper motion of the companion with its host at more than 5$\sigma$ significance.
Furthermore, they detected a linearly decreasing separation between the stellar companion and the primary implying a nearly edge-on orbital solution, which we could confirm with our data.

\subsubsection{WASP-7}
\label{subsubsec:analysis_WASP-7}

\begin{figure}
\resizebox{\hsize}{!}{\includegraphics{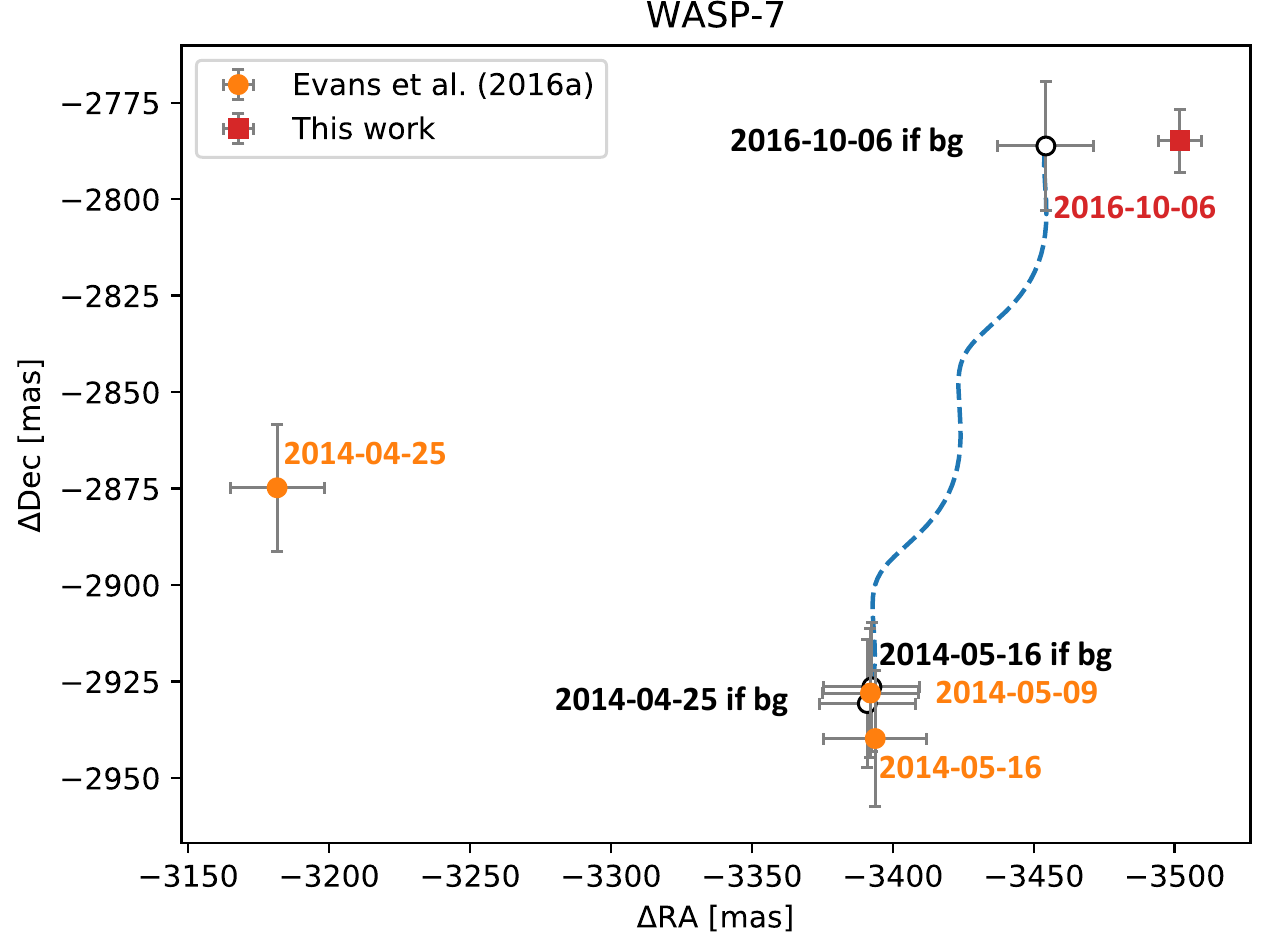}}
\caption{
Proper motion analysis of CC 1 around WASP-7.
The dashed blue line represents the trajectory of a static background (bg) object.
}
\label{fig:WASP-7_CPM}
\end{figure}

\citet{Evans2016a} reported a candidate companion around WASP-7 at a separation of $4\farcs414\pm0\farcs011$ and a PA of $228\fdg73\pm0\fdg12$.
However, no extensive analysis was performed whether this candidate is actually bound to the exoplanet host star.
The separation and PA presented in \citet{Evans2016a} are an average of three individual epochs obtained on April 25, May 9, and May 16, 2014.
As presented in Fig.~\ref{fig:WASP-7_CPM} the astrometry based on the data from April 25, 2014, does not agree with the two later epochs.
Instead of averaging over all three datapoints, we used the data from May 9, 2014, as baseline for a further proper motion analysis\footnote{
Note that we present the common proper motion tests in a plot that displays the candidate's differential offsets in Right Ascension and Declination to the host, henceforth.
Using one datapoint as baseline, we simulate the trajectory of a static background object based on the exoplanet host star's parallax and proper motion.
Several measurements of the candidate companion's astrometry help to discern whether it is orbiting the primary or a background contaminant.
}.

We also detected the candidate in our IRDIS data with a separation of $4\farcs474\pm0\farcs007$ at a PA of $231\fdg51\pm0\fdg11$.
Including this new epoch in a proper motion analysis, as presented in Fig.~\ref{fig:WASP-7_CPM}, clearly showed that the object better agrees with the background trajectory than with being a bound companion.

\subsubsection{WASP-8}
\label{subsubsec:analysis_WASP-8}

We re-detected WASP-8~B south of the primary at a separation $4\farcs520\pm0\farcs005$ and with a PA of $170\fdg9\pm0\fdg1$.
This stellar companion was first detected by \citet{2010A+A...517L...1Q} who classified it as an M-type dwarf.
Further studies by \citet{Ngo2015} and \citet{Evans2016a} confirmed the companionship status by common proper motion at more than 5$\sigma$ significance.
This was consolidated by additional Gaia DR2 astrometric measurements, which provide parallaxes of $11.09\pm0.04$\,mas and $11.02\pm0.04$\,mas as well as proper motions of $(\mu_{\alpha}^A,\mu_{\delta}^A)=(109.75\pm0.06,7.61\pm0.06)$\,mas per year $(\mu_{\alpha}^B,\mu_{\delta}^B)=(110.26\pm0.06,5.57\pm0.06)$\,mas per year for primary A and secondary B, respectively.

\subsubsection{WASP-20}
\label{subsubsec:analysis_WASP-20}

Using the same SPHERE data as presented in this article, \citet{Evans2016b} reported the detection of a bright, close-in binary to WASP-20. 
Our new evaluation of these data showed, however, that the companion's position angle given in \citet{Evans2016b} is not correct. 
We found this to be because \citet{Evans2016b} treated the data as being collected in field stabilized imaging mode, whereas it was actually obtained in pupil-stabilized mode. 
Our new analysis of the data yielded measurements of the separation and magnitude contrast that agree within the uncertainties with the values derived by \citet{Evans2016b}; 
the correct position angle of WASP-20~B is $216\fdg0\pm0\fdg6$.

Furthermore, we inferred a slightly higher effective temperature estimate for WASP-20~B that is, however, consistent within the uncertainties with the value of $5060\pm250\,$K as presented in \citet{Evans2016b}.
This discrepancy can be explained by the ATLAS9 \citep{Castelli1994} models used by \citet{Evans2016b} in comparison to the more recent BT-Settl models that we used instead.
Unfortunately, no precise parallax measurement of the host was provided by Gaia DR2 -- probably due to the binary nature of the system.
This resulted in the rather large uncertainties in effective temperature as presented in Table~\ref{tbl:cc_astrometry_photometry}, which may be constrained by better distance estimates based on future Gaia data releases.

As the object was only observed in a single epoch, \citet{Evans2016b} could not perform any assessment of common proper motion.
Furthermore, the CC is not detected in Gaia DR2, so we evaluated the companionship with TRILEGAL instead.
This analysis provided a probability of 0.004\% for the CC to be a background contaminant.

\subsubsection{WASP-54}
\label{subsubsec:analysis_WASP-54}

A companion candidate around WASP-54 was first detected by \citet{Evans2016a}.
Further proper motion analysis presented in \citet{Evans2018} led to the preliminary conclusion that the object is a bound companion.
The authors, however, state that additional measurements are required to confirm this hypothesis.

\begin{figure}
\resizebox{\hsize}{!}{\includegraphics{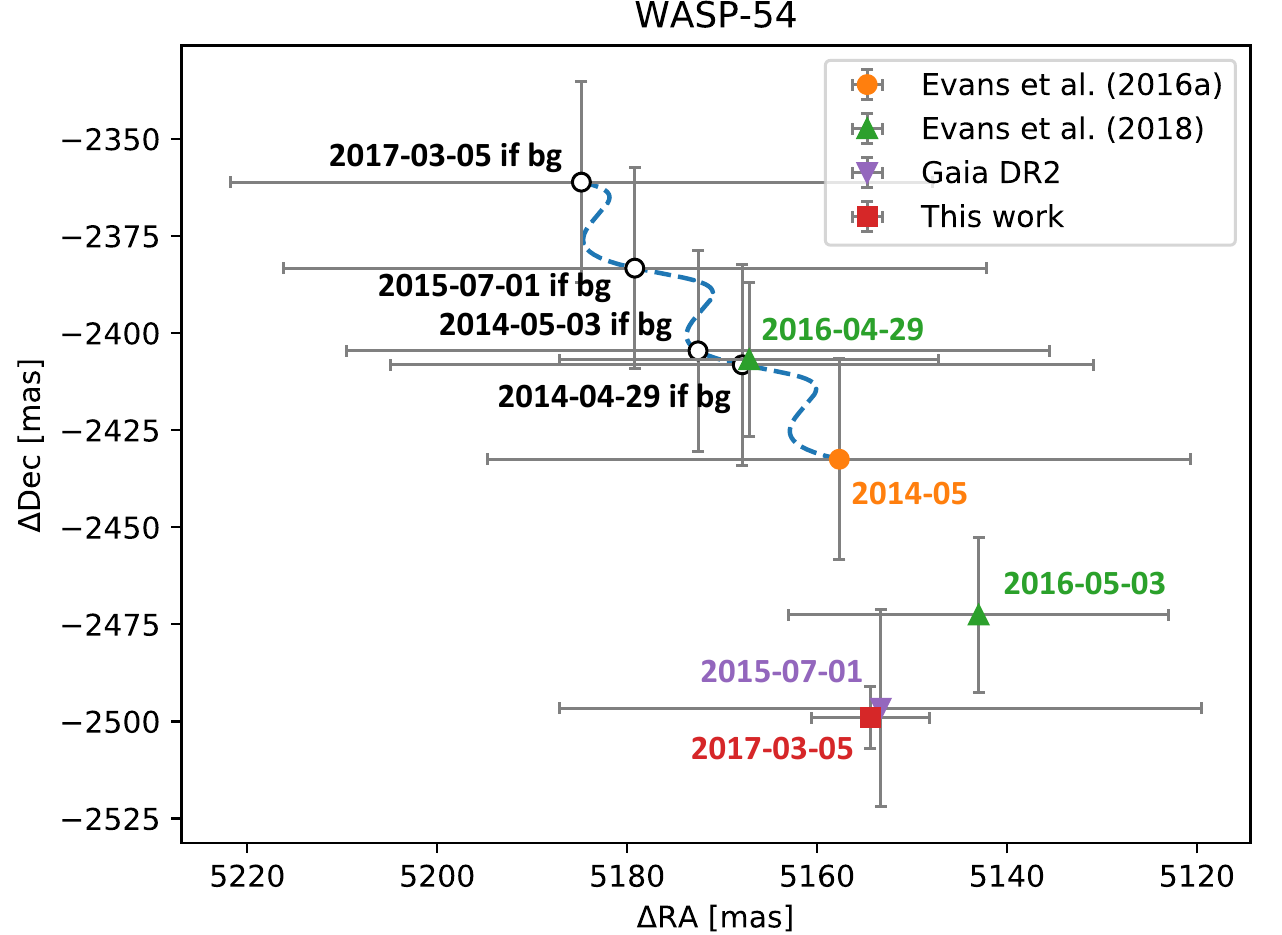}}
\caption{
Proper motion analysis of CC 1 around WASP-54.
The first measurement from \citet{Evans2016a} (orange circle) is the average of four individual epochs, collected from May 6 until May 8, 2014.
The dashed blue line represents the trajectory of a static background (bg) object.
}
\label{fig:WASP-54_CPM}
\end{figure}

We combined the data presented in \citet{Evans2016a} and \citet{Evans2018} with the latest SPHERE epoch and additional astrometric data from Gaia DR2.
The latter only provided coordinates of the CC and no proper motion that could be used for confirming its companionship.
In Fig.~\ref{fig:WASP-54_CPM} we analysed these data in a proper motion diagram.
The data presented in \citet{Evans2016a} consist of five individual epochs obtained around May, 2014.
The individual measurements had an intrinsic scatter larger than the provided uncertainties.
For that reason, we averaged the single measurements using the standard deviation of the datapoints as an uncertainty of the combined measurement.
One of these datapoints, obtained on 2014 April 18, deviated by more than 3$\sigma$ from the average of the remaining measurements.
We thus removed this datapoint from our combined astrometry solution for this first epoch.

In \citet{Evans2018} two additional epochs, 2015 April 29 and 2016 May 3, were presented.
As shown in Fig.~\ref{fig:WASP-54_CPM} the first of these epochs agrees well with the expected position of a static background object.
The second epoch, however, assigns the companion a position in the opposite direction as expected from a background object.
Because both epochs do not agree within their uncertainties, it is likely that the results of \citet{Evans2018} were subject to a source of systematic error not accounted for in the quoted uncertainties.

No clear conclusion could be drawn from these data alone, but adding Gaia and our latest SPHERE measurements facilitated an unambiguous classification of the potential companion.
Both additional datapoints were not compatible with the trajectory of a static background object but are consistent with a co-moving companion.
Therefore, we conclude that WASP-54~B is actually a stellar binary to WASP~54~A.
From our $K_s$-band photometry we derived a mass of $0.19^{+0.01}_{-0.01}$\,M$_\sun$.

\subsubsection{WASP-68}
\label{subsubsec:analysis_WASP-68}

CC 1 presented in \citet{Evans2018}, at a separation of approximately 13\farcs1 and with a position angle of 119\fdg7, was confirmed as a co-moving stellar companion by Gaia DR2 parallaxes of $4.39\pm0.03$\,mas and $4.19\pm0.15$\,mas for primary and secondary, respectively.
Additional proper motion measurements of $(\mu_{\alpha}^A,\mu_{\delta}^A)=(-11.17\pm0.06,-6.21\pm0.04)$\,mas per year $(\mu_{\alpha}^B,\mu_{\delta}^B)=(-11.45\pm0.24,-6.24\pm0.17)$\,mas per year strengthened the claim that the CC is actually WASP-68~B, a stellar companion to WASP-68~A.
However, we did not detect any CCs around WASP-68 within the IRDIS field of view.

\subsubsection{WASP-70}
\label{subsubsec:analysis_WASP-70}

A K3 stellar companion was found to exoplanet host WASP-70 by \citet{2014MNRAS.445.1114A} and we also detected the object in our SPHERE data.
Previous studies \citep[e.g.][]{Wollert2015b,Evans2016a,Evans2018} stated common proper motion of the companion at 5$\sigma$ significance.
This was also confirmed by Gaia DR2, which provided parallaxes of $4.47\pm0.06$\,mas and $4.35\pm0.03$\,mas as well as proper motions of $(\mu_{\alpha}^A,\mu_{\delta}^A)=(33.24\pm0.08,-30.04\pm0.05)$\,mas per year $(\mu_{\alpha}^B,\mu_{\delta}^B)=(44.77\pm0.05,-30.11\pm0.03)$\,mas per year.
From our $K_s$-band photometry we derived a mass of $0.70^{+0.06}_{-0.05}$\,M$_\sun$ for WASP-70~B.

\subsubsection{WASP-72}
\label{subsubsec:analysis_WASP-72}

We detected a candidate companion to WASP-72 at a separation of $0\farcs639\pm0\farcs003$ and a position angle of $331\fdg9\pm0\fdg3$ that was previously unknown.
By stellar population synthesis models we derived a probability of 0.02\% that the CC is an unassociated background or foreground object.
For the case of confirmed common proper motion, we calculated a mass estimate of $0.66^{+0.02}_{-0.02}$\,M$_\sun$.

\subsubsection{WASP-76}
\label{subsubsec:analysis_WASP-76}

\begin{figure}
\resizebox{\hsize}{!}{\includegraphics{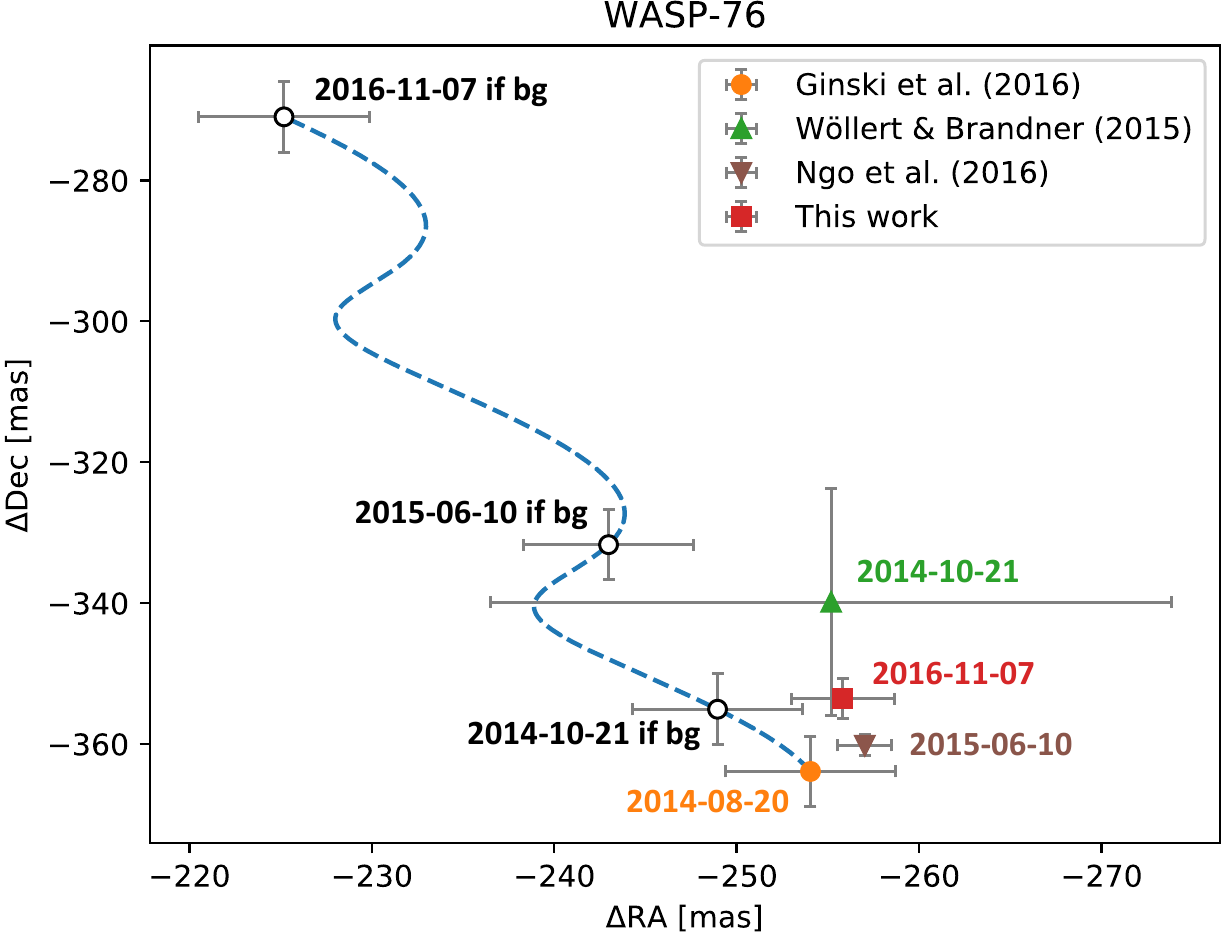}}
\caption{
Proper motion analysis of CC 1 around WASP-76.
The dashed blue line represents the trajectory of a static background (bg) object.
}
\label{fig:WASP-76_CPM}
\end{figure}

We re-detected the stellar candidate companion to WASP-76 that was first detected by \citet{Wollert2015b}.
Follow-up studies led by \citet{Ginski2016a} and \citet{Ngo2016} suggested that the companion shows common proper motions with its host.
We confirmed this trend with our additional SPHERE epoch as presented in Fig.~\ref{fig:WASP-76_CPM}; 
a background object could be ruled out at 5$\sigma$ significance.
For the stellar companion WASP-76~B we estimated a mass of $0.78^{+0.03}_{-0.03}$\,M$_\sun$ based on our $K_s$-band photometry.

\subsubsection{WASP-80}
\label{subsubsec:analysis_WASP-80}

We report the detection of a new candidate companion around WASP-80 at a separation of $2\farcs132\pm0\farcs010$ and a position angle of $275\fdg5\pm0\fdg3$.
Although the system was explored by previous studies of \citet{Wollert2015b}, \citet{Evans2016a}, and \citet{Evans2018} no candidate companions were revealed by these programs.
This is in good agreement with the large magnitude contrast of $9.25\pm0.28$\,mag at which we detected the companion just above the noise level. This is below the detection threshold of previous surveys, which explains why it remained previously undetected.
From our TRILEGAL analysis we derived a probability of 3.29\% that the CC is not associated with WASP-80.
Assuming the object is gravitationally bound to the exoplanet host we estimated a mass of $0.07^{+0.01}_{-0.01}$\,M$_\sun$ based on the $K_s$ magnitude.

\subsubsection{WASP-87}
\label{subsubsec:analysis_WASP-87}

In the discovery paper reporting a hot Jupiter around WASP-87, \citet{Anderson2014a} also detected a potential stellar companion south east of the star at a separation of 8\farcs2.
\citet{Evans2018} suggested that the proper motion analysis presented in \citet{Anderson2014a} based on UCAC4 data \citep{Zacharias2013} is not supported by other catalogues.
Based on its colour, \citet{Evans2018} concluded that the two components are nevertheless bound.
This assumption was confirmed by Gaia DR2 parallaxes of $3.32\pm0.04$\,mas and $3.19\pm0.04$\,mas for WASP-87~A and WASP-87~B, respectively.
Furthermore, the proper motions of $(\mu_{\alpha}^A,\mu_{\delta}^A)=(-1.36\pm0.06,3.92\pm0.04)$\,mas per year and $(\mu_{\alpha}^B,\mu_{\delta}^B)=(-1.73\pm0.04,4.20\pm0.04)$\,mas per year were absolutely compatible with a gravitationally bound binary system.

\begin{figure}
\resizebox{\hsize}{!}{\includegraphics{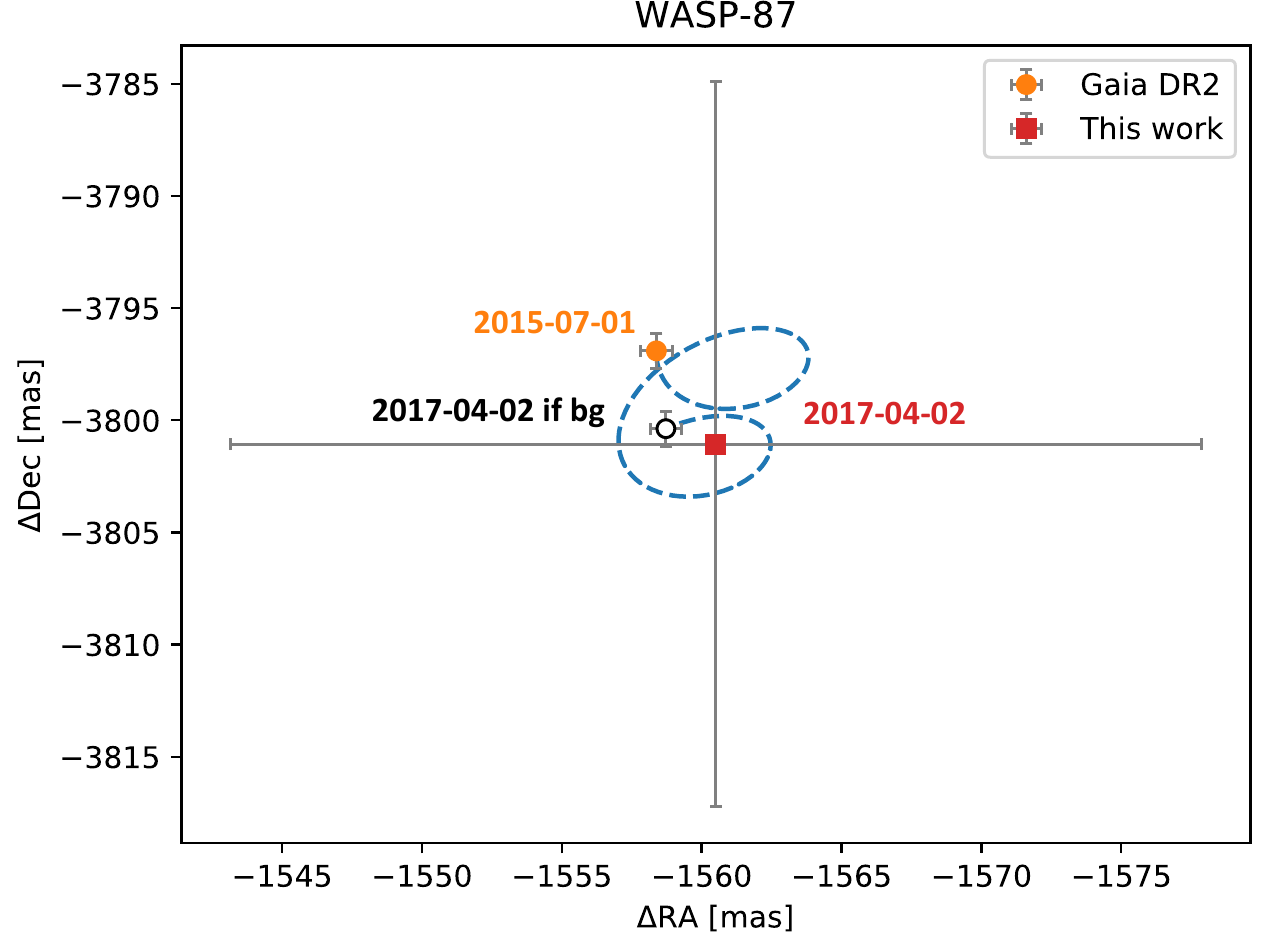}}
\caption{
Proper motion analysis of CC 1 around WASP-87.
The dashed blue line represents the trajectory of a static background (bg) object.
}
\label{fig:WASP-87_CPM}
\end{figure}

Within the IRDIS field of view, we detected two additional point sources southeast of the star.
Both were also detected by Gaia DR2, but only for CC 2 the catalogue provided a parallax estimate, whereas for CC 1 just the celestial position was measured.
From the parallax measurement of $0.02\pm0.14$\,mas for CC 2 we could clearly confirm this object as a background source.
As for CC 1 only the position was provided by Gaia DR2 we could perform a common proper motion analysis as presented in Fig.~\ref{fig:WASP-87_CPM}.
This analysis placed CC 1 close to the expected position of a stationary background object.
Due to the large magnitude contrast of CC 1, however, the SPHERE detection was only marginal.
Therefore, the derived astrometric precision had too large uncertainties to either confirm CC 1 as a co-moving companion or to show that it is a background object.
Our TRILEGAL analysis provided a probability of 19.83\% that CC 1 is not associated to WASP-87.

\subsubsection{WASP-88}
\label{subsubsec:analysis_WASP-88}

We report the detection of a new candidate companion north of WASP-88.
It is rather faint with a magnitude contrast of $7.60\pm0.53$\,mag.
From our stellar population synthesis model analysis, we derived a probability of 1.65\% that this CC is a background object rather than bound to WASP-88.

\subsubsection{WASP-108}
\label{subsubsec:analysis_WASP-108}

The system was explored within the scope of one previous multiplicity study of exoplanet host stars \citep{Evans2018}.
These authors reported several CCs, however only two of them have colours consistent with being bound to the planet host star.
As WASP-108 lies within a crowded field, \citet{Evans2018} did not rule out the possibility that both sources are background stars.
Instead they explicitly stated the necessity of additional tests.
\citet{Evans2018} estimated that the first object at 19\farcs4563 to the northeast is likely to be background, based on differing proper motion from the host reported in UCAC4, NOMAD, and PPMXL catalogues.
This was confirmed by the latest Gaia astrometry that provided a parallax of $0.18\pm0.03$\,mas in contradiction to the measured value for WASP-108 itself of $3.84\pm0.05$\,mas.
For the second CC discussed by \citet{Evans2018} no proper motion data were available by the time of their analysis.
The latest Gaia astrometry proved that the object is in good agreement with a co-moving companion.
\citet{GAIA2018} reported a parallax of $2.93\pm0.47$\,mas for the companion to which we will refer as WASP-108~B henceforth.
Also the proper motions of $(\mu_{\alpha}^A,\mu_{\delta}^A)=(25.80\pm0.13,-22.57\pm0.08)$\,mas per year and $(\mu_{\alpha}^B,\mu_{\delta}^B)=(24.76\pm0.97,-21.13\pm0.69)$\,mas per year confirmed the hypothesis of a gravitationally bound binary.

\begin{figure}
\resizebox{\hsize}{!}{\includegraphics{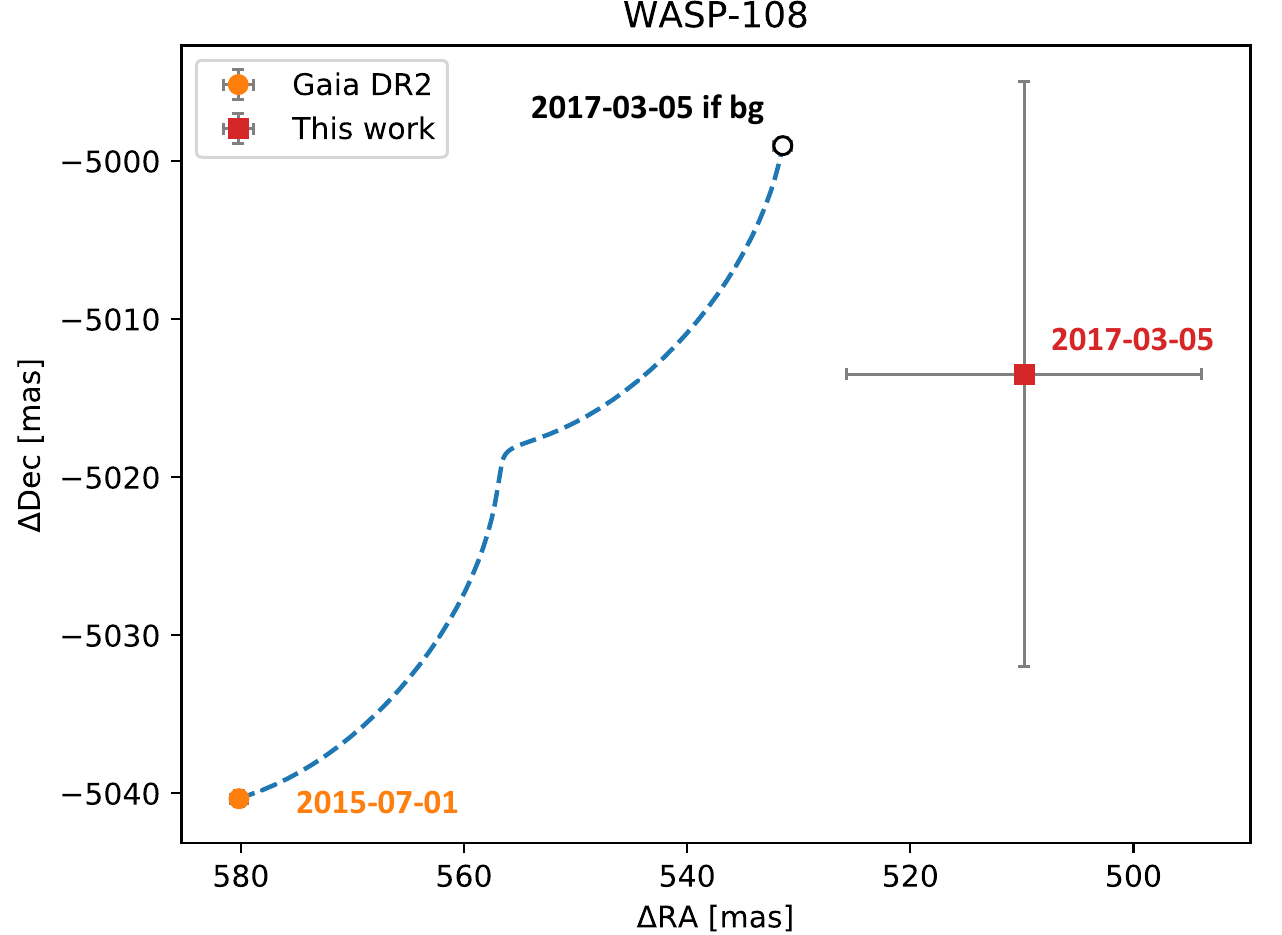}}
\caption{
Proper motion analysis of CC 2 around WASP-108.
The dashed blue line represents the trajectory of a static background (bg) object.
}
\label{fig:WASP-108_CPM}
\end{figure}

In addition, we found two CCs within the IRDIS field of view.
CC 1 is very close to WASP-108 at a magnitude contrast of $\Delta K_s=3.90\pm0.06$\,mag.
Due to its proximity it is likely to be gravitationally bound to the primary.
This agrees very well with our TRILEGAL analysis that provided a probability of 0.02\% that CC 1 is rather an unrelated background or foreground contaminant.
The second CC in the IRDIS data was detected south of the star at a separation of $5\farcs039\pm0\farcs005$.
We performed a proper motion check based on Gaia DR2 and our SPHERE data as presented in Fig.~\ref{fig:WASP-108_CPM}.
This analysis indicated that CC 2 is compatible with a background object that has a non-zero proper motion;
this hypothesis was supported by a background probability of 32.82\% based on our TRILEGAL analysis.
Due to the large uncertainties in the SPHERE astrometry, however, further tests are necessary to confirm this theory.

\subsubsection{WASP-111}
\label{subsubsec:analysis_WASP-111}

In the IRDIS data we re-detected the companion that was first identified by \citet{Evans2018} east of WASP-111 at a separation of $5\farcs039\pm0\farcs005$.
Gaia DR2 data confirmed that the companion is bound as WASP-111~A and WASP-111~B were measured to be co-moving with $(\mu_{\alpha}^A,\mu_{\delta}^A)=(12.88\pm0.10,-4.31\pm0.11)$\,mas per year and $(\mu_{\alpha}^B,\mu_{\delta}^B)=(13.35\pm0.10,-5.15\pm0.10)$\,mas per year and co-distant with parallaxes of $3.33\pm0.07$\,mas and $3.39\pm0.07$\,mas.

\subsubsection{WASP-118}
\label{subsubsec:analysis_WASP-118}

We detected a new CC around WASP-118 at a separation of $1\farcs251\pm0\farcs004$ and with a position angle of $246\fdg5\pm0\fdg2$.
TRILEGAL analysis provided a probability of 0.09\% that this CC is not associated to WASP-118.
For the case that the CC is actually gravitationally bound to the host, we derived a mass of $0.15^{+0.01}_{-0.01}\,$M$_\sun$.

\subsubsection{WASP-120}
\label{subsubsec:analysis_WASP-120}

The IRDIS data revealed a potential binary companion east of WASP-120 at a separation of approximately 2\farcs2.
Our simulated stellar population around the position of the primary predicted background probabilities of 0.47\% and 0.51\% for CC 1 and 2, respectively.
This supports the hypothesis that WASP-120 is a hierarchical triple system WASP-120~ABC.
Further astrometric measurements are required to confirm this theory.

\subsubsection{WASP-122}
\label{subsubsec:analysis_WASP-122}

We detected a new CC north of WASP-122 at a separation of approximately $0\farcs8$.
The TRILEGAL analysis yielded a probability of 0.50\% that this CC is not associated with the exoplanet host star.
We derived a mass estimate of $0.23^{+0.04}_{-0.04}$\,M$_\sun$, for the case that the CC is actually co-moving with WASP-122.

\subsubsection{WASP-123}
\label{subsubsec:analysis_WASP-123}

\begin{figure}
\resizebox{\hsize}{!}{\includegraphics{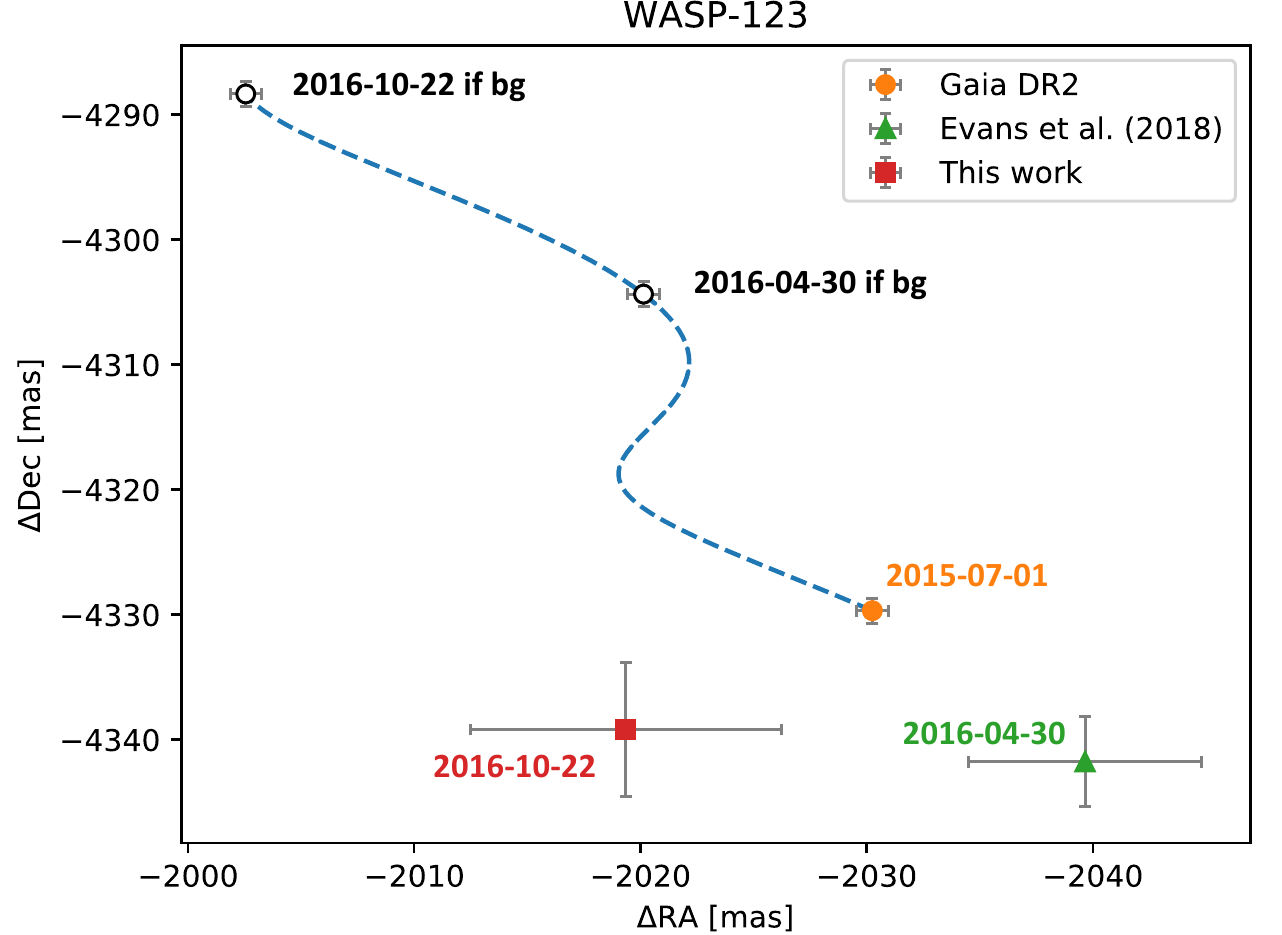}}
\caption{
Proper motion analysis of CC 1 around WASP-123.
The dashed blue line represents the trajectory of a static background (bg) object.
}
\label{fig:WASP-123_CPM}
\end{figure}

\citet{Evans2018} detected a CC south of WASP-123 at a separation of $4\farcs8$ that is marginally consistent with a bound object based on its colour.
But a conclusive result whether this companion is actually co-moving was not presented.
By combining the data from \citet{Evans2018}, Gaia DR2 astrometry, and our IRDIS data we analysed the proper motion of the CC as presented in Fig.~\ref{fig:WASP-123_CPM}.
This clearly demonstrated that the CC is not compatible with a stationary background object with a significance greater than 5$\sigma$.
Therefore, we conclude that the CC is actually WASP-123~B, a stellar companion to WASP-123~A with a mass of approximately $0.40^{+0.02}_{-0.02}$\,M$_\sun$.

\subsubsection{WASP-130}
\label{subsubsec:analysis_WASP-130}

We detected a bright CC east of WASP-130 at a separation of $0\farcs6$.
Although the target was also included in previous exoplanet host star multiplicity surveys, no companion was detected by any of these \citep{Evans2018}.
The TRILEGAL analysis yielded a probability of 0.22\% that this CC is a background or foreground contaminant.
Assuming the object is gravitationally bound to WASP-130, we derived a mass estimate of $0.30^{+0.03}_{-0.02}$\,M$_\sun$.

\subsubsection{WASP-131}
\label{subsubsec:analysis_WASP-131}

We detected a very close-in CC to WASP-131 at a separation of $0\farcs189\pm0.003$ and with a position angle of $111\fdg5\pm0\fdg9$ that had not been detected by any previous surveys.
Due to the proximity and no other objects in the field of view, it is very likely to orbit the primary.
This assumption is in good agreement with a background probability of only 0.01\%, based on our synthetic stellar population models around the host.
If confirmed, WASP-131~B, would be a stellar companion with a mass of $0.62^{+0.05}_{-0.04}$\,M$_\sun$.

\subsubsection{WASP-137}
\label{subsubsec:analysis_WASP-137}

We report the first detection of a CC south of WASP-137.
Our TRILEGAL analysis suggested a probability of only 0.14\% that this object is not associated with the exoplanet host.
From the $K_s$-band photometry we estimated a mass of $0.17^{+0.02}_{-0.02}$\,M$_\sun$ for the CC, assuming it is gravitationally associated.

\subsubsection{Non-detection of confirmed companions}
\label{subsubsec:analysis_non_detections}

As the IRDIS field of view is limited to approximately 5\farcs5 in radial separation, there are some companions to stars from our sample that were not detected within the scope of this survey.
These confirmed multiple systems are K2-02 \citep{2015ApJ...800...59V,Evans2018} and WASP-94 \citep{2014A+A...572A..49N,Evans2018}.
Furthermore, we could confirm previous candidate companions outside the IRDIS field of view around WASP-68 \citep[][and section~\ref{subsubsec:analysis_WASP-68} of this work]{Evans2018}, WASP-87 \citep[][and section~\ref{subsubsec:analysis_WASP-87} of this work]{Evans2018}, and WASP-108 \citep[][and section~\ref{subsubsec:analysis_WASP-108} of this work]{Evans2018} as actual co-moving companions based on Gaia DR2 astrometry.

\subsection{Multiplicity rate}
\label{subsec:analysis_multiplicity_rate}

For our sample of 45 observed exoplanet host stars, we reported nine targets (HAT-P-41, HAT-P-57, WASP-2, WASP-8, WASP-54, WASP-70, WASP-76, WASP-111, WASP-123) which harbour at least one companion within the IRDIS field of view that shows clear common proper motion with the primary from several epochs of observations.
Furthermore, five additional stars from the sample were confirmed multiple systems with binary components lying outside the IRDIS field of view:
the confirmation of these binaries was either performed by previous studies (K2-2, WASP-94) or by evaluation of Gaia DR2 astrometric measurements for former candidate companions within this work (WASP-68, WASP-87, WASP-108).
In addition we found 12 systems that show ambiguous candidate companions, where future checks to prove common proper motion at 5$\sigma$ significance are necessary\footnote{Note that WASP-87 and WASP-108 -- though harbouring CCs within the IRDIS field of view -- are already proven to be multiple systems with companions at farther separations (see Sections~\ref{subsubsec:analysis_WASP-87} and \ref{subsubsec:analysis_WASP-108}).} (K2-38, WASP-20, WASP-72, WASP-80, WASP-87, WASP-88, WASP-118, WASP-120, WASP-122, WASP-130, WASP-131, WASP-137).

We simulated the stellar multiplicity rate of the exoplanet host stars in our sample as
\begin{align}
    \eta_i = \frac{1}{N}\sum^{N}_{j=1}\left(\bigvee^{n_j}_{k=1} B_{ijk}(n=1,p^\mathrm{C}_{jk})\right)\;,
\end{align}
where $i$ describes the index of the simulation (to be repeated $10^6$ times), $N$ denotes the sample size of 45 exoplanet host stars, $n_j$ is the number of CCs around target $j$, and $B_{ijk}$ describes a draw from a binomial distribution with $n=1$ and $p^\mathrm{C}_{jk}$, where the latter refers to the probability that CC $k$ around target $j$ is actually bound to its host.
CCs that were confirmed to be gravitationally bound (labelled 'C' in Table~\ref{tbl:cc_astrometry_photometry} plus five additional confirmed companions outside the IRDIS field of view) were assigned $p^\mathrm{C}=1$\,.
Targets without any CCs -- or CCs that were proven to be background -- were assigned $p^\mathrm{C}=0$, accordingly.
The remaining ambiguous cases were assigned $p^\mathrm{C}=1-p^\mathrm{B}$, with $p^\mathrm{B}$ denoting the previously determined probability of being a background contaminant based on our TRILEGAL analysis (equation~\ref{eqn:background_probability}).

The outcome of $B_{ijk}$ is either 0 or 1, so we calculated the logical disjunction over all CCs of an individual target to simulate whether this host is part of a multiple system or not.
Making $10^6$ independent draws for each CC and accounting for the sample size of $N=45$ resulted in a multiplicity rate of $55.4^{+5.9}_{-9.4}\,\%$.
The uncertainties were obtained as the 68\% confidence level around the average of the simulated $\eta_i$.
However, this analysis only addresses the statistical errors that might occur due to our inconclusive characterization of some CCs and the limited size of the sample.
Of course there might be other intrinsic biases caused by sample selection, or size of the used field of view, that were not considered in this multiplicity estimate.

\subsection{Detection limits}
\label{subsec:detection_limits}

To assess the sensitivity we achieved around each target as a function of angular separation we estimated the contrast in our reduced IRDIS images.
For this purpose we used the non-coronagraphic flux frames and fitted a two-dimensional Gaussian function to the unsaturated PSF.
We took the best-fit amplitude of this function as an estimate of the stellar flux and scaled it to account for exposure time difference to the science images and attenuation by potential neutral density filters.
The noise was estimated directly from the post-processed coronagraphic images in radial annuli with a width of 55\,mas.
The annuli were centred around the position of the star behind the coronagraphic mask and we chose 100 discrete steps of equidistant radii -- growing from the inner working angle of approximately 100\,mas (Wilby et al. in prep.) up to the edge of the detector.
Afterwards, we determined the standard deviation inside each annulus to obtain an estimate for the noise as a function of separation.

\begin{figure}
\resizebox{\hsize}{!}{\includegraphics{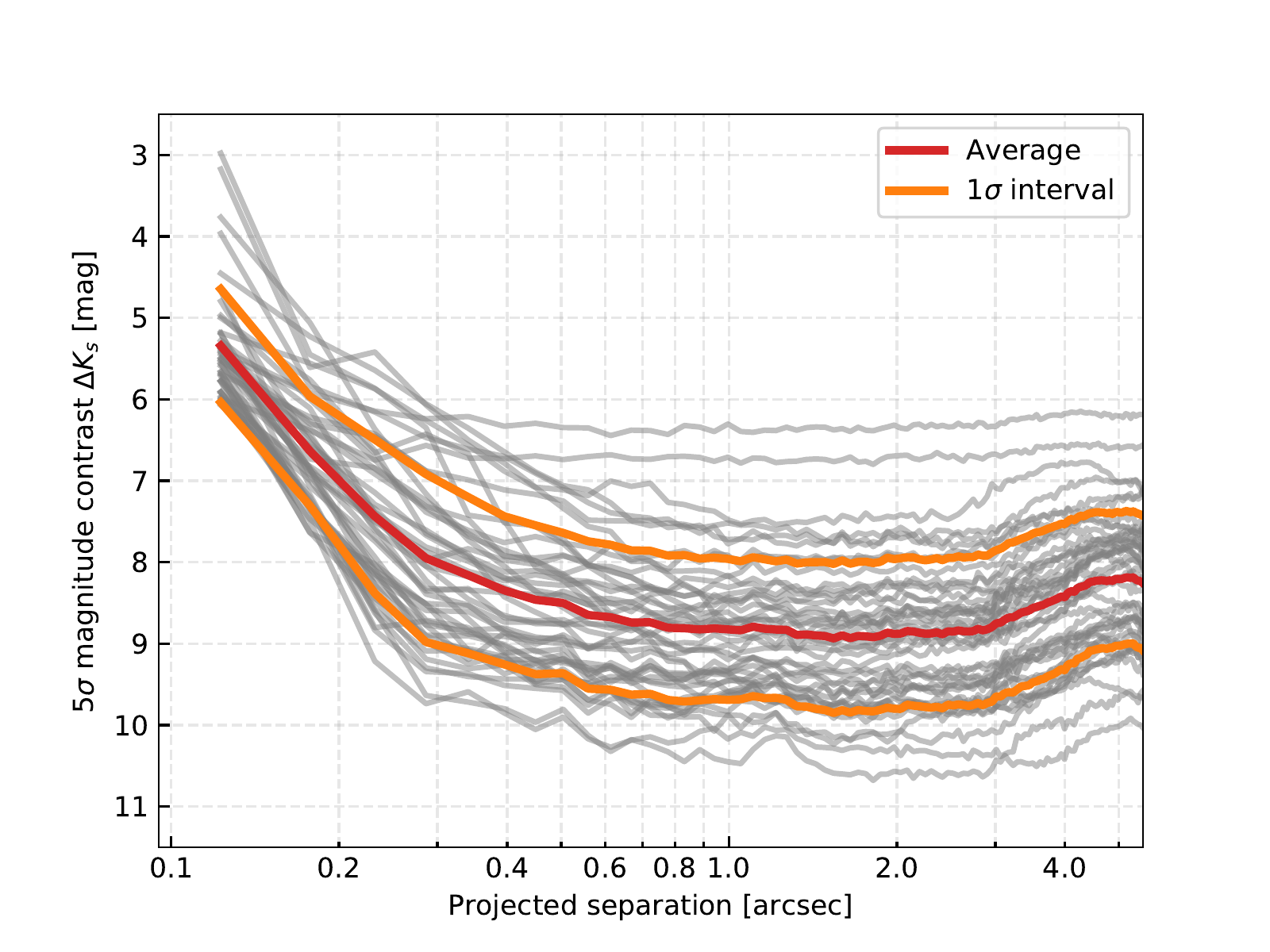}}
\caption{
Detection limits of our SPHERE survey for detection of stellar companions to known exoplanet host stars.
The grey lines represent all individual targets and epochs as presented in Table~\ref{tbl:observational_setup} and the red curve and orange curves indicate the average contrast performance and the corresponding 1$\sigma$ interval.
}
\label{fig:detection_limits_all}
\end{figure}

For HAT-P-57, where two epochs of the target were obtained, we continued analysing just the slightly deeper contrast that was obtained on the night of October 9, 2016.
The 5$\sigma$ detection limits for all datasets are presented in Fig.~\ref{fig:detection_limits_all}.
The spread in contrast performance between different datasets can be explained by the strongly varying atmospheric conditions for different observations of the programme as presented in Table~\ref{tbl:observational_setup}.
On average we reached a magnitude contrast of $7.0\pm0.8$\,mag at a separation of 200\,mas and we were background limited with an average magnitude contrast of $8.9\pm0.9$\,mag at separations larger than 1\arcsec.
Due to the missing sky frames and the imperfect background subtraction, a slight decrease of the contrast performance was observed for all datasets.
This was the case for separations larger than 3\arcsec and the strength of the effect in of the order of half a magnitude.
The detailed contrast performance for each individual target evaluated at discrete separations of 0\farcs2, 0\farcs5, 1\farcs0, 2\farcs0, and 5\farcs0 is presented in Table~\ref{tbl:detection_limits}.
\begin{table*}
\caption{
Contrast performance for all datasets evaluated at discrete separations.
The magnitude contrast is converted to a mass limit using AMES-Cond, AMES-Dusty, and BT-Settl models as described in Sect.~\ref{subsec:characterization_of_ccs}.
}
\label{tbl:detection_limits}
\tiny
\def\arraystretch{1.2}
\setlength{\tabcolsep}{10pt}
\centering
\begin{tabular}{@{}lllllllllll@{}}
\hline\hline
Star & \multicolumn{2}{c}{Contrast at 0\farcs2} & \multicolumn{2}{c}{Contrast at 0\farcs5} & \multicolumn{2}{c}{Contrast at 1\farcs0} & \multicolumn{2}{c}{Contrast at 2\farcs5} & \multicolumn{2}{c}{Contrast at 5\farcs0} \\  
& (mag) & (M$_\mathrm{Jup}$) & (mag) & (M$_\mathrm{Jup}$) & (mag) & (M$_\mathrm{Jup}$) & (mag) & (M$_\mathrm{Jup}$) & (mag) & (M$_\mathrm{Jup}$) \\
\hline
HAT-P-41 & $6.59\pm0.12$ & $154^{+8}_{-8}$ & $8.16\pm0.08$ & $89^{+1}_{-1}$ & $8.62\pm0.07$ & $83^{+1}_{-1}$ & $8.61\pm0.06$ & $83^{+1}_{-1}$ & $7.74\pm0.05$ & $97^{+1}_{-1}$ \\
HAT-P-57 & $7.91\pm0.11$ & $90^{+7}_{-2}$ & $9.00\pm0.07$ & $73^{+17}_{-3}$ & $9.02\pm0.06$ & $73^{+16}_{-3}$ & $8.79\pm0.04$ & $76^{+13}_{-3}$ & $8.07\pm0.04$ & $87^{+6}_{-1}$ \\
K2-02 & $5.62\pm0.11$ & $99^{+2}_{-2}$ & $8.21\pm0.07$ & $72^{+5}_{-1}$ & $9.36\pm0.05$ & $69^{+11}_{-1}$ & $9.84\pm0.03$ & $70^{+15}_{-3}$ & $9.18\pm0.02$ & $69^{+9}_{-1}$ \\
K2-24 & $5.97\pm0.11$ & $133^{+8}_{-8}$ & $7.07\pm0.08$ & $91^{+1}_{-1}$ & $7.77\pm0.06$ & $82^{+1}_{-1}$ & $7.69\pm0.05$ & $83^{+1}_{-1}$ & $7.01\pm0.04$ & $92^{+1}_{-1}$ \\
K2-38 & $7.66\pm0.11$ & $83^{+4}_{-1}$ & $8.90\pm0.08$ & $71^{+12}_{-2}$ & $8.96\pm0.06$ & $71^{+12}_{-2}$ & $8.95\pm0.05$ & $71^{+12}_{-2}$ & $8.18\pm0.04$ & $78^{+6}_{-1}$ \\
K2-39 & $7.75\pm0.11$ & $144^{+8}_{-8}$ & $9.19\pm0.07$ & $88^{+1}_{-1}$ & $9.45\pm0.05$ & $85^{+1}_{-1}$ & $9.51\pm0.03$ & $84^{+1}_{-1}$ & $9.03\pm0.02$ & $91^{+1}_{-1}$ \\
K2-99 & $5.40\pm0.11$ & $393^{+26}_{-27}$ & $7.05\pm0.07$ & $185^{+7}_{-7}$ & $7.72\pm0.05$ & $142^{+6}_{-6}$ & $7.63\pm0.04$ & $147^{+5}_{-5}$ & $7.20\pm0.03$ & $175^{+5}_{-5}$ \\
KELT-10 & $7.67\pm0.11$ & $84^{+2}_{-1}$ & $8.77\pm0.07$ & $73^{+7}_{-1}$ & $8.78\pm0.05$ & $73^{+7}_{-1}$ & $8.79\pm0.04$ & $73^{+7}_{-1}$ & $8.25\pm0.03$ & $78^{+4}_{-1}$ \\
WASP-2 & $6.51\pm0.11$ & $89^{+1}_{-2}$ & $7.92\pm0.08$ & $75^{+1}_{-1}$ & $8.17\pm0.06$ & $74^{+1}_{-1}$ & $8.24\pm0.05$ & $74^{+1}_{-1}$ & $7.71\pm0.04$ & $77^{+1}_{-1}$ \\
WASP-7 & $7.37\pm0.11$ & $98^{+1}_{-3}$ & $9.44\pm0.07$ & $71^{+3}_{-1}$ & $9.82\pm0.05$ & $68^{+4}_{-2}$ & $9.86\pm0.03$ & $68^{+4}_{-2}$ & $9.19\pm0.03$ & $74^{+2}_{-1}$ \\
WASP-8 & $8.21\pm0.11$ & $75^{+8}_{-1}$ & $9.92\pm0.07$ & $66^{+18}_{-3}$ & $10.16\pm0.05$ & $65^{+18}_{-3}$ & $10.36\pm0.03$ & $63^{+18}_{-4}$ & $9.67\pm0.02$ & $67^{+16}_{-2}$ \\
WASP-16 & $7.35\pm0.12$ & $85^{+1}_{-1}$ & $8.71\pm0.08$ & $74^{+1}_{-1}$ & $8.82\pm0.06$ & $74^{+1}_{-1}$ & $8.71\pm0.05$ & $74^{+1}_{-1}$ & $7.99\pm0.05$ & $79^{+1}_{-1}$ \\
WASP-20 & $6.24\pm0.12$ & $147^{+19}_{-18}$ & $7.70\pm0.08$ & $89^{+3}_{-4}$ & $8.43\pm0.07$ & $81^{+2}_{-2}$ & $8.56\pm0.06$ & $80^{+3}_{-1}$ & $7.87\pm0.06$ & $87^{+3}_{-3}$ \\
WASP-21 & $6.37\pm0.13$ & $115^{+9}_{-9}$ & $7.24\pm0.10$ & $90^{+1}_{-2}$ & $7.52\pm0.09$ & $86^{+1}_{-1}$ & $7.79\pm0.08$ & $83^{+1}_{-1}$ & $7.47\pm0.08$ & $87^{+1}_{-1}$ \\
WASP-29 & $7.98\pm0.11$ & $74^{+1}_{-1}$ & $9.53\pm0.07$ & $69^{+1}_{-1}$ & $9.66\pm0.05$ & $69^{+2}_{-1}$ & $9.77\pm0.03$ & $74^{+2}_{-1}$ & $9.00\pm0.03$ & $71^{+1}_{-1}$ \\
WASP-30 & $6.01\pm0.13$ & $167^{+11}_{-11}$ & $6.34\pm0.10$ & $146^{+9}_{-9}$ & $6.31\pm0.09$ & $148^{+8}_{-8}$ & $6.34\pm0.08$ & $146^{+8}_{-8}$ & $6.20\pm0.07$ & $155^{+8}_{-8}$ \\
WASP-54 & $6.65\pm0.11$ & $153^{+9}_{-9}$ & $9.19\pm0.07$ & $78^{+1}_{-1}$ & $9.31\pm0.06$ & $77^{+1}_{-1}$ & $9.31\pm0.04$ & $77^{+1}_{-1}$ & $8.54\pm0.04$ & $84^{+1}_{-1}$ \\
WASP-68 & $6.81\pm0.11$ & $134^{+7}_{-7}$ & $8.40\pm0.07$ & $84^{+1}_{-1}$ & $8.73\pm0.05$ & $81^{+1}_{-1}$ & $8.63\pm0.03$ & $82^{+1}_{-1}$ & $7.97\pm0.03$ & $90^{+1}_{-1}$ \\
WASP-69 & $7.85\pm0.11$ & $74^{+1}_{-1}$ & $9.82\pm0.07$ & $74^{+1}_{-4}$ & $10.45\pm0.05$ & $72^{+1}_{-1}$ & $10.59\pm0.03$ & $71^{+1}_{-1}$ & $10.01\pm0.02$ & $73^{+1}_{-1}$ \\
WASP-70 & $7.04\pm0.11$ & $95^{+2}_{-1}$ & $7.93\pm0.08$ & $82^{+1}_{-1}$ & $7.93\pm0.06$ & $82^{+1}_{-1}$ & $7.87\pm0.05$ & $83^{+1}_{-1}$ & $7.57\pm0.04$ & $86^{+1}_{-1}$ \\
WASP-71 & $7.75\pm0.11$ & $116^{+9}_{-9}$ & $9.14\pm0.07$ & $83^{+1}_{-1}$ & $9.32\pm0.05$ & $81^{+1}_{-1}$ & $8.92\pm0.03$ & $86^{+1}_{-1}$ & $8.22\pm0.03$ & $97^{+1}_{-1}$ \\
WASP-72 & $7.16\pm0.11$ & $160^{+9}_{-9}$ & $8.21\pm0.07$ & $100^{+2}_{-2}$ & $8.46\pm0.06$ & $95^{+2}_{-1}$ & $8.41\pm0.04$ & $97^{+1}_{-1}$ & $7.83\pm0.04$ & $117^{+5}_{-5}$ \\
WASP-73 & $7.78\pm0.11$ & $114^{+8}_{-8}$ & $9.44\pm0.07$ & $81^{+1}_{-1}$ & $9.60\pm0.05$ & $79^{+1}_{-1}$ & $9.49\pm0.04$ & $80^{+1}_{-1}$ & $8.61\pm0.03$ & $91^{+1}_{-1}$ \\
WASP-74 & $7.43\pm0.11$ & $96^{+1}_{-3}$ & $9.21\pm0.07$ & $75^{+1}_{-1}$ & $9.59\pm0.05$ & $73^{+1}_{-1}$ & $9.72\pm0.03$ & $72^{+1}_{-1}$ & $9.13\pm0.02$ & $76^{+1}_{-1}$ \\
WASP-76 & $7.74\pm0.11$ & $102^{+4}_{-8}$ & $8.92\pm0.07$ & $83^{+1}_{-1}$ & $9.88\pm0.05$ & $74^{+1}_{-1}$ & $9.84\pm0.03$ & $74^{+1}_{-1}$ & $9.12\pm0.03$ & $81^{+1}_{-1}$ \\
WASP-80 & $7.13\pm0.11$ & $74^{+1}_{-1}$ & $8.76\pm0.07$ & $69^{+1}_{-1}$ & $9.40\pm0.05$ & $72^{+4}_{-1}$ & $9.55\pm0.03$ & $72^{+4}_{-1}$ & $9.00\pm0.02$ & $74^{+2}_{-1}$ \\
WASP-87 & $6.31\pm0.12$ & $166^{+9}_{-9}$ & $7.76\pm0.09$ & $94^{+2}_{-2}$ & $7.94\pm0.07$ & $91^{+1}_{-1}$ & $7.93\pm0.06$ & $91^{+1}_{-1}$ & $7.21\pm0.06$ & $109^{+4}_{-5}$ \\
WASP-88 & $6.53\pm0.12$ & $180^{+9}_{-9}$ & $6.74\pm0.09$ & $168^{+7}_{-7}$ & $6.72\pm0.07$ & $169^{+6}_{-6}$ & $6.70\pm0.06$ & $170^{+6}_{-6}$ & $6.59\pm0.06$ & $177^{+5}_{-5}$ \\
WASP-94 & $7.97\pm0.11$ & $89^{+1}_{-2}$ & $9.57\pm0.07$ & $73^{+1}_{-1}$ & $9.63\pm0.05$ & $73^{+1}_{-1}$ & $9.67\pm0.03$ & $73^{+1}_{-1}$ & $8.81\pm0.03$ & $80^{+1}_{-1}$ \\
WASP-95 & $7.45\pm0.11$ & $88^{+1}_{-1}$ & $9.33\pm0.07$ & $73^{+2}_{-1}$ & $9.58\pm0.05$ & $72^{+2}_{-1}$ & $9.74\pm0.04$ & $71^{+2}_{-1}$ & $8.91\pm0.03$ & $74^{+1}_{-1}$ \\
WASP-97 & $7.78\pm0.11$ & $81^{+1}_{-1}$ & $9.40\pm0.07$ & $71^{+4}_{-1}$ & $9.37\pm0.05$ & $71^{+4}_{-1}$ & $9.40\pm0.04$ & $71^{+4}_{-1}$ & $8.76\pm0.03$ & $74^{+2}_{-1}$ \\
WASP-99 & $7.65\pm0.11$ & $97^{+1}_{-2}$ & $9.39\pm0.07$ & $76^{+1}_{-1}$ & $9.89\pm0.05$ & $72^{+1}_{-1}$ & $10.13\pm0.03$ & $71^{+2}_{-1}$ & $9.55\pm0.02$ & $74^{+1}_{-1}$ \\
WASP-108 & $7.19\pm0.12$ & $95^{+2}_{-1}$ & $8.41\pm0.09$ & $79^{+1}_{-1}$ & $8.48\pm0.08$ & $79^{+1}_{-1}$ & $8.51\pm0.07$ & $78^{+1}_{-1}$ & $7.73\pm0.06$ & $86^{+1}_{-1}$ \\
WASP-109 & $5.53\pm0.11$ & $196^{+9}_{-9}$ & $7.11\pm0.08$ & $101^{+2}_{-2}$ & $7.48\pm0.06$ & $93^{+1}_{-2}$ & $7.46\pm0.05$ & $94^{+1}_{-1}$ & $7.02\pm0.05$ & $103^{+1}_{-2}$ \\
WASP-111 & $7.80\pm0.11$ & $102^{+3}_{-6}$ & $8.95\pm0.07$ & $83^{+1}_{-1}$ & $9.19\pm0.05$ & $81^{+1}_{-1}$ & $9.08\pm0.04$ & $82^{+1}_{-1}$ & $8.50\pm0.03$ & $89^{+1}_{-1}$ \\
WASP-117 & $7.35\pm0.11$ & $91^{+2}_{-2}$ & $9.23\pm0.07$ & $73^{+1}_{-1}$ & $9.46\pm0.05$ & $72^{+1}_{-1}$ & $9.52\pm0.04$ & $72^{+1}_{-1}$ & $8.75\pm0.03$ & $76^{+1}_{-1}$ \\
WASP-118 & $6.85\pm0.11$ & $149^{+11}_{-11}$ & $8.14\pm0.08$ & $92^{+2}_{-2}$ & $8.17\pm0.06$ & $92^{+2}_{-2}$ & $8.08\pm0.05$ & $93^{+2}_{-2}$ & $7.55\pm0.04$ & $104^{+2}_{-6}$ \\
WASP-120 & $6.92\pm0.12$ & $140^{+8}_{-8}$ & $8.37\pm0.08$ & $87^{+1}_{-1}$ & $8.39\pm0.07$ & $87^{+1}_{-1}$ & $8.32\pm0.06$ & $88^{+1}_{-1}$ & $7.66\pm0.05$ & $100^{+1}_{-1}$ \\
WASP-121 & $6.78\pm0.11$ & $134^{+8}_{-8}$ & $8.13\pm0.08$ & $87^{+2}_{-1}$ & $8.87\pm0.06$ & $79^{+3}_{-1}$ & $8.92\pm0.05$ & $78^{+3}_{-1}$ & $7.88\pm0.05$ & $91^{+1}_{-1}$ \\
WASP-122 & $6.30\pm0.12$ & $150^{+7}_{-7}$ & $7.75\pm0.08$ & $90^{+1}_{-1}$ & $8.74\pm0.06$ & $79^{+1}_{-1}$ & $8.84\pm0.05$ & $78^{+1}_{-1}$ & $7.83\pm0.05$ & $88^{+1}_{-1}$ \\
WASP-123 & $5.69\pm0.11$ & $160^{+8}_{-8}$ & $7.32\pm0.07$ & $89^{+1}_{-1}$ & $8.51\pm0.06$ & $77^{+1}_{-1}$ & $8.58\pm0.04$ & $77^{+1}_{-1}$ & $7.86\pm0.04$ & $83^{+1}_{-1}$ \\
WASP-130 & $5.60\pm0.12$ & $141^{+8}_{-8}$ & $8.54\pm0.08$ & $73^{+14}_{-1}$ & $8.75\pm0.07$ & $71^{+16}_{-1}$ & $8.64\pm0.06$ & $72^{+15}_{-1}$ & $7.78\pm0.05$ & $80^{+6}_{-1}$ \\
WASP-131 & $6.12\pm0.11$ & $184^{+8}_{-8}$ & $8.75\pm0.07$ & $82^{+1}_{-1}$ & $9.12\pm0.05$ & $79^{+1}_{-1}$ & $9.32\pm0.03$ & $77^{+1}_{-1}$ & $8.74\pm0.02$ & $82^{+1}_{-1}$ \\
WASP-136 & $7.00\pm0.11$ & $158^{+9}_{-9}$ & $9.07\pm0.07$ & $83^{+1}_{-1}$ & $9.34\pm0.06$ & $81^{+1}_{-1}$ & $9.41\pm0.04$ & $80^{+1}_{-1}$ & $8.72\pm0.04$ & $88^{+1}_{-1}$ \\
WASP-137 & $6.85\pm0.11$ & $131^{+8}_{-8}$ & $8.28\pm0.07$ & $86^{+1}_{-1}$ & $8.31\pm0.06$ & $86^{+1}_{-1}$ & $8.26\pm0.04$ & $86^{+1}_{-1}$ & $7.67\pm0.04$ & $96^{+1}_{-1}$ \\
\hline
\end{tabular}
\end{table*}
We converted the magnitude contrast to mass limits by the same metric as illustrated in Sect.~\ref{subsec:characterization_of_ccs} using AMES-Cond, AMES-Dusty, and BT-Settl models \citep{Allard2001,Baraffe2003}.
The corresponding contrast curves for each individual target are presented in Appendix~\ref{sec:individual_detection_limits}, instead.

For almost all targets within the sample we were sensitive to stellar companions with masses larger than $0.1$\,M$_\sun$ at separations larger than 0\farcs5 and for most of those we even reached the threshold to the regime of brown dwarfs around $0.08$\,M$_\sun$.
In the five cases where we do not achieve this sensitivity, this was caused by the large distances to the corresponding targets of more than 350\,pc and/or poor AO conditions.
It is clear that the sensitivity achieved in only 256\,s of integration with SPHERE in mediocre conditions outperformed similar studies based on lucky imaging or conducted with other AO-assisted instruments.

\section{Discussion}
\label{sec:discussion}

\subsection{Multiplicity rate}
\label{subsec:discussion_multiplicity_rate}

We derived a multiplicity rate of $55.4^{+5.9}_{-9.4}\,\%$ from our sample of exoplanet host stars.
This value seems to be higher than estimates of many previous near infrared surveys targeting transiting exoplanet host stars to search for stellar companions, which derive multiplicity fractions of $21\pm12\,\%$ \citep{Daemgen2009}, $38\pm15\,\%$ \citep{Faedi2013}, $29\pm12\,\%$ \citep{Bergfors2013}, and $33\pm15\,\%$ \citep{Adams2013} among their samples.
Though the sample sizes of these studies were considerably smaller than the number of targets studied within the scope of this survey, this discrepancy in multiplicity rates most likely originates from the incompleteness of these previous surveys.
As most of these programmes were carried out using lucky-imaging strategies or with the first generation of AO-assisted imagers, the sensitivity achieved at small separation to the host stars was lower than that achievable with SPHERE.
A more accurate assessment of this incompleteness was presented by \citet{Ngo2015}, who derived a raw multiplicity fraction of $34\pm7\,\%$ for their sample of 50 transiting exoplanet hosts.
After simulating the population of binaries that were missed due to the instrument's sensitivity and limited field of view, they presented a corrected fraction of $49\pm9\,\%$, instead.
This value is in very good agreement with the rate derived from our sample, as we already considered previously detected companions outside of SPHERE's field of view for the statistical analysis.

\subsection{Hot Jupiter host stars}
\label{subsec:discussion_hot_jupiter_host_stars}

A large sub-sample of the targets studied within this survey are host stars to transiting hot Jupiters.
To study all stars from our sample that harbour giant planets with masses larger than $0.1\,M_\mathrm{jup}$ and semi-major axes smaller than 0.1\,au, we only needed to dismiss K2-2, K2-24, K2-38, K2-99, and WASP-130 from the original set.
Reiterating the analysis as described in Sect.~\ref{subsec:analysis_multiplicity_rate}, provided a multiplicity rate of $54.8^{+6.3}_{-9.9}\,\%$ for this sub-sample of hot Jupiter hosts.
Consequently, we aimed to assess whether this sub-sample of 40 targets is representative for the general population of host Jupiter host stars.

\begin{figure*}
\resizebox{\hsize}{!}{\includegraphics{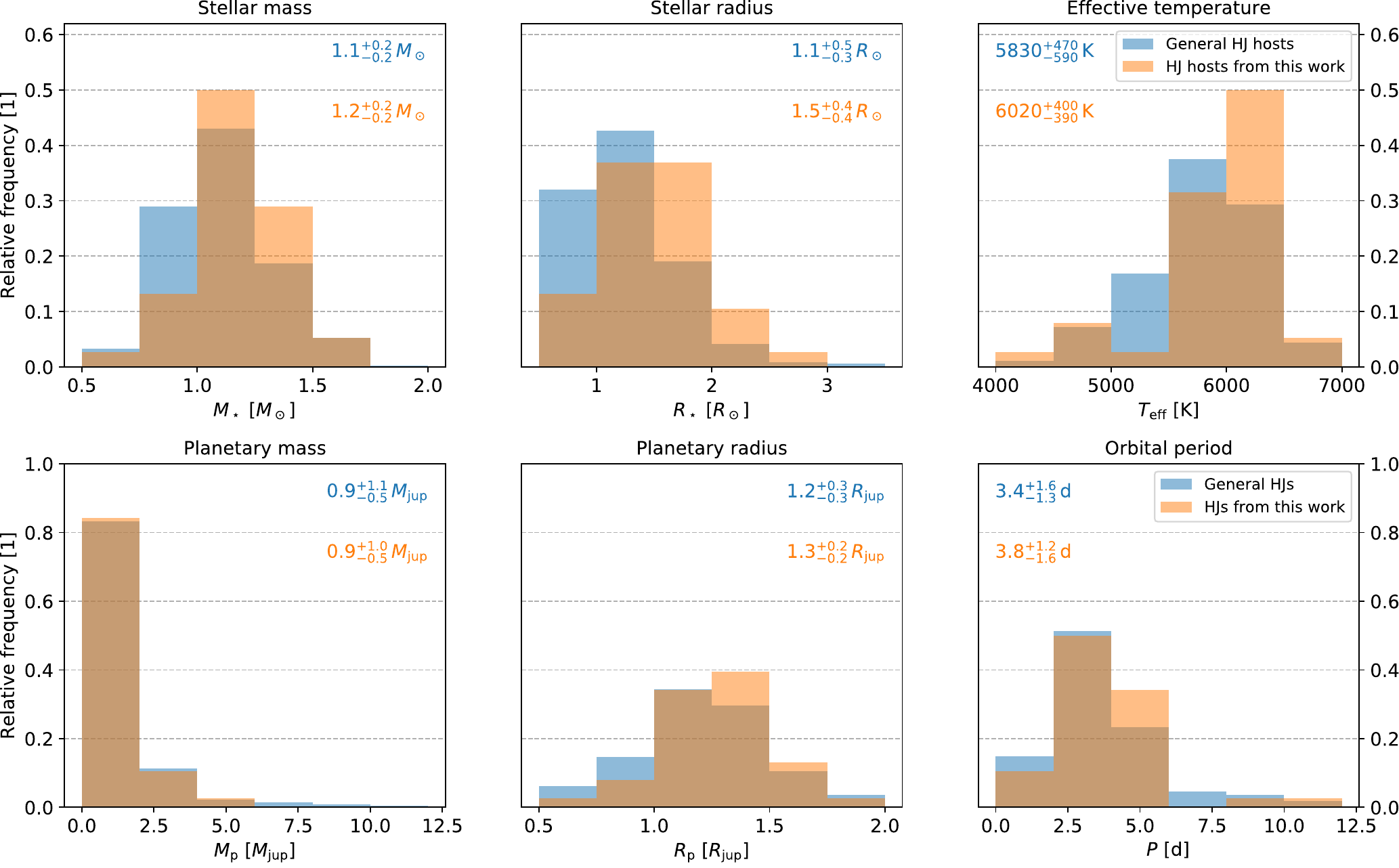}}
\caption{
Histograms of hot Jupiter (HJ) system properties.
We compare the targets analysed within the scope of this study (orange bars) to a general sample of hot Jupiter environments (blue bars).
In the \textit{top panel} the relative frequency distributions of stellar masses $M_\star$, radii $R_\star$, and effective temperatures $T_\mathrm{eff}$ amongst both samples are presented.
The \textit{lower panel} shows properties of the transiting companions such as planetary masses $M_\mathrm{p}$, planetary radii $R_\mathrm{p}$, and orbital periods $P$ instead.
In the upper part of each plot, we present the 68\,\% confidence intervals around the medians of the corresponding distributions.
}
\label{fig:comparison_general_hjs}
\end{figure*}

As described in Sect.~\ref{sec:introduction}, our target selection was purely restricted by the position on sky - as we required the objects to be observable with the VLT - and the targets' $R$ band magnitude to enable AO-assisted imaging.
All hot Jupiter host stars that met these criteria were observed within this survey, even if these had been considered in previous studies.
To further evaluate the quality of our sub-sample, we compiled a control group of 366 objects from the Exoplanet Orbit Database \citep{Han2014}, considering all hosts to transiting planets with masses larger than $0.1\,M_\mathrm{jup}$ and semi-major axes smaller than 0.1\,au.
We compared our sub-sample of hot Jupiters to the control group using six observables, of which three are describing properties of the hosts and three are characterising the transiting giant planets.
These parameters are the stellar masses $M_\star$, stellar radii $R_\star$, effective temperatures $T_\mathrm{eff}$, planetary masses $M_\mathrm{p}$, planetary radii $R_\mathrm{p}$, and orbital periods $P$.
In Fig.~\ref{fig:comparison_general_hjs} we present the relative frequency distributions of these observables among control group and targets used for this study.
There seems to be a trend towards slightly higher mass stars in our sample with respect to the general population of hot Jupiter hosts.
This is in good agreement with the applied magnitude cutoff inducing a marginal bias towards brighter and thus more massive host stars.
The same trend is marginally detected for the planetary properties as well.
Nevertheless, the distributions of all observables presented in Fig.~\ref{fig:comparison_general_hjs} are in good agreement between our sample and the control group and the 68\,\% confidence intervals we determined for both samples intersect significantly for each of the parameter distributions.
We thus argue that the targets analysed within the scope of this study can be considered a good representation of typical hot Jupiter systems.

\subsection{Correlation between stellar multiplicity and exoplanet eccentricities}
\label{subsec:correlation_multiplicity_expolanet_properties}

Nine systems in our sample harbour a transiting exoplanet that shows a non-zero eccentricity.
To test theories on the formation of these particular systems we evaluated the multiplicity rates among these environments and in comparison to the systems that do not have any known eccentric transiting planets.
For that purpose, we repeated the analysis from Section~\ref{subsec:analysis_multiplicity_rate} for the two sub-samples of eccentric and non-eccentric planet host stars.
From this analysis we obtained multiplicity rates of  $44^{+15}_{-19}\,\%$ and $58^{+6}_{-11}\,\%$ for the systems that host eccentric planetary companions and those that do not, respectively.
The large uncertainties on especially the former value arise from the very limited sample size of nine systems with the required properties.
Nevertheless, there is no statistically significant difference between the multiplicity rates amongst eccentric and non-eccentric sub-samples.
This is in good agreement with previous results from \citet{Ngo2015} and \citet{Ngo2017}.

\section{Conclusions}
\label{sec:conclusions}

We have observed a sample of 45 transiting exoplanet host stars with VLT/SPHERE/IRDIS to search for stellar companions:
\begin{itemize}
    \item 
    We have detected 11 candidate companions that had been identified by previous studies around 10 targets of our sample.
    For these candidate companions, we could confirm nine as co-moving binaries with common proper motion, proving HAT-P-41, HAT-P-57, WASP-2, WASP-8, WASP-54, WASP-70, WASP-76, and WASP-111 to be multiple systems.
    One candidate around WASP-7 has been confirmed to be a background object, instead.
    The status of a very bright and close companion to WASP-20 is still ambiguous, as only one epoch of astrometric data was available.
    Synthetic stellar population models, however, suggest that WASP-20~B is a gravitationally bound binary, which is in agreement with the conclusions from \citet{Evans2016b}.
    \item 
    We have detected 16 candidates that had not been reported by previous studies.
    These candidates are distributed among 13 different systems.
    By combination of SPHERE and Gaia astrometry we could show that WASP-123 is a binary system, whereas we could prove candidate companions around WASP-87 (CC 2) and WASP-108 (CC 2) to be background objects.
    For new candidate companions detected around K2-38, WASP-72, WASP-80, WASP-88, WASP-108 (CC 1), WASP-118, WASP-120, WASP-122, WASP-120, WASP-131 and WASP-137 too few astrometric measurements were available to prove common proper motion at 5$\sigma$ significance.
    Based on stellar population synthesis models, we derived the probability that the candidates are instead background contaminants.
    The most promising candidates with background probabilities smaller than 0.1\% were detected around WASP-131, WASP-72, and WASP-118.
    \item
    Additional proper motion checks need to be performed to test the companionship of these newly identified candidates and WASP-20~B.
    \item 
    We have derived detection limits for all of our targets and have shown that we reach an average magnitude contrast of $7.0\pm0.8$\,mag at a separation of 0\farcs2, while we are background limited for separations about 1\farcs0 with an average magnitude contrast of $8.9\pm0.9$\,mag. 
    For each individual target we have converted the derived contrast into a threshold of detectable mass by applying AMES-Cond, AMES-Dusty, and BT-Settl models depending on the effective temperature of the object.
    For 40 targets, we have been able to exclude companions with masses larger than $0.1$\,M$_\sun$ for separations that are larger than 0\farcs5, and in 20 cases we have reached the lower mass limit for potential stellar companions of approximately $0.08$\,M$_\sun$.
    \item 
    Based on our results, we have derived a stellar multiplicity rate of $55.4^{+5.9}_{-9.4}\,\%$ among our sample, which is in good agreement with results from previous surveys. 
    For the representative sub-sample of 40 host stars to transiting hot Jupiters, the derived multiplicity fraction is $54.8^{+6.3}_{-9.9}\,\%$.
    \item
    We have not detected any correlation between multiplicity of stellar systems and the eccentricity of planets that are detected around these stars.
\end{itemize}

We have shown that SPHERE is a great instrument to carry out studies like this, and the precision of the Gaia mission -- especially the claimed performance of future data releases -- is also a valuable tool to find stellar companions to exoplanet host stars.

In a companion work \citep{paper2} we will revisit the systems for which we have identified relatively bright nearby companions in the current work. 
We will use new and existing photometric and spectroscopic observations to redetermine the properties of the systems, corrected for the light contributed by the nearby companion stars.

\begin{acknowledgements}

We thank the anonymous referee for the feedback that helped improving he quality of the manuscript.

The  research  of  A.~J.~Bohn and F.~Snik leading  to  these results has received funding from the European Research Council under ERC Starting Grant agreement 678194 (FALCONER). 

This research has made use of the Exoplanet Orbit Database and the Exoplanet Data Explorer at \url{http://www.exoplanets.org/}.

To achieve the scientific results presented in this article we used the \emph{Python} programming language\footnote{Python Software Foundation, \url{https://www.python.org/}}, especially the \emph{SciPy} \citep{Virtanen2019}, \emph{NumPy} \citep{numpy}, \emph{Matplotlib} \citep{Matplotlib}, \emph{scikit-image} \citep{scikit-image}, \emph{scikit-learn} \citep{scikit-learn}, \emph{photutils} \citep{photutils}, and \emph{astropy} \citep{astropy_1,astropy_2} packages.

\end{acknowledgements}

\bibliographystyle{aa} 
\bibliography{mybib} 

\begin{thebibliography}{120}
\expandafter\ifx\csname natexlab\endcsname\relax\def\natexlab#1{#1}\fi

\bibitem[{{Adams} {et~al.}(2013){Adams}, {Dupree}, {Kulesa}, \&
  {McCarthy}}]{Adams2013}
{Adams}, E.~R., {Dupree}, A.~K., {Kulesa}, C., \& {McCarthy}, D. 2013, \aj,
  146, 9

\bibitem[{{Allard} {et~al.}(2001){Allard}, {Hauschildt}, {Alexander},
  {Tamanai}, \& {Schweitzer}}]{Allard2001}
{Allard}, F., {Hauschildt}, P.~H., {Alexander}, D.~R., {Tamanai}, A., \&
  {Schweitzer}, A. 2001, \apj, 556, 357

\bibitem[{{Anderson} {et~al.}(2018){Anderson}, {Bouchy}, {Brown}, {Collier
  Cameron}, {Delrez}, {Gillon}, {Gonz{\'a}lez Hern{\'a}ndez}, {Hellier},
  {Jehin}, {Lendl}, {Maxted}, {Neveu-VanMalle}, {Nielsen}, {Pepe}, {Perger},
  {Pollacco}, {Queloz}, {Rey}, {S{\'e}gransan}, {Smalley}, {Toledo-Padr{\'o}n},
  {Triaud}, {Turner}, {Udry}, \& {West}}]{arXiv181209264}
{Anderson}, D.~R., {Bouchy}, F., {Brown}, D.~J.~A., {et~al.} 2018, arXiv
  e-prints [\eprint[arXiv]{1812.09264}]

\bibitem[{{Anderson} {et~al.}(2014{\natexlab{a}}){Anderson}, {Brown}, {Collier
  Cameron}, {Delrez}, {Fumel}, {Gillon}, {Hellier}, {Jehin}, {Lendl}, {Maxted},
  {Neveu-VanMalle}, {Pepe}, {Pollacco}, {Queloz}, {Rojo}, {Segransan},
  {Serenelli}, {Smalley}, {Smith}, {Southworth}, {Triaud}, {Turner}, {Udry}, \&
  {West}}]{Anderson2014a}
{Anderson}, D.~R., {Brown}, D.~J.~A., {Collier Cameron}, A., {et~al.}
  2014{\natexlab{a}}, arXiv e-prints, arXiv:1410.3449

\bibitem[{{Anderson} {et~al.}(2014{\natexlab{b}}){Anderson}, {Collier Cameron},
  {Delrez}, {Doyle}, {Faedi}, {Fumel}, {Gillon}, {G{\'o}mez Maqueo Chew},
  {Hellier}, {Jehin}, {Lendl}, {Maxted}, {Pepe}, {Pollacco}, {Queloz},
  {S{\'e}gransan}, {Skillen}, {Smalley}, {Smith}, {Southworth}, {Triaud},
  {Turner}, {Udry}, \& {West}}]{2014MNRAS.445.1114A}
{Anderson}, D.~R., {Collier Cameron}, A., {Delrez}, L., {et~al.}
  2014{\natexlab{b}}, \mnras, 445, 1114

\bibitem[{{Anderson} {et~al.}(2011){Anderson}, {Collier Cameron}, {Hellier},
  {Lendl}, {Maxted}, {Pollacco}, {Queloz}, {Smalley}, {Smith}, {Todd}, \&
  {Others}}]{2011ApJ...726L..19A}
{Anderson}, D.~R., {Collier Cameron}, A., {Hellier}, C., {et~al.} 2011, ApJ,
  726, L19

\bibitem[{{Astropy Collaboration} {et~al.}(2018){Astropy Collaboration},
  {Price-Whelan}, {Sip{\H o}cz}, {G{\"u}nther}, {Lim}, {Crawford}, {Conseil},
  {Shupe}, {Craig}, {Dencheva}, {Ginsburg}, {VanderPlas}, {Bradley},
  {P{\'e}rez-Su{\'a}rez}, {de Val-Borro}, {Aldcroft}, {Cruz}, {Robitaille},
  {Tollerud}, {Ardelean}, {Babej}, {Bach}, {Bachetti}, {Bakanov}, {Bamford},
  {Barentsen}, {Barmby}, {Baumbach}, {Berry}, {Biscani}, {Boquien}, {Bostroem},
  {Bouma}, {Brammer}, {Bray}, {Breytenbach}, {Buddelmeijer}, {Burke},
  {Calderone}, {Cano Rodr{\'{\i}}guez}, {Cara}, {Cardoso}, {Cheedella},
  {Copin}, {Corrales}, {Crichton}, {D'Avella}, {Deil}, {Depagne}, {Dietrich},
  {Donath}, {Droettboom}, {Earl}, {Erben}, {Fabbro}, {Ferreira}, {Finethy},
  {Fox}, {Garrison}, {Gibbons}, {Goldstein}, {Gommers}, {Greco}, {Greenfield},
  {Groener}, {Grollier}, {Hagen}, {Hirst}, {Homeier}, {Horton}, {Hosseinzadeh},
  {Hu}, {Hunkeler}, {Ivezi{\'c}}, {Jain}, {Jenness}, {Kanarek}, {Kendrew},
  {Kern}, {Kerzendorf}, {Khvalko}, {King}, {Kirkby}, {Kulkarni}, {Kumar},
  {Lee}, {Lenz}, {Littlefair}, {Ma}, {Macleod}, {Mastropietro}, {McCully},
  {Montagnac}, {Morris}, {Mueller}, {Mumford}, {Muna}, {Murphy}, {Nelson},
  {Nguyen}, {Ninan}, {N{\"o}the}, {Ogaz}, {Oh}, {Parejko}, {Parley}, {Pascual},
  {Patil}, {Patil}, {Plunkett}, {Prochaska}, {Rastogi}, {Reddy Janga},
  {Sabater}, {Sakurikar}, {Seifert}, {Sherbert}, {Sherwood-Taylor}, {Shih},
  {Sick}, {Silbiger}, {Singanamalla}, {Singer}, {Sladen}, {Sooley},
  {Sornarajah}, {Streicher}, {Teuben}, {Thomas}, {Tremblay}, {Turner},
  {Terr{\'o}n}, {van Kerkwijk}, {de la Vega}, {Watkins}, {Weaver}, {Whitmore},
  {Woillez}, {Zabalza}, \& {Astropy Contributors}}]{astropy_2}
{Astropy Collaboration}, {Price-Whelan}, A.~M., {Sip{\H o}cz}, B.~M., {et~al.}
  2018, \aj, 156, 123

\bibitem[{{Astropy Collaboration} {et~al.}(2013){Astropy Collaboration},
  {Robitaille}, {Tollerud}, {Greenfield}, {Droettboom}, {Bray}, {Aldcroft},
  {Davis}, {Ginsburg}, {Price-Whelan}, {Kerzendorf}, {Conley}, {Crighton},
  {Barbary}, {Muna}, {Ferguson}, {Grollier}, {Parikh}, {Nair}, {Unther},
  {Deil}, {Woillez}, {Conseil}, {Kramer}, {Turner}, {Singer}, {Fox}, {Weaver},
  {Zabalza}, {Edwards}, {Azalee Bostroem}, {Burke}, {Casey}, {Crawford},
  {Dencheva}, {Ely}, {Jenness}, {Labrie}, {Lim}, {Pierfederici}, {Pontzen},
  {Ptak}, {Refsdal}, {Servillat}, \& {Streicher}}]{astropy_1}
{Astropy Collaboration}, {Robitaille}, T.~P., {Tollerud}, E.~J., {et~al.} 2013,
  \aap, 558, A33

\bibitem[{{Auvergne} {et~al.}(2009){Auvergne}, {Bodin}, {Boisnard}, {Buey},
  {Chaintreuil}, {Epstein}, {Jouret}, {Lam-Trong}, {Levacher}, {Magnan},
  {Perez}, {Plasson}, {Plesseria}, {Peter}, {Steller}, {Tiph{\`e}ne}, {Baglin},
  {Agogu{\'e}}, {Appourchaux}, {Barbet}, {Beaufort}, {Bellenger}, {Berlin},
  {Bernardi}, {Blouin}, {Boumier}, {Bonneau}, {Briet}, {Butler}, {Cautain},
  {Chiavassa}, {Costes}, {Cuvilho}, {Cunha-Parro}, {de Oliveira Fialho},
  {Decaudin}, {Defise}, {Djalal}, {Docclo}, {Drummond}, {Dupuis}, {Exil},
  {Faur{\'e}}, {Gaboriaud}, {Gamet}, {Gavalda}, {Grolleau}, {Gueguen},
  {Guivarc'h}, {Guterman}, {Hasiba}, {Huntzinger}, {Hustaix}, {Imbert},
  {Jeanville}, {Johlander}, {Jorda}, {Journoud}, {Karioty}, {Kerjean},
  {Lafond}, {Lapeyrere}, {Landiech}, {Larqu{\'e}}, {Laudet}, {Le Merrer},
  {Leporati}, {Leruyet}, {Levieuge}, {Llebaria}, {Martin}, {Mazy}, {Mesnager},
  {Michel}, {Moalic}, {Monjoin}, {Naudet}, {Neukirchner}, {Nguyen-Kim},
  {Ollivier}, {Orcesi}, {Ottacher}, {Oulali}, {Parisot}, {Perruchot},
  {Piacentino}, {Pinheiro da Silva}, {Platzer}, {Pontet}, {Pradines},
  {Quentin}, {Rohbeck}, {Rolland}, {Rollenhagen}, {Romagnan}, {Russ}, {Samadi},
  {Schmidt}, {Schwartz}, {Sebbag}, {Smit}, {Sunter}, {Tello}, {Toulouse},
  {Ulmer}, {Vandermarcq}, {Vergnault}, {Wallner}, {Waultier}, \&
  {Zanatta}}]{Auvergne2009}
{Auvergne}, M., {Bodin}, P., {Boisnard}, L., {et~al.} 2009, \aap, 506, 411

\bibitem[{{Bailer-Jones} {et~al.}(2018){Bailer-Jones}, {Rybizki}, {Fouesneau},
  {Mantelet}, \& {Andrae}}]{BailerJones18}
{Bailer-Jones}, C.~A.~L., {Rybizki}, J., {Fouesneau}, M., {Mantelet}, G., \&
  {Andrae}, R. 2018, \aj, 156, 58

\bibitem[{{Bakos} {et~al.}(2004){Bakos}, {Noyes}, {Kov{\'a}cs}, {Stanek},
  {Sasselov}, \& {Domsa}}]{Bakos2004}
{Bakos}, G., {Noyes}, R.~W., {Kov{\'a}cs}, G., {et~al.} 2004, \pasp, 116, 266

\bibitem[{{Baraffe} {et~al.}(2003){Baraffe}, {Chabrier}, {Barman}, {Allard}, \&
  {Hauschildt}}]{Baraffe2003}
{Baraffe}, I., {Chabrier}, G., {Barman}, T.~S., {Allard}, F., \& {Hauschildt},
  P.~H. 2003, \aap, 402, 701

\bibitem[{{Baranne} {et~al.}(1996){Baranne}, {Queloz}, {Mayor}, {Adrianzyk},
  {Knispel}, {Kohler}, {Lacroix}, {Meunier}, {Rimbaud}, \& {Vin}}]{Baranne1996}
{Baranne}, A., {Queloz}, D., {Mayor}, M., {et~al.} 1996, \aaps, 119, 373

\bibitem[{{Batygin}(2012)}]{Batygin2012}
{Batygin}, K. 2012, \nat, 491, 418

\bibitem[{{Batygin} {et~al.}(2016){Batygin}, {Bodenheimer}, \&
  {Laughlin}}]{Batygin2016}
{Batygin}, K., {Bodenheimer}, P.~H., \& {Laughlin}, G.~P. 2016, \apj, 829, 114

\bibitem[{{Bergfors} {et~al.}(2013){Bergfors}, {Brandner}, {Daemgen}, {Biller},
  {Hippler}, {Janson}, {Kudryavtseva}, {Gei{\ss}ler}, {Henning}, \&
  {K{\"o}hler}}]{Bergfors2013}
{Bergfors}, C., {Brandner}, W., {Daemgen}, S., {et~al.} 2013, \mnras, 428, 182

\bibitem[{{Beuzit} {et~al.}(2019){Beuzit}, {Vigan}, {Mouillet}, {Dohlen},
  {Gratton}, {Boccaletti}, {Sauvage}, {Schmid}, {Langlois}, {Petit},
  {Baruffolo}, {Feldt}, {Milli}, {Wahhaj}, {Abe}, {Anselmi}, {Antichi},
  {Barette}, {Baudrand}, {Baudoz}, {Bazzon}, {Bernardi}, {Blanchard}, {Brast},
  {Bruno}, {Buey}, {Carbillet}, {Carle}, {Cascone}, {Chapron}, {Charton},
  {Chauvin}, {Claudi}, {Costille}, {De Caprio}, {de Boer}, {Delboulb{\'e}},
  {Desidera}, {Dominik}, {Downing}, {Dupuis}, {Fabron}, {Fantinel}, {Farisato},
  {Feautrier}, {Fedrigo}, {Fusco}, {Gigan}, {Ginski}, {Girard}, {Giro},
  {Gisler}, {Gluck}, {Gry}, {Henning}, {Hubin}, {Hugot}, {Incorvaia}, {Jaquet},
  {Kasper}, {Lagadec}, {Lagrange}, {Le Coroller}, {Le Mignant}, {Le Ruyet},
  {Lessio}, {Lizon}, {Llored}, {Lundin}, {Madec}, {Magnard}, {Marteaud},
  {Martinez}, {Maurel}, {M{\'e}nard}, {Mesa}, {M{\"o}ller-Nilsson}, {Moulin},
  {Moutou}, {Orign{\'e}}, {Parisot}, {Pavlov}, {Perret}, {Pragt}, {Puget},
  {Rabou}, {Ramos}, {Reess}, {Rigal}, {Rochat}, {Roelfsema}, {Rousset}, {Roux},
  {Saisse}, {Salasnich}, {Santambrogio}, {Scuderi}, {Segransan}, {Sevin},
  {Siebenmorgen}, {Soenke}, {Stadler}, {Suarez}, {Tiph{\`e}ne}, {Turatto},
  {Udry}, {Vakili}, {Waters}, {Weber}, {Wildi}, {Zins}, \&
  {Zurlo}}]{Beuzit2019}
{Beuzit}, J.~L., {Vigan}, A., {Mouillet}, D., {et~al.} 2019, \aap, 631, A155

\bibitem[{{Bodenheimer} {et~al.}(2000){Bodenheimer}, {Hubickyj}, \&
  {Lissauer}}]{Bodenheimer2000}
{Bodenheimer}, P., {Hubickyj}, O., \& {Lissauer}, J.~J. 2000, \icarus, 143, 2

\bibitem[{{Boley} {et~al.}(2016){Boley}, {Granados Contreras}, \&
  {Gladman}}]{Boley2016}
{Boley}, A.~C., {Granados Contreras}, A.~P., \& {Gladman}, B. 2016, \apjl, 817,
  L17

\bibitem[{{Borucki} {et~al.}(2010){Borucki}, {Koch}, {Basri}, {Batalha},
  {Brown}, {Caldwell}, {Caldwell}, {Christensen-Dalsgaard}, {Cochran},
  {DeVore}, {Dunham}, {Dupree}, {Gautier}, {Geary}, {Gilliland}, {Gould},
  {Howell}, {Jenkins}, {Kondo}, {Latham}, {Marcy}, {Meibom}, {Kjeldsen},
  {Lissauer}, {Monet}, {Morrison}, {Sasselov}, {Tarter}, {Boss}, {Brownlee},
  {Owen}, {Buzasi}, {Charbonneau}, {Doyle}, {Fortney}, {Ford}, {Holman},
  {Seager}, {Steffen}, {Welsh}, {Rowe}, {Anderson}, {Buchhave}, {Ciardi},
  {Walkowicz}, {Sherry}, {Horch}, {Isaacson}, {Everett}, {Fischer}, {Torres},
  {Johnson}, {Endl}, {MacQueen}, {Bryson}, {Dotson}, {Haas}, {Kolodziejczak},
  {Van Cleve}, {Chandrasekaran}, {Twicken}, {Quintana}, {Clarke}, {Allen},
  {Li}, {Wu}, {Tenenbaum}, {Verner}, {Bruhweiler}, {Barnes}, \&
  {Prsa}}]{Borucki2010}
{Borucki}, W.~J., {Koch}, D., {Basri}, G., {et~al.} 2010, Science, 327, 977

\bibitem[{{Bouchy} {et~al.}(2010){Bouchy}, {Hebb}, {Skillen}, {Collier
  Cameron}, {Smalley}, {Udry}, {Anderson}, {Boisse}, {Enoch}, {Haswell}, \&
  {Others}}]{2010A+A...519A..98B}
{Bouchy}, F., {Hebb}, L., {Skillen}, I., {et~al.} 2010, A\&A, 519, A98

\bibitem[{{Bradley} {et~al.}(2016){Bradley}, {Sipocz}, {Robitaille},
  {Tollerud}, {Deil}, {Vin{\'\i}cius}, {Barbary}, {G{\"u}nther}, {Bostroem},
  {Droettboom}, {Bray}, {Bratholm}, {Pickering}, {Craig}, {Pascual}, {Greco},
  {Donath}, {Kerzendorf}, {Littlefair}, {Barentsen}, {D'Eugenio}, \&
  {Weaver}}]{photutils}
{Bradley}, L., {Sipocz}, B., {Robitaille}, T., {et~al.} 2016, {Photutils:
  Photometry tools}

\bibitem[{{Brown} {et~al.}(2017){Brown}, {Triaud}, {Doyle}, {Gillon}, {Lendl},
  {Anderson}, {Collier Cameron}, {H{\'e}brard}, {Hellier}, {Lovis}, {Maxted},
  {Pepe}, {Pollacco}, {Queloz}, \& {Smalley}}]{2017MNRAS.464..810B}
{Brown}, D.~J.~A., {Triaud}, A.~H.~M.~J., {Doyle}, A.~P., {et~al.} 2017,
  \mnras, 464, 810

\bibitem[{{Butler} {et~al.}(1997){Butler}, {Marcy}, {Williams}, {Hauser}, \&
  {Shirts}}]{Butler1997}
{Butler}, R.~P., {Marcy}, G.~W., {Williams}, E., {Hauser}, H., \& {Shirts}, P.
  1997, \apj, 474, L115

\bibitem[{{Carbillet} {et~al.}(2011){Carbillet}, {Bendjoya}, {Abe}, {Guerri},
  {Boccaletti}, {Daban}, {Dohlen}, {Ferrari}, {Robbe-Dubois}, {Douet}, \&
  {Vakili}}]{Carbillet2011}
{Carbillet}, M., {Bendjoya}, P., {Abe}, L., {et~al.} 2011, Experimental
  Astronomy, 30, 39

\bibitem[{{Castelli} \& {Kurucz}(1994)}]{Castelli1994}
{Castelli}, F. \& {Kurucz}, R.~L. 1994, \aap, 281, 817

\bibitem[{{Chatterjee} {et~al.}(2008){Chatterjee}, {Ford}, {Matsumura}, \&
  {Rasio}}]{Chatterjee2008}
{Chatterjee}, S., {Ford}, E.~B., {Matsumura}, S., \& {Rasio}, F.~A. 2008, \apj,
  686, 580

\bibitem[{{Ciceri} {et~al.}(2013){Ciceri}, {Mancini}, {Southworth}, {Nikolov},
  {Bozza}, {Bruni}, {Calchi Novati}, {D'Ago}, \&
  {Henning}}]{2013A+A...557A..30C}
{Ciceri}, S., {Mancini}, L., {Southworth}, J., {et~al.} 2013, A\&A, 557, A30

\bibitem[{{Claudi} {et~al.}(2008){Claudi}, {Turatto}, {Gratton}, {Antichi},
  {Bonavita}, {Bruno}, {Cascone}, {De Caprio}, {Desidera}, {Giro}, {Mesa},
  {Scuderi}, {Dohlen}, {Beuzit}, \& {Puget}}]{Claudi2008}
{Claudi}, R.~U., {Turatto}, M., {Gratton}, R.~G., {et~al.} 2008, in Society of
  Photo-Optical Instrumentation Engineers (SPIE) Conference Series, Vol. 7014,
  Ground-based and Airborne Instrumentation for Astronomy II, 70143E

\bibitem[{{Collier Cameron} {et~al.}(2007){Collier Cameron}, {Bouchy},
  {H{\'e}brard}, {Maxted}, {Pollacco}, {Pont}, {Skillen}, {Smalley}, {Street},
  {West}, {Wilson}, \& {Others}}]{2007MNRAS.375..951C}
{Collier Cameron}, A., {Bouchy}, F., {H{\'e}brard}, G., {et~al.} 2007, MNRAS,
  375, 951

\bibitem[{{Cosentino} {et~al.}(2012){Cosentino}, {Lovis}, {Pepe}, {Collier
  Cameron}, {Latham}, {Molinari}, {Udry}, {Bezawada}, {Black}, {Born},
  {Buchschacher}, {Charbonneau}, {Figueira}, {Fleury}, {Galli}, {Gallie},
  {Gao}, {Ghedina}, {Gonzalez}, {Gonzalez}, {Guerra}, {Henry}, {Horne},
  {Hughes}, {Kelly}, {Lodi}, {Lunney}, {Maire}, {Mayor}, {Micela}, {Ordway},
  {Peacock}, {Phillips}, {Piotto}, {Pollacco}, {Queloz}, {Rice}, {Riverol},
  {Riverol}, {San Juan}, {Sasselov}, {Segransan}, {Sozzetti}, {Sosnowska},
  {Stobie}, {Szentgyorgyi}, {Vick}, \& {Weber}}]{Cosentino2012}
{Cosentino}, R., {Lovis}, C., {Pepe}, F., {et~al.} 2012, in Society of
  Photo-Optical Instrumentation Engineers (SPIE) Conference Series, Vol. 8446,
  Ground-based and Airborne Instrumentation for Astronomy IV, 84461V

\bibitem[{{Cutri} {et~al.}(2012){Cutri}, {Skrutskie}, {van Dyk}, {Beichman},
  {Carpenter}, {Chester}, {Cambresy}, {Evans}, {Fowler}, {Gizis}, {Howard},
  {Huchra}, {Jarrett}, {Kopan}, {Kirkpatrick}, {Light}, {Marsh}, {McCallon},
  {Schneider}, {Stiening}, {Sykes}, {Weinberg}, {Wheaton}, {Wheelock}, \&
  {Zacharias}}]{Cutri2012a}
{Cutri}, R.~M., {Skrutskie}, M.~F., {van Dyk}, S., {et~al.} 2012, VizieR Online
  Data Catalog, II/281

\bibitem[{{Daemgen} {et~al.}(2009){Daemgen}, {Hormuth}, {Brandner}, {Bergfors},
  {Janson}, {Hippler}, \& {Henning}}]{Daemgen2009}
{Daemgen}, S., {Hormuth}, F., {Brandner}, W., {et~al.} 2009, \aap, 498, 567

\bibitem[{{Delrez} {et~al.}(2016){Delrez}, {Santerne}, {Almenara}, {Anderson},
  {Collier-Cameron}, {D{\'{\i}}az}, {Gillon}, {Hellier}, {Jehin}, {Lendl}, \&
  {Others}}]{2016MNRAS.458.4025D}
{Delrez}, L., {Santerne}, A., {Almenara}, J.-M., {et~al.} 2016, MNRAS, 458,
  4025

\bibitem[{{Delrez} {et~al.}(2014){Delrez}, {Van Grootel}, {Anderson},
  {Collier-Cameron}, {Doyle}, {Fumel}, {Gillon}, {Hellier}, {Jehin}, {Lendl},
  {Neveu-VanMalle}, {Maxted}, {Pepe}, {Pollacco}, {Queloz}, {S{\'e}gransan},
  {Smalley}, {Smith}, {Southworth}, {Triaud}, {Udry}, \&
  {West}}]{2014A+A...563A.143D}
{Delrez}, L., {Van Grootel}, V., {Anderson}, D.~R., {et~al.} 2014, \aap, 563,
  A143

\bibitem[{{Dietrich} \& {Ginski}(2018)}]{Dietrich2018}
{Dietrich}, J. \& {Ginski}, C. 2018, \aap, 620, A102

\bibitem[{{Dohlen} {et~al.}(2008){Dohlen}, {Langlois}, {Saisse}, {Hill},
  {Origne}, {Jacquet}, {Fabron}, {Blanc}, {Llored}, {Carle}, {Moutou}, {Vigan},
  {Boccaletti}, {Carbillet}, {Mouillet}, \& {Beuzit}}]{Dohlen2008}
{Dohlen}, K., {Langlois}, M., {Saisse}, M., {et~al.} 2008, in \procspie, Vol.
  7014, Ground-based and Airborne Instrumentation for Astronomy II, 70143L

\bibitem[{{Eggleton} \& {Kiseleva-Eggleton}(2001)}]{Eggleton2001}
{Eggleton}, P.~P. \& {Kiseleva-Eggleton}, L. 2001, \apj, 562, 1012

\bibitem[{{Evans} {et~al.}(2016{\natexlab{a}}){Evans}, {Southworth}, {Maxted},
  {Skottfelt}, {Hundertmark}, {J{\o}rgensen}, {Dominik}, {Alsubai}, {Andersen},
  {Bozza}, {Bramich}, {Burgdorf}, {Ciceri}, {D'Ago}, {Figuera Jaimes}, {Gu},
  {Haugb{\o}lle}, {Hinse}, {Juncher}, {Kains}, {Kerins}, {Korhonen},
  {Kuffmeier}, {Mancini}, {Peixinho}, {Popovas}, {Rabus}, {Rahvar}, {Schmidt},
  {Snodgrass}, {Starkey}, {Surdej}, {Tronsgaard}, {von Essen}, {Wang}, \&
  {Wertz}}]{Evans2016a}
{Evans}, D.~F., {Southworth}, J., {Maxted}, P.~F.~L., {et~al.}
  2016{\natexlab{a}}, \aap, 589, A58

\bibitem[{{Evans} {et~al.}(2016{\natexlab{b}}){Evans}, {Southworth}, \&
  {Smalley}}]{Evans2016b}
{Evans}, D.~F., {Southworth}, J., \& {Smalley}, B. 2016{\natexlab{b}}, \apj,
  833, L19

\bibitem[{{Evans} {et~al.}(2018){Evans}, {Southworth}, {Smalley},
  {J{\o}rgensen}, {Dominik}, {Andersen}, {Bozza}, {Bramich}, {Burgdorf},
  {Ciceri}, {D'Ago}, {Figuera Jaimes}, {Gu}, {Hinse}, {Henning}, {Hundertmark},
  {Kains}, {Kerins}, {Korhonen}, {Kokotanekova}, {Kuffmeier}, {Longa-Pe{\~n}a},
  {Mancini}, {MacKenzie}, {Popovas}, {Rabus}, {Rahvar}, {Sajadian},
  {Snodgrass}, {Skottfelt}, {Surdej}, {Tronsgaard}, {Unda-Sanzana}, {von
  Essen}, {Wang}, \& {Wertz}}]{Evans2018}
{Evans}, D.~F., {Southworth}, J., {Smalley}, B., {et~al.} 2018, \aap, 610, A20

\bibitem[{{Fabrycky} \& {Tremaine}(2007)}]{Fabrycky2007}
{Fabrycky}, D. \& {Tremaine}, S. 2007, \apj, 669, 1298

\bibitem[{{Faedi} {et~al.}(2013{\natexlab{a}}){Faedi}, {Pollacco}, {Barros},
  {Brown}, {Collier Cameron}, {Doyle}, {Enoch}, {Gillon}, {G{\'o}mez Maqueo
  Chew}, {H{\'e}brard}, \& {Others}}]{2013A+A...551A..73F}
{Faedi}, F., {Pollacco}, D., {Barros}, S.~C.~C., {et~al.} 2013{\natexlab{a}},
  A\&A, 551, A73

\bibitem[{{Faedi} {et~al.}(2013{\natexlab{b}}){Faedi}, {Staley}, {G{\'o}mez
  Maqueo Chew}, {Pollacco}, {Dhital}, {Barros}, {Skillen}, {Hebb}, {Mackay}, \&
  {Watson}}]{Faedi2013}
{Faedi}, F., {Staley}, T., {G{\'o}mez Maqueo Chew}, Y., {et~al.}
  2013{\natexlab{b}}, \mnras, 433, 2097

\bibitem[{Fienup(1997)}]{Fienup1997}
Fienup, J.~R. 1997, Applied optics, 36, 8352

\bibitem[{{Fischer} {et~al.}(1999){Fischer}, {Marcy}, {Butler}, {Vogt}, \&
  {Apps}}]{Fischer1999}
{Fischer}, D.~A., {Marcy}, G.~W., {Butler}, R.~P., {Vogt}, S.~S., \& {Apps}, K.
  1999, \pasp, 111, 50

\bibitem[{{Fusco} {et~al.}(2006){Fusco}, {Rousset}, {Sauvage}, {Petit},
  {Beuzit}, {Dohlen}, {Mouillet}, {Charton}, {Nicolle}, {Kasper}, {Baudoz}, \&
  {Puget}}]{Fusco2006}
{Fusco}, T., {Rousset}, G., {Sauvage}, J.-F., {et~al.} 2006, Optics Express,
  14, 7515

\bibitem[{{Gaia Collaboration} {et~al.}(2018){Gaia Collaboration}, {Brown},
  {Vallenari}, {Prusti}, {de Bruijne}, {Babusiaux}, {Bailer-Jones}, {Biermann},
  {Evans}, {Eyer}, {Jansen}, {Jordi}, {Klioner}, {Lammers}, {Lindegren},
  {Luri}, {Mignard}, {Panem}, {Pourbaix}, {Randich}, {Sartoretti}, {Siddiqui},
  {Soubiran}, {van Leeuwen}, {Walton}, {Arenou}, {Bastian}, {Cropper},
  {Drimmel}, {Katz}, {Lattanzi}, {Bakker}, {Cacciari}, {Casta{\~n}eda},
  {Chaoul}, {Cheek}, {De Angeli}, {Fabricius}, {Guerra}, {Holl}, {Masana},
  {Messineo}, {Mowlavi}, {Nienartowicz}, {Panuzzo}, {Portell}, {Riello},
  {Seabroke}, {Tanga}, {Th{\'e}venin}, {Gracia-Abril}, {Comoretto},
  {Garcia-Reinaldos}, {Teyssier}, {Altmann}, {Andrae}, {Audard},
  {Bellas-Velidis}, {Benson}, {Berthier}, {Blomme}, {Burgess}, {Busso},
  {Carry}, {Cellino}, {Clementini}, {Clotet}, {Creevey}, {Davidson}, {De
  Ridder}, {Delchambre}, {Dell'Oro}, {Ducourant},
  {Fern{\'a}ndez-Hern{\'a}ndez}, {Fouesneau}, {Fr{\'e}mat}, {Galluccio},
  {Garc{\'\i}a-Torres}, {Gonz{\'a}lez-N{\'u}{\~n}ez}, {Gonz{\'a}lez-Vidal},
  {Gosset}, {Guy}, {Halbwachs}, {Hambly}, {Harrison}, {Hern{\'a}ndez},
  {Hestroffer}, {Hodgkin}, {Hutton}, {Jasniewicz}, {Jean-Antoine-Piccolo},
  {Jordan}, {Korn}, {Krone-Martins}, {Lanzafame}, {Lebzelter}, {L{\"o}ffler},
  {Manteiga}, {Marrese}, {Mart{\'\i}n-Fleitas}, {Moitinho}, {Mora}, {Muinonen},
  {Osinde}, {Pancino}, {Pauwels}, {Petit}, {Recio-Blanco}, {Richards},
  {Rimoldini}, {Robin}, {Sarro}, {Siopis}, {Smith}, {Sozzetti}, {S{\"u}veges},
  {Torra}, {van Reeven}, {Abbas}, {Abreu Aramburu}, {Accart}, {Aerts},
  {Altavilla}, {{\'A}lvarez}, {Alvarez}, {Alves}, {Anderson}, {Andrei},
  {Anglada Varela}, {Antiche}, {Antoja}, {Arcay}, {Astraatmadja}, {Bach},
  {Baker}, {Balaguer-N{\'u}{\~n}ez}, {Balm}, {Barache}, {Barata}, {Barbato},
  {Barblan}, {Barklem}, {Barrado}, {Barros}, {Barstow}, {Bartholom{\'e}
  Mu{\~n}oz}, {Bassilana}, {Becciani}, {Bellazzini}, {Berihuete}, {Bertone},
  {Bianchi}, {Bienaym{\'e}}, {Blanco-Cuaresma}, {Boch}, {Boeche}, {Bombrun},
  {Borrachero}, {Bossini}, {Bouquillon}, {Bourda}, {Bragaglia}, {Bramante},
  {Breddels}, {Bressan}, {Brouillet}, {Br{\"u}semeister}, {Brugaletta},
  {Bucciarelli}, {Burlacu}, {Busonero}, {Butkevich}, {Buzzi}, {Caffau},
  {Cancelliere}, {Cannizzaro}, {Cantat-Gaudin}, {Carballo}, {Carlucci},
  {Carrasco}, {Casamiquela}, {Castellani}, {Castro-Ginard}, {Charlot},
  {Chemin}, {Chiavassa}, {Cocozza}, {Costigan}, {Cowell}, {Crifo}, {Crosta},
  {Crowley}, {Cuypers}, {Dafonte}, {Damerdji}, {Dapergolas}, {David}, {David},
  {de Laverny}, {De Luise}, {De March}, {de Martino}, {de Souza}, {de Torres},
  {Debosscher}, {del Pozo}, {Delbo}, {Delgado}, {Delgado}, {Di Matteo},
  {Diakite}, {Diener}, {Distefano}, {Dolding}, {Drazinos}, {Dur{\'a}n},
  {Edvardsson}, {Enke}, {Eriksson}, {Esquej}, {Eynard Bontemps}, {Fabre},
  {Fabrizio}, {Faigler}, {Falc{\~a}o}, {Farr{\`a}s Casas}, {Federici},
  {Fedorets}, {Fernique}, {Figueras}, {Filippi}, {Findeisen}, {Fonti},
  {Fraile}, {Fraser}, {Fr{\'e}zouls}, {Gai}, {Galleti}, {Garabato},
  {Garc{\'\i}a-Sedano}, {Garofalo}, {Garralda}, {Gavel}, {Gavras}, {Gerssen},
  {Geyer}, {Giacobbe}, {Gilmore}, {Girona}, {Giuffrida}, {Glass}, {Gomes},
  {Granvik}, {Gueguen}, {Guerrier}, {Guiraud}, {Guti{\'e}rrez-S{\'a}nchez},
  {Haigron}, {Hatzidimitriou}, {Hauser}, {Haywood}, {Heiter}, {Helmi}, {Heu},
  {Hilger}, {Hobbs}, {Hofmann}, {Holland}, {Huckle}, {Hypki}, {Icardi},
  {Jan{\ss}en}, {Jevardat de Fombelle}, {Jonker}, {Juh{\'a}sz}, {Julbe},
  {Karampelas}, {Kewley}, {Klar}, {Kochoska}, {Kohley}, {Kolenberg},
  {Kontizas}, {Kontizas}, {Koposov}, {Kordopatis}, {Kostrzewa-Rutkowska},
  {Koubsky}, {Lambert}, {Lanza}, {Lasne}, {Lavigne}, {Le Fustec}, {Le
  Poncin-Lafitte}, {Lebreton}, {Leccia}, {Leclerc}, {Lecoeur-Taibi},
  {Lenhardt}, {Leroux}, {Liao}, {Licata}, {Lindstr{\o}m}, {Lister}, {Livanou},
  {Lobel}, {L{\'o}pez}, {Managau}, {Mann}, {Mantelet}, {Marchal}, {Marchant},
  {Marconi}, {Marinoni}, {Marschalk{\'o}}, {Marshall}, {Martino}, {Marton},
  {Mary}, {Massari}, {Matijevi{\v{c}}}, {Mazeh}, {McMillan}, {Messina},
  {Michalik}, {Millar}, {Molina}, {Molinaro}, {Moln{\'a}r}, {Montegriffo},
  {Mor}, {Morbidelli}, {Morel}, {Morris}, {Mulone}, {Muraveva}, {Musella},
  {Nelemans}, {Nicastro}, {Noval}, {O'Mullane}, {Ord{\'e}novic},
  {Ord{\'o}{\~n}ez-Blanco}, {Osborne}, {Pagani}, {Pagano}, {Pailler},
  {Palacin}, {Palaversa}, {Panahi}, {Pawlak}, {Piersimoni}, {Pineau}, {Plachy},
  {Plum}, {Poggio}, {Poujoulet}, {Pr{\v{s}}a}, {Pulone}, {Racero}, {Ragaini},
  {Rambaux}, {Ramos-Lerate}, {Regibo}, {Reyl{\'e}}, {Riclet}, {Ripepi}, {Riva},
  {Rivard}, {Rixon}, {Roegiers}, {Roelens}, {Romero-G{\'o}mez}, {Rowell},
  {Royer}, {Ruiz-Dern}, {Sadowski}, {Sagrist{\`a} Sell{\'e}s}, {Sahlmann},
  {Salgado}, {Salguero}, {Sanna}, {Santana-Ros}, {Sarasso}, {Savietto},
  {Schultheis}, {Sciacca}, {Segol}, {Segovia}, {S{\'e}gransan}, {Shih},
  {Siltala}, {Silva}, {Smart}, {Smith}, {Solano}, {Solitro}, {Sordo}, {Soria
  Nieto}, {Souchay}, {Spagna}, {Spoto}, {Stampa}, {Steele},
  {Steidelm{\"u}ller}, {Stephenson}, {Stoev}, {Suess}, {Surdej}, {Szabados},
  {Szegedi-Elek}, {Tapiador}, {Taris}, {Tauran}, {Taylor}, {Teixeira},
  {Terrett}, {Teyssand ier}, {Thuillot}, {Titarenko}, {Torra Clotet}, {Turon},
  {Ulla}, {Utrilla}, {Uzzi}, {Vaillant}, {Valentini}, {Valette}, {van Elteren},
  {Van Hemelryck}, {van Leeuwen}, {Vaschetto}, {Vecchiato}, {Veljanoski},
  {Viala}, {Vicente}, {Vogt}, {von Essen}, {Voss}, {Votruba}, {Voutsinas},
  {Walmsley}, {Weiler}, {Wertz}, {Wevers}, {Wyrzykowski}, {Yoldas},
  {{\v{Z}}erjal}, {Ziaeepour}, {Zorec}, {Zschocke}, {Zucker}, {Zurbach}, \&
  {Zwitter}}]{GAIA2018}
{Gaia Collaboration}, {Brown}, A.~G.~A., {Vallenari}, A., {et~al.} 2018, \aap,
  616, A1

\bibitem[{{Gibson} {et~al.}(2013){Gibson}, {Aigrain}, {Barstow}, {Evans},
  {Fletcher}, \& {Irwin}}]{2013MNRAS.428.3680G}
{Gibson}, N.~P., {Aigrain}, S., {Barstow}, J.~K., {et~al.} 2013, MNRAS, 428,
  3680

\bibitem[{{Gillon} {et~al.}(2013){Gillon}, {Anderson}, {Collier-Cameron},
  {Doyle}, {Fumel}, {Hellier}, {Jehin}, {Lendl}, {Maxted}, {Montalb{\'a}n}, \&
  {Others}}]{2013A+A...552A..82G}
{Gillon}, M., {Anderson}, D.~R., {Collier-Cameron}, A., {et~al.} 2013, A\&A,
  552, A82

\bibitem[{{Ginski} {et~al.}(2016){Ginski}, {Mugrauer}, {Seeliger}, {Buder},
  {Errmann}, {Avenhaus}, {Mouillet}, {Maire}, \& {Raetz}}]{Ginski2016a}
{Ginski}, C., {Mugrauer}, M., {Seeliger}, M., {et~al.} 2016, \mnras, 457, 2173

\bibitem[{{Girardi} {et~al.}(2005){Girardi}, {Groenewegen}, {Hatziminaoglou},
  \& {da Costa}}]{Girardi2005}
{Girardi}, L., {Groenewegen}, M.~A.~T., {Hatziminaoglou}, E., \& {da Costa}, L.
  2005, \aap, 436, 895

\bibitem[{{Guerri} {et~al.}(2011){Guerri}, {Daban}, {Robbe-Dubois}, {Douet},
  {Abe}, {Baudrand}, {Carbillet}, {Boccaletti}, {Bendjoya}, {Gouvret}, \&
  {Vakili}}]{Guerri2011}
{Guerri}, G., {Daban}, J.-B., {Robbe-Dubois}, S., {et~al.} 2011, Experimental
  Astronomy, 30, 59

\bibitem[{Guizar-Sicairos {et~al.}(2008)Guizar-Sicairos, Thurman, \&
  Fienup}]{Guizar2008}
Guizar-Sicairos, M., Thurman, S.~T., \& Fienup, J.~R. 2008, Optics letters, 33,
  156

\bibitem[{{Han} {et~al.}(2014){Han}, {Wang}, {Wright}, {Feng}, {Zhao},
  {Fakhouri}, {Brown}, \& {Hancock}}]{Han2014}
{Han}, E., {Wang}, S.~X., {Wright}, J.~T., {et~al.} 2014, \pasp, 126, 827

\bibitem[{{Hartman} {et~al.}(2012){Hartman}, {Bakos}, {B{\'e}ky}, {Torres},
  {Latham}, {Csubry}, {Penev}, {Shporer}, {Fulton}, {Buchhave}, {Johnson},
  {Howard}, {Marcy}, {Fischer}, {Kov{\'a}cs}, {Noyes}, {Esquerdo}, {Everett},
  {Szklen{\'a}r}, {Quinn}, {Bieryla}, {Knox}, {Hinz}, {Sasselov}, {F{\H
  u}r{\'e}sz}, {Stefanik}, {L{\'a}z{\'a}r}, {Papp}, \&
  {S{\'a}ri}}]{2012AJ....144..139H}
{Hartman}, J.~D., {Bakos}, G.~{\'A}., {B{\'e}ky}, B., {et~al.} 2012, \aj, 144,
  139

\bibitem[{{Hartman} {et~al.}(2015){Hartman}, {Bakos}, {Buchhave}, {Torres},
  {Latham}, {Kov{\'a}cs}, {Bhatti}, {Csubry}, {de Val-Borro}, {Penev}, {Huang},
  {B{\'e}ky}, {Bieryla}, {Quinn}, {Howard}, {Marcy}, {Johnson}, {Isaacson},
  {Fischer}, {Noyes}, {Falco}, {Esquerdo}, {Knox}, {Hinz}, {L{\'a}z{\'a}r},
  {Papp}, \& {S{\'a}ri}}]{2015AJ....150..197H}
{Hartman}, J.~D., {Bakos}, G.~{\'A}., {Buchhave}, L.~A., {et~al.} 2015, \aj,
  150, 197

\bibitem[{{Hay} {et~al.}(2016){Hay}, {Collier-Cameron}, {Doyle}, {H{\'e}brard},
  {Skillen}, {Anderson}, {Barros}, {Brown}, {Bouchy}, {Busuttil}, {Delorme},
  {Delrez}, {Demangeon}, {D{\'{\i}}az}, {Gillon}, {G{\'o}mez Maqueo Chew},
  {Gonz{\`a}lez}, {Hellier}, {Holmes}, {Jarvis}, {Jehin}, {Joshi}, {Kolb},
  {Lendl}, {Maxted}, {McCormac}, {Miller}, {Mortier}, {Pall{\'e}}, {Pollacco},
  {Prieto-Arranz}, {Queloz}, {S{\'e}gransan}, {Simpson}, {Smalley},
  {Southworth}, {Triaud}, {Turner}, {Udry}, {Vanhuysse}, {West}, \&
  {Wilson}}]{2016MNRAS.463.3276H}
{Hay}, K.~L., {Collier-Cameron}, A., {Doyle}, A.~P., {et~al.} 2016, \mnras,
  463, 3276

\bibitem[{{Hellier} {et~al.}(2017){Hellier}, {Anderson}, {Cameron}, {Delrez},
  {Gillon}, {Jehin}, {Lendl}, {Maxted}, {Neveu-VanMalle}, {Pepe}, {Pollacco},
  {Queloz}, {S{\'e}gransan}, {Smalley}, {Southworth}, {Triaud}, {Udry}, {Wagg},
  \& {West}}]{2017MNRAS.465.3693H}
{Hellier}, C., {Anderson}, D.~R., {Cameron}, A.~C., {et~al.} 2017, \mnras, 465,
  3693

\bibitem[{{Hellier} {et~al.}(2014){Hellier}, {Anderson}, {Cameron}, {Delrez},
  {Gillon}, {Jehin}, {Lendl}, {Maxted}, {Pepe}, {Pollacco}, {Queloz},
  {S{\'e}gransan}, {Smalley}, {Smith}, {Southworth}, {Triaud}, {Udry}, \&
  {West}}]{2014MNRAS.440.1982H}
{Hellier}, C., {Anderson}, D.~R., {Cameron}, A.~C., {et~al.} 2014, MNRAS, 440,
  1982

\bibitem[{{Hellier} {et~al.}(2015){Hellier}, {Anderson}, {Collier Cameron},
  {Delrez}, {Gillon}, {Jehin}, {Lendl}, {Maxted}, {Pepe}, {Pollacco}, {Queloz},
  {S{\'e}gransan}, {Smalley}, {Smith}, {Southworth}, {Triaud}, {Turner},
  {Udry}, \& {West}}]{2015AJ....150...18H}
{Hellier}, C., {Anderson}, D.~R., {Collier Cameron}, A., {et~al.} 2015, \aj,
  150, 18

\bibitem[{{Hellier} {et~al.}(2010){Hellier}, {Anderson}, {Collier Cameron},
  {Gillon}, {Lendl}, {Maxted}, {Queloz}, {Smalley}, {Triaud}, {West}, {Brown},
  {Enoch}, {Lister}, {Pepe}, {Pollacco}, {S{\'e}gransan}, \&
  {Udry}}]{2010ApJ...723L..60H}
{Hellier}, C., {Anderson}, D.~R., {Collier Cameron}, A., {et~al.} 2010, \apjl,
  723, L60

\bibitem[{{Hellier} {et~al.}(2009){Hellier}, {Anderson}, {Gillon}, {Lister},
  {Maxted}, {Queloz}, {Smalley}, {Triaud}, {West}, {Wilson}, \&
  {Others}}]{2009ApJ...690L..89H}
{Hellier}, C., {Anderson}, D.~R., {Gillon}, M., {et~al.} 2009, ApJ, 690, L89

\bibitem[{{Hunter}(2007)}]{Matplotlib}
{Hunter}, J.~D. 2007, Computing in Science and Engineering, 9, 90

\bibitem[{{Kley} \& {Nelson}(2012)}]{Kley2012}
{Kley}, W. \& {Nelson}, R.~P. 2012, \araa, 50, 211

\bibitem[{{Kuhn} {et~al.}(2016){Kuhn}, {Rodriguez}, {Collins}, {Lund},
  {Siverd}, {Col{\'o}n}, {Pepper}, {Stassun}, {Cargile}, {James}, {Penev},
  {Zhou}, {Bayliss}, {Tan}, {Curtis}, {Udry}, {Segransan}, {Mawet}, {Dhital},
  {Soutter}, {Hart}, {Carter}, {Gaudi}, {Myers}, {Beatty}, {Eastman},
  {Reichart}, {Haislip}, {Kielkopf}, {Bieryla}, {Latham}, {Jensen}, {Oberst},
  \& {Stevens}}]{2016MNRAS.459.4281K}
{Kuhn}, R.~B., {Rodriguez}, J.~E., {Collins}, K.~A., {et~al.} 2016, \mnras,
  459, 4281

\bibitem[{{Lai} {et~al.}(2011){Lai}, {Foucart}, \& {Lin}}]{Lai2011}
{Lai}, D., {Foucart}, F., \& {Lin}, D. N.~C. 2011, in IAU Symposium, Vol. 276,
  The Astrophysics of Planetary Systems: Formation, Structure, and Dynamical
  Evolution, ed. A.~{Sozzetti}, M.~G. {Lattanzi}, \& A.~P. {Boss}, 295--299

\bibitem[{{Lam} {et~al.}(2017){Lam}, {Faedi}, {Brown}, {Anderson}, {Delrez},
  {Gillon}, {H{\'e}brard}, {Lendl}, {Mancini}, {Southworth}, {Smalley},
  {Triaud}, {Turner}, {Hay}, {Armstrong}, {Barros}, {Bonomo}, {Bouchy},
  {Boumis}, {Collier Cameron}, {Doyle}, {Hellier}, {Henning}, {Jehin}, {King},
  {Kirk}, {Louden}, {Maxted}, {McCormac}, {Osborn}, {Palle}, {Pepe},
  {Pollacco}, {Prieto-Arranz}, {Queloz}, {Rey}, {S{\'e}gransan}, {Udry},
  {Walker}, {West}, \& {Wheatley}}]{2017A+A...599A...3L}
{Lam}, K.~W.~F., {Faedi}, F., {Brown}, D.~J.~A., {et~al.} 2017, \aap, 599, A3

\bibitem[{{Law} {et~al.}(2014){Law}, {Morton}, {Baranec}, {Riddle},
  {Ravichandran}, {Ziegler}, {Johnson}, {Tendulkar}, {Bui}, {Burse}, {Das},
  {Dekany}, {Kulkarni}, {Punnadi}, \& {Ramaprakash}}]{Law2014}
{Law}, N.~M., {Morton}, T., {Baranec}, C., {et~al.} 2014, \apj, 791, 35

\bibitem[{{Lendl} {et~al.}(2014){Lendl}, {Triaud}, {Anderson}, {Collier
  Cameron}, {Delrez}, {Doyle}, {Gillon}, {Hellier}, {Jehin}, {Maxted},
  {Neveu-VanMalle}, {Pepe}, {Pollacco}, {Queloz}, {S{\'e}gransan}, {Smalley},
  {Smith}, {Udry}, {Van Grootel}, \& {West}}]{2014A+A...568A..81L}
{Lendl}, M., {Triaud}, A.~H.~M.~J., {Anderson}, D.~R., {et~al.} 2014, \aap,
  568, A81

\bibitem[{{Lillo-Box} {et~al.}(2014){Lillo-Box}, {Barrado}, \&
  {Bouy}}]{Lillo-Box2014}
{Lillo-Box}, J., {Barrado}, D., \& {Bouy}, H. 2014, \aap, 566, A103

\bibitem[{{Lin} {et~al.}(1996){Lin}, {Bodenheimer}, \& {Richardson}}]{Lin1996}
{Lin}, D.~N.~C., {Bodenheimer}, P., \& {Richardson}, D.~C. 1996, \nat, 380, 606

\bibitem[{{Lister} {et~al.}(2009){Lister}, {Anderson}, {Gillon}, {Hebb},
  {Smalley}, {Triaud}, {Collier Cameron}, {Wilson}, {West}, {Bentley}, \&
  {Others}}]{2009ApJ...703..752L}
{Lister}, T.~A., {Anderson}, D.~R., {Gillon}, M., {et~al.} 2009, ApJ, 703, 752

\bibitem[{{Maire} {et~al.}(2016){Maire}, {Langlois}, {Dohlen}, {Lagrange},
  {Gratton}, {Chauvin}, {Desidera}, {Girard}, {Milli}, {Vigan}, {Zins},
  {Delorme}, {Beuzit}, {Claudi}, {Feldt}, {Mouillet}, {Puget}, {Turatto}, \&
  {Wildi}}]{Maire2016}
{Maire}, A.-L., {Langlois}, M., {Dohlen}, K., {et~al.} 2016, in Ground-based
  and Airborne Instrumentation for Astronomy VI, Vol. 9908, 990834

\bibitem[{{Mancini} {et~al.}(2014){Mancini}, {Southworth}, {Ciceri}, {Dominik},
  {Henning}, {J{\o}rgensen}, {Lanza}, {Rabus}, {Snodgrass}, {Vilela}, \&
  {Others}}]{2014A+A...562A.126M}
{Mancini}, L., {Southworth}, J., {Ciceri}, S., {et~al.} 2014, A\&A, 562, A126

\bibitem[{{Mancini} {et~al.}(2019){Mancini}, {Southworth}, {Molli{\`e}re},
  {Tregloan-Reed}, {Juvan}, {Chen}, {Sarkis}, {Bruni}, {Ciceri}, {Andersen},
  {Bozza}, {Bramich}, {Burgdorf}, {D'Ago}, {Dominik}, {Evans}, {Figuera
  Jaimes}, {Fossati}, {Henning}, {Hinse}, {Hundertmark}, {J{\o}rgensen},
  {Kerins}, {Korhonen}, {K{\"u}ffmeier}, {Longa}, {Peixinho}, {Popovas},
  {Rabus}, {Rahvar}, {Skottfelt}, {Snodgrass}, {Tronsgaard}, {Wang}, \&
  {Wertz}}]{Mancini2019}
{Mancini}, L., {Southworth}, J., {Molli{\`e}re}, P., {et~al.} 2019, \mnras,
  485, 5168

\bibitem[{{Maxted} {et~al.}(2015){Maxted}, {Serenelli}, \&
  {Southworth}}]{Maxted2015}
{Maxted}, P.~F.~L., {Serenelli}, A.~M., \& {Southworth}, J. 2015, \aap, 577,
  A90

\bibitem[{{Mayor} {et~al.}(2003){Mayor}, {Pepe}, {Queloz}, {Bouchy},
  {Rupprecht}, {Lo Curto}, {Avila}, {Benz}, {Bertaux}, {Bonfils}, {Dall},
  {Dekker}, {Delabre}, {Eckert}, {Fleury}, {Gilliotte}, {Gojak}, {Guzman},
  {Kohler}, {Lizon}, {Longinotti}, {Lovis}, {Megevand}, {Pasquini}, {Reyes},
  {Sivan}, {Sosnowska}, {Soto}, {Udry}, {van Kesteren}, {Weber}, \&
  {Weilenmann}}]{Mayor2003}
{Mayor}, M., {Pepe}, F., {Queloz}, D., {et~al.} 2003, The Messenger, 114, 20

\bibitem[{{Mayor} \& {Queloz}(1995)}]{Mayor1995}
{Mayor}, M. \& {Queloz}, D. 1995, \nat, 378, 355

\bibitem[{{Mo{\v c}nik} {et~al.}(2017){Mo{\v c}nik}, {Hellier}, {Anderson},
  {Clark}, \& {Southworth}}]{2017MNRAS.469.1622M}
{Mo{\v c}nik}, T., {Hellier}, C., {Anderson}, D.~R., {Clark}, B.~J.~M., \&
  {Southworth}, J. 2017, MNRAS, 469, 1622

\bibitem[{{Nagasawa} {et~al.}(2008){Nagasawa}, {Ida}, \&
  {Bessho}}]{Nagasawa2008}
{Nagasawa}, M., {Ida}, S., \& {Bessho}, T. 2008, \apj, 678, 498

\bibitem[{{Neveu-VanMalle} {et~al.}(2014){Neveu-VanMalle}, {Queloz},
  {Anderson}, {Charbonnel}, {Collier Cameron}, {Delrez}, {Gillon}, {Hellier},
  {Jehin}, {Lendl}, \& {Others}}]{2014A+A...572A..49N}
{Neveu-VanMalle}, M., {Queloz}, D., {Anderson}, D.~R., {et~al.} 2014, A\&A,
  572, A49

\bibitem[{{Ngo} {et~al.}(2017){Ngo}, {Knutson}, {Bryan}, {Blunt}, {Nielsen},
  {Batygin}, {Bowler}, {Crepp}, {Hinkley}, {Howard}, \& {Mawet}}]{Ngo2017}
{Ngo}, H., {Knutson}, H.~A., {Bryan}, M.~L., {et~al.} 2017, \aj, 153, 242

\bibitem[{{Ngo} {et~al.}(2016){Ngo}, {Knutson}, {Hinkley}, {Bryan}, {Crepp},
  {Batygin}, {Crossfield}, {Hansen}, {Howard}, {Johnson}, {Mawet}, {Morton},
  {Muirhead}, \& {Wang}}]{Ngo2016}
{Ngo}, H., {Knutson}, H.~A., {Hinkley}, S., {et~al.} 2016, \apj, 827, 8

\bibitem[{{Ngo} {et~al.}(2015){Ngo}, {Knutson}, {Hinkley}, {Crepp}, {Bechter},
  {Batygin}, {Howard}, {Johnson}, {Morton}, \& {Muirhead}}]{Ngo2015}
{Ngo}, H., {Knutson}, H.~A., {Hinkley}, S., {et~al.} 2015, \apj, 800, 138

\bibitem[{Oliphant(2006)}]{numpy}
Oliphant, T.~E. 2006, A guide to NumPy, Vol.~1 (Trelgol Publishing USA)

\bibitem[{{Pedregosa} {et~al.}(2012){Pedregosa}, {Varoquaux}, {Gramfort},
  {Michel}, {Thirion}, {Grisel}, {Blondel}, {M{\"u}ller}, {Nothman}, {Louppe},
  {Prettenhofer}, {Weiss}, {Dubourg}, {Vanderplas}, {Passos}, {Cournapeau},
  {Brucher}, {Perrot}, \& {Duchesnay}}]{scikit-learn}
{Pedregosa}, F., {Varoquaux}, G., {Gramfort}, A., {et~al.} 2012, arXiv
  e-prints, arXiv:1201.0490

\bibitem[{{Petigura} {et~al.}(2018){Petigura}, {Benneke}, {Batygin}, {Fulton},
  {Werner}, {Krick}, {Gorjian}, {Sinukoff}, {Deck}, {Mills}, \&
  {Deming}}]{2018AJ....156...89P}
{Petigura}, E.~A., {Benneke}, B., {Batygin}, K., {et~al.} 2018, \aj, 156, 89

\bibitem[{{Petigura} {et~al.}(2016){Petigura}, {Howard}, {Lopez}, {Deck},
  {Fulton}, {Crossfield}, {Ciardi}, {Chiang}, {Lee}, {Isaacson}, {Beichman},
  {Hansen}, {Schlieder}, \& {Sinukoff}}]{2016ApJ...818...36P}
{Petigura}, E.~A., {Howard}, A.~W., {Lopez}, E.~D., {et~al.} 2016, \apj, 818,
  36

\bibitem[{{Petigura} {et~al.}(2017){Petigura}, {Sinukoff}, {Lopez},
  {Crossfield}, {Howard}, {Brewer}, {Fulton}, {Isaacson}, {Ciardi}, {Howell},
  {Everett}, {Horch}, {Hirsch}, {Weiss}, \& {Schlieder}}]{2017AJ....153..142P}
{Petigura}, E.~A., {Sinukoff}, E., {Lopez}, E.~D., {et~al.} 2017, \aj, 153, 142

\bibitem[{{Pollacco} {et~al.}(2006){Pollacco}, {Skillen}, {Collier Cameron},
  {Christian}, {Hellier}, {Irwin}, {Lister}, {Street}, {West}, {Anderson},
  {Clarkson}, {Deeg}, {Enoch}, {Evans}, {Fitzsimmons}, {Haswell}, {Hodgkin},
  {Horne}, {Kane}, {Keenan}, {Maxted}, {Norton}, {Osborne}, {Parley}, {Ryans},
  {Smalley}, {Wheatley}, \& {Wilson}}]{Pollacco2006}
{Pollacco}, D.~L., {Skillen}, I., {Collier Cameron}, A., {et~al.} 2006, \pasp,
  118, 1407

\bibitem[{{Pollack} {et~al.}(1996){Pollack}, {Hubickyj}, {Bodenheimer},
  {Lissauer}, {Podolak}, \& {Greenzweig}}]{Pollack1996}
{Pollack}, J.~B., {Hubickyj}, O., {Bodenheimer}, P., {et~al.} 1996, \icarus,
  124, 62

\bibitem[{{Queloz} {et~al.}(2010){Queloz}, {Anderson}, {Collier Cameron},
  {Gillon}, {Hebb}, {Hellier}, {Maxted}, {Pepe}, {Pollacco}, {S{\'e}gransan},
  {Smalley}, {Triaud}, {Udry}, \& {West}}]{2010A+A...517L...1Q}
{Queloz}, D., {Anderson}, D., {Collier Cameron}, A., {et~al.} 2010, A\&A, 517,
  L1

\bibitem[{{Rasio} \& {Ford}(1996)}]{Rasio1996}
{Rasio}, F.~A. \& {Ford}, E.~B. 1996, Science, 274, 954

\bibitem[{{Serenelli} {et~al.}(2013){Serenelli}, {Bergemann}, {Ruchti}, \&
  {Casagrande}}]{Serenelli2013}
{Serenelli}, A.~M., {Bergemann}, M., {Ruchti}, G., \& {Casagrande}, L. 2013,
  \mnras, 429, 3645

\bibitem[{{Sinukoff} {et~al.}(2016){Sinukoff}, {Howard}, {Petigura},
  {Schlieder}, {Crossfield}, {Ciardi}, {Fulton}, {Isaacson}, {Aller},
  {Baranec}, {Beichman}, {Hansen}, {Knutson}, {Law}, {Liu}, {Riddle}, \&
  {Dressing}}]{2016ApJ...827...78S}
{Sinukoff}, E., {Howard}, A.~W., {Petigura}, E.~A., {et~al.} 2016, \apj, 827,
  78

\bibitem[{{Smith} {et~al.}(2013){Smith}, {Anderson}, {Bouchy}, {Collier
  Cameron}, {Doyle}, {Fumel}, {Gillon}, {H{\'e}brard}, {Hellier}, {Jehin},
  {Lendl}, {Maxted}, {Moutou}, {Pepe}, {Pollacco}, {Queloz}, {Santerne},
  {Segransan}, {Smalley}, {Southworth}, {Triaud}, {Udry}, \&
  {West}}]{2013A+A...552A.120S}
{Smith}, A.~M.~S., {Anderson}, D.~R., {Bouchy}, F., {et~al.} 2013, \aap, 552,
  A120

\bibitem[{{Smith} {et~al.}(2017){Smith}, {Gandolfi}, {Barrag{\'a}n}, {Bowler},
  {Csizmadia}, {Endl}, {Fridlund}, {Grziwa}, {Guenther}, {Hatzes}, {Nowak},
  {Albrecht}, {Alonso}, {Cabrera}, {Cochran}, {Deeg}, {Cusano},
  {Eigm{\"u}ller}, {Erikson}, {Hidalgo}, {Hirano}, {Johnson}, {Korth}, {Mann},
  {Narita}, {Nespral}, {Palle}, {P{\"a}tzold}, {Prieto-Arranz}, {Rauer},
  {Ribas}, {Tingley}, \& {Wolthoff}}]{2017MNRAS.464.2708S}
{Smith}, A.~M.~S., {Gandolfi}, D., {Barrag{\'a}n}, O., {et~al.} 2017, \mnras,
  464, 2708

\bibitem[{{Socrates} {et~al.}(2012){Socrates}, {Katz}, {Dong}, \&
  {Tremaine}}]{Socrates2012}
{Socrates}, A., {Katz}, B., {Dong}, S., \& {Tremaine}, S. 2012, \apj, 750, 106

\bibitem[{{Soummer}(2005)}]{Soummer2005}
{Soummer}, R. 2005, \apj, 618, L161

\bibitem[{{Southworth}(2011)}]{Southworth2011}
{Southworth}, J. 2011, \mnras, 417, 2166

\bibitem[{{Southworth}(2012)}]{2012MNRAS.426.1291S}
{Southworth}, J. 2012, MNRAS, 426, 1291

\bibitem[{{Southworth} {et~al.}(2020){Southworth}, {Bohn}, {Ginski},
  {Kenworthy}, \& {Mancini}}]{paper2}
{Southworth}, J., {Bohn}, A., {Ginski}, C., {Kenworthy}, M., \& {Mancini}, L.
  2020, \aap, submitted

\bibitem[{{Southworth} {et~al.}(2013){Southworth}, {Mancini}, {Browne},
  {Burgdorf}, {Calchi Novati}, {Dominik}, {Gerner}, {Hinse}, {J{\o}rgensen},
  {Kains}, {Ricci}, {Sch{\"a}fer}, {Sch{\"o}nebeck}, {Tregloan-Reed},
  {Alsubai}, {Bozza}, {Chen}, {Dodds}, {Dreizler}, {Fang}, {Finet}, {Gu},
  {Hardis}, {Harps{\o}e}, {Henning}, {Hundertmark}, {Jessen-Hansen}, {Kerins},
  {Kjeldsen}, {Liebig}, {Lund}, {Lundkvist}, {Mathiasen}, {Nikolov}, {Penny},
  {Proft}, {Rahvar}, {Sahu}, {Scarpetta}, {Skottfelt}, {Snodgrass}, {Surdej},
  \& {Wertz}}]{2013MNRAS.434.1300S}
{Southworth}, J., {Mancini}, L., {Browne}, P., {et~al.} 2013, MNRAS, 434, 1300

\bibitem[{{Stolker} {et~al.}(2019){Stolker}, {Bonse}, {Quanz}, {Amara},
  {Cugno}, {Bohn}, \& {Boehle}}]{Stolker2019}
{Stolker}, T., {Bonse}, M.~J., {Quanz}, S.~P., {et~al.} 2019, \aap, 621, A59

\bibitem[{{Triaud} {et~al.}(2013{\natexlab{a}}){Triaud}, {Anderson}, {Collier
  Cameron}, {Doyle}, {Fumel}, {Gillon}, {Hellier}, {Jehin}, {Lendl}, {Lovis},
  {Maxted}, {Pepe}, {Pollacco}, {Queloz}, {S{\'e}gransan}, {Smalley}, {Smith},
  {Udry}, {West}, \& {Wheatley}}]{2013A+A...551A..80T}
{Triaud}, A.~H.~M.~J., {Anderson}, D.~R., {Collier Cameron}, A., {et~al.}
  2013{\natexlab{a}}, \aap, 551, A80

\bibitem[{{Triaud} {et~al.}(2013{\natexlab{b}}){Triaud}, {Hebb}, {Anderson},
  {Cargile}, {Collier Cameron}, {Doyle}, {Faedi}, {Gillon}, {Gomez Maqueo
  Chew}, {Hellier}, \& {Others}}]{2013A+A...549A..18T}
{Triaud}, A.~H.~M.~J., {Hebb}, L., {Anderson}, D.~R., {et~al.}
  2013{\natexlab{b}}, A\&A, 549, A18

\bibitem[{{Turner} {et~al.}(2016){Turner}, {Anderson}, {Collier Cameron},
  {Delrez}, {Evans}, {Gillon}, {Hellier}, {Jehin}, {Lendl}, {Maxted}, {Pepe},
  {Pollacco}, {Queloz}, {S{\'e}gransan}, {Smalley}, {Smith}, {Triaud}, {Udry},
  \& {West}}]{2016PASP..128f4401T}
{Turner}, O.~D., {Anderson}, D.~R., {Collier Cameron}, A., {et~al.} 2016,
  \pasp, 128, 064401

\bibitem[{{Udry} \& {Santos}(2007)}]{Udry2007}
{Udry}, S. \& {Santos}, N.~C. 2007, \araa, 45, 397

\bibitem[{Van~der Walt {et~al.}(2014)Van~der Walt, Sch{\"o}nberger,
  Nunez-Iglesias, Boulogne, Warner, Yager, Gouillart, \& Yu}]{scikit-image}
Van~der Walt, S., Sch{\"o}nberger, J.~L., Nunez-Iglesias, J., {et~al.} 2014,
  PeerJ, 2, e453

\bibitem[{{Van Eylen} {et~al.}(2016){Van Eylen}, {Albrecht}, {Gandolfi}, {Dai},
  {Winn}, {Hirano}, {Narita}, {Bruntt}, {Prieto-Arranz}, {B{\'e}jar}, {Nowak},
  {Lund}, {Palle}, {Ribas}, {Sanchis-Ojeda}, {Yu}, {Arriagada}, {Butler},
  {Crane}, {Handberg}, {Deeg}, {Jessen-Hansen}, {Johnson}, {Nespral}, {Rogers},
  {Ryu}, {Shectman}, {Shrotriya}, {Slumstrup}, {Takeda}, {Teske}, {Thompson},
  {Vanderburg}, \& {Wittenmyer}}]{2016AJ....152..143V}
{Van Eylen}, V., {Albrecht}, S., {Gandolfi}, D., {et~al.} 2016, \aj, 152, 143

\bibitem[{{Vanderburg} {et~al.}(2015){Vanderburg}, {Montet}, {Johnson},
  {Buchhave}, {Zeng}, {Pepe}, {Collier Cameron}, {Latham}, {Molinari}, {Udry},
  {Lovis}, {Matthews}, {Cameron}, {Law}, {Bowler}, {Angus}, {Baranec},
  {Bieryla}, {Boschin}, {Charbonneau}, {Cosentino}, {Dumusque}, {Figueira},
  {Guenther}, {Harutyunyan}, {Hellier}, {Kuschnig}, {Lopez-Morales}, {Mayor},
  {Micela}, {Moffat}, {Pedani}, {Phillips}, {Piotto}, {Pollacco}, {Queloz},
  {Rice}, {Riddle}, {Rowe}, {Rucinski}, {Sasselov}, {S{\'e}gransan},
  {Sozzetti}, {Szentgyorgyi}, {Watson}, \& {Weiss}}]{2015ApJ...800...59V}
{Vanderburg}, A., {Montet}, B.~T., {Johnson}, J.~A., {et~al.} 2015, \apj, 800,
  59

\bibitem[{{Vigan} {et~al.}(2010){Vigan}, {Moutou}, {Langlois}, {Allard},
  {Boccaletti}, {Carbillet}, {Mouillet}, \& {Smith}}]{Vigan2010}
{Vigan}, A., {Moutou}, C., {Langlois}, M., {et~al.} 2010, \mnras, 407, 71

\bibitem[{{Virtanen} {et~al.}(2019){Virtanen}, {Gommers}, {Oliphant},
  {Haberland}, {Reddy}, {Cournapeau}, {Burovski}, {Peterson}, {Weckesser},
  {Bright}, {van der Walt}, {Brett}, {Wilson}, {Jarrod Millman}, {Mayorov},
  {Nelson}, {Jones}, {Kern}, {Larson}, {Carey}, {Polat}, {Feng}, {Moore}, {Vand
  erPlas}, {Laxalde}, {Perktold}, {Cimrman}, {Henriksen}, {Quintero}, {Harris},
  {Archibald}, {Ribeiro}, {Pedregosa}, {van Mulbregt}, \&
  {Contributors}}]{Virtanen2019}
{Virtanen}, P., {Gommers}, R., {Oliphant}, T.~E., {et~al.} 2019, arXiv
  e-prints, arXiv:1907.10121

\bibitem[{{Winn} {et~al.}(2010){Winn}, {Fabrycky}, {Albrecht}, \&
  {Johnson}}]{Winn2010}
{Winn}, J.~N., {Fabrycky}, D., {Albrecht}, S., \& {Johnson}, J.~A. 2010, \apj,
  718, L145

\bibitem[{{W{\"o}llert} \& {Brandner}(2015)}]{Wollert2015b}
{W{\"o}llert}, M. \& {Brandner}, W. 2015, \aap, 579, A129

\bibitem[{{W{\"o}llert} {et~al.}(2015){W{\"o}llert}, {Brandner}, {Bergfors}, \&
  {Henning}}]{Wollert2015a}
{W{\"o}llert}, M., {Brandner}, W., {Bergfors}, C., \& {Henning}, T. 2015, \aap,
  575, A23

\bibitem[{{Wu} \& {Lithwick}(2011)}]{Wu2011}
{Wu}, Y. \& {Lithwick}, Y. 2011, \apj, 735, 109

\bibitem[{{Wu} \& {Murray}(2003)}]{Wu2003}
{Wu}, Y. \& {Murray}, N. 2003, \apj, 589, 605

\bibitem[{{Zacharias} {et~al.}(2013){Zacharias}, {Finch}, {Girard}, {Henden},
  {Bartlett}, {Monet}, \& {Zacharias}}]{Zacharias2013}
{Zacharias}, N., {Finch}, C.~T., {Girard}, T.~M., {et~al.} 2013, \aj, 145, 44

\end{thebibliography}

\begin{appendix}

\section{Individual detection limits}
\label{sec:individual_detection_limits}

\begin{figure*}
\centering

\begin{subfigure}[b]{0.3\textwidth}
\includegraphics[width=\textwidth]{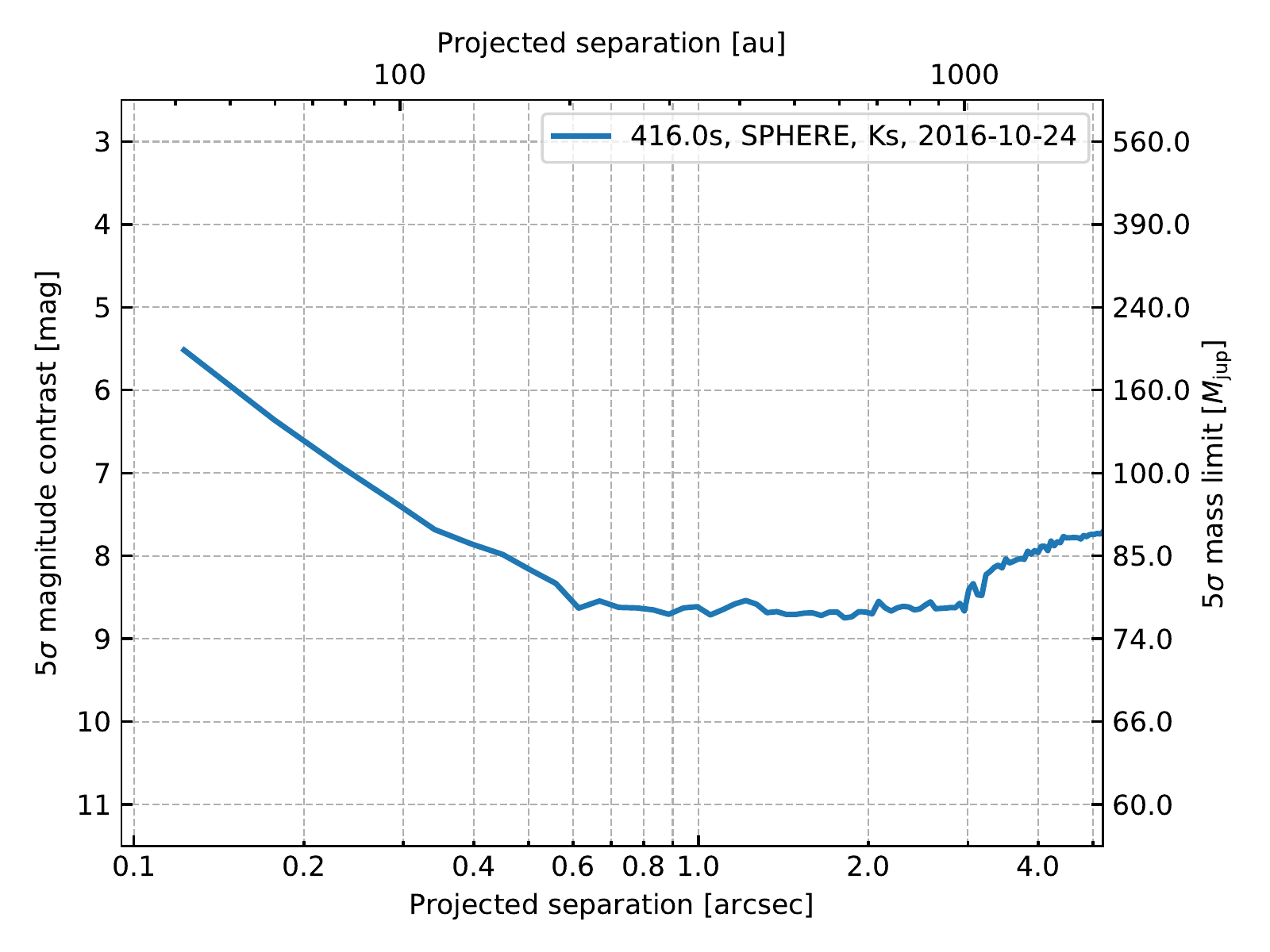}
\subcaption{HAT-P-41}
\end{subfigure}
\begin{subfigure}[b]{0.3\textwidth}
\includegraphics[width=\textwidth]{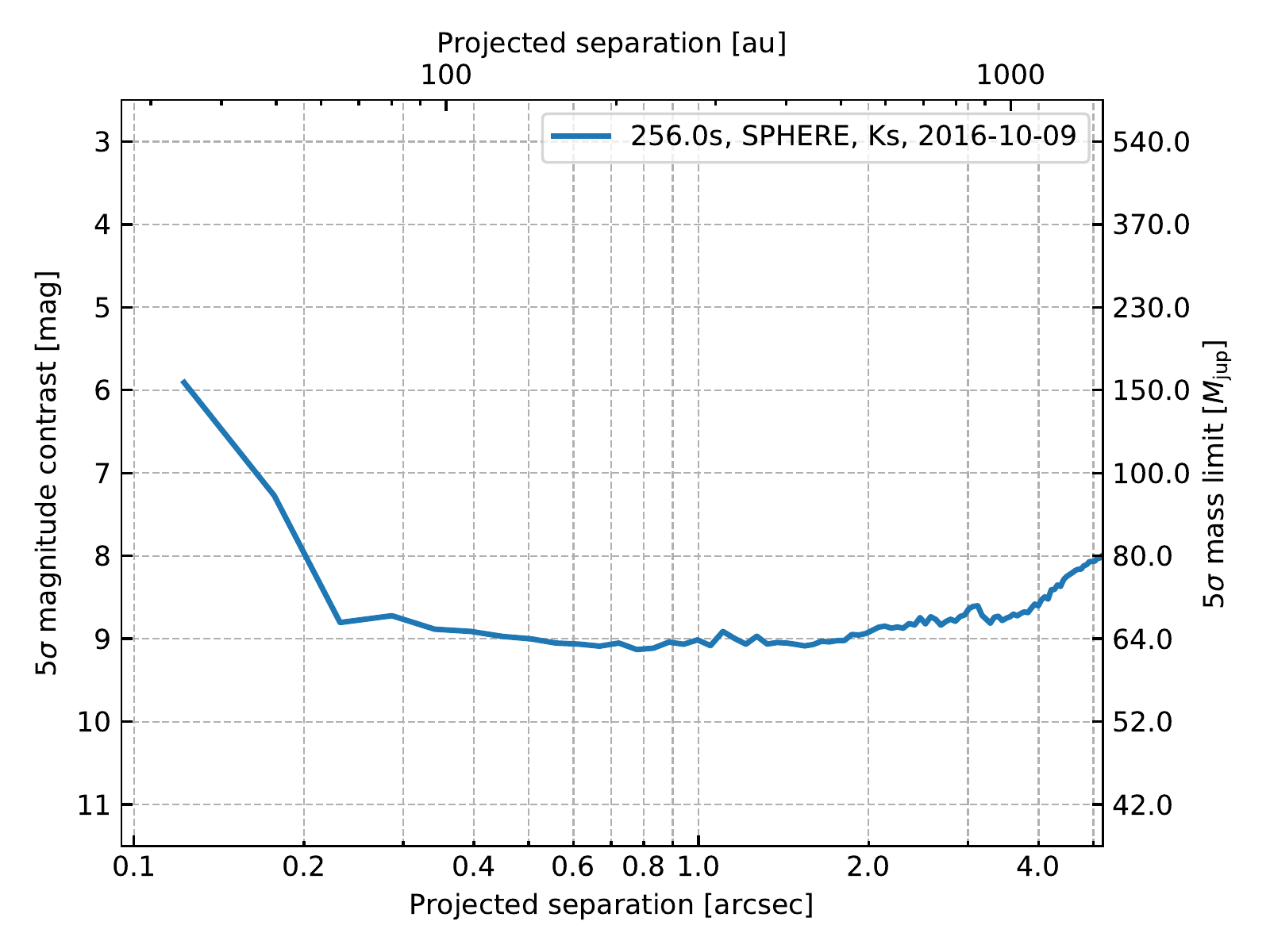}
\subcaption{HAT-P-57}
\end{subfigure}
\begin{subfigure}[b]{0.3\textwidth}
\includegraphics[width=\textwidth]{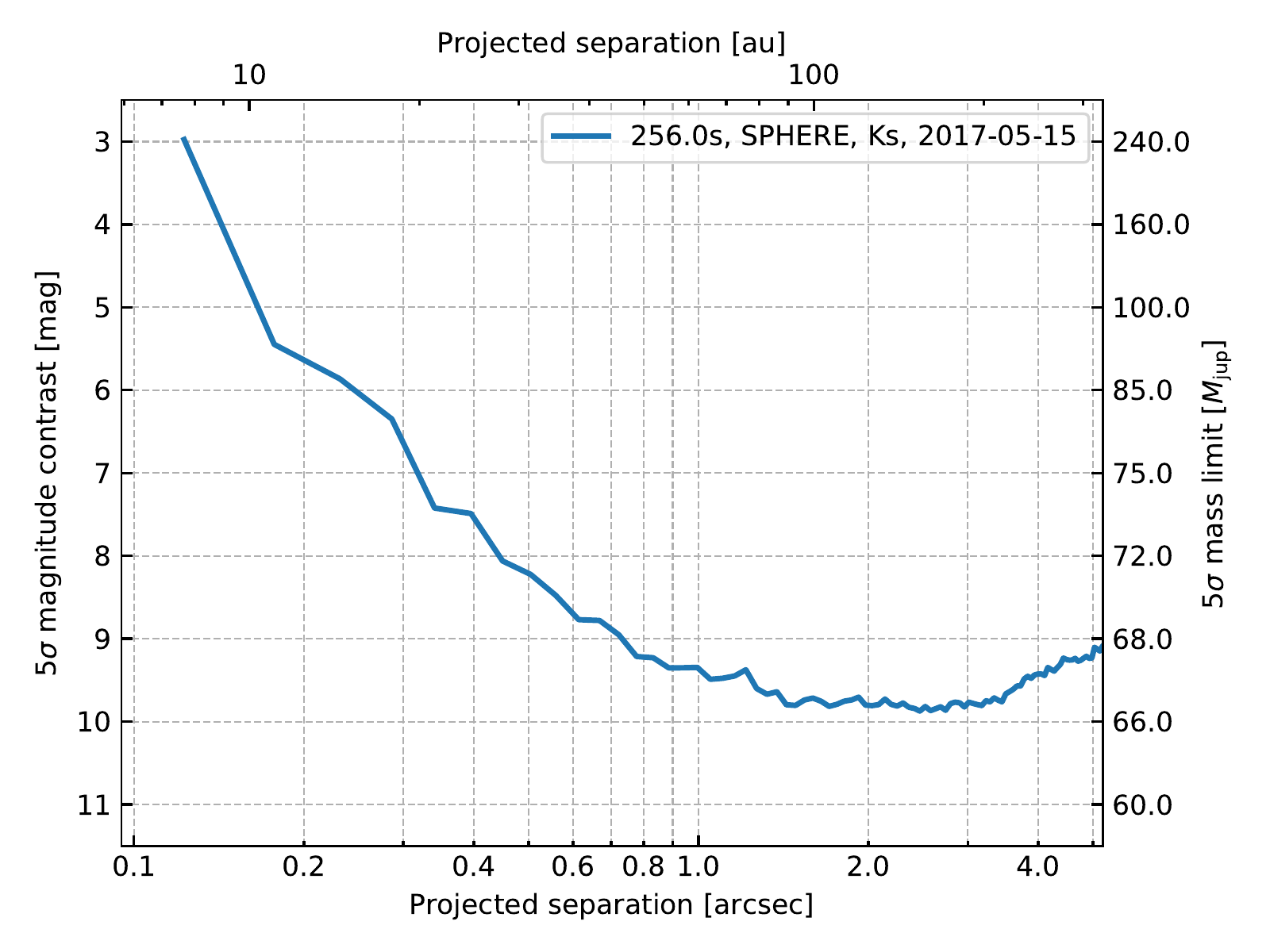}
\subcaption{K2-02}
\end{subfigure}

\begin{subfigure}[b]{0.3\textwidth}
\includegraphics[width=\textwidth]{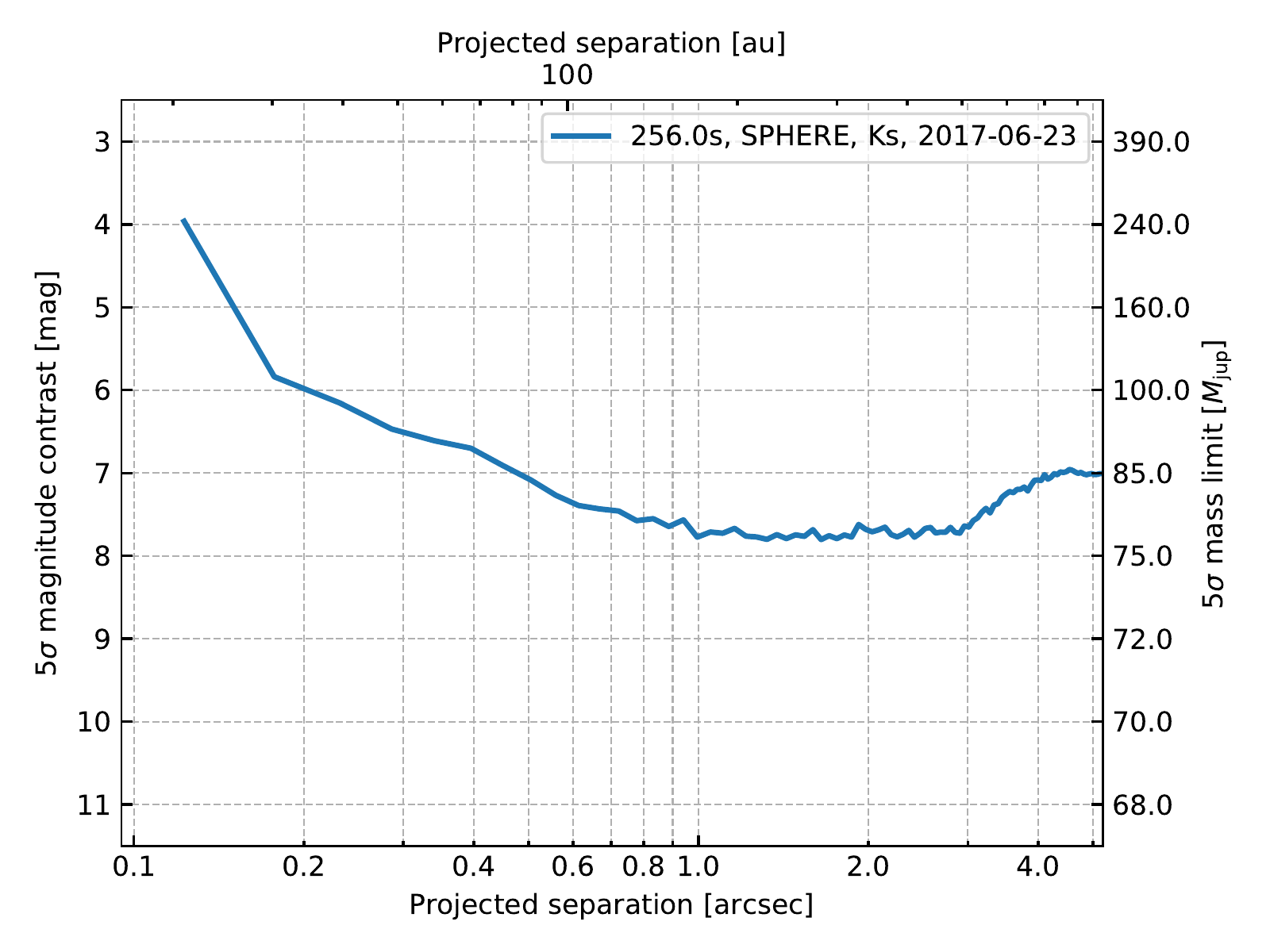}
\subcaption{K2-24}
\end{subfigure}
\begin{subfigure}[b]{0.3\textwidth}
\includegraphics[width=\textwidth]{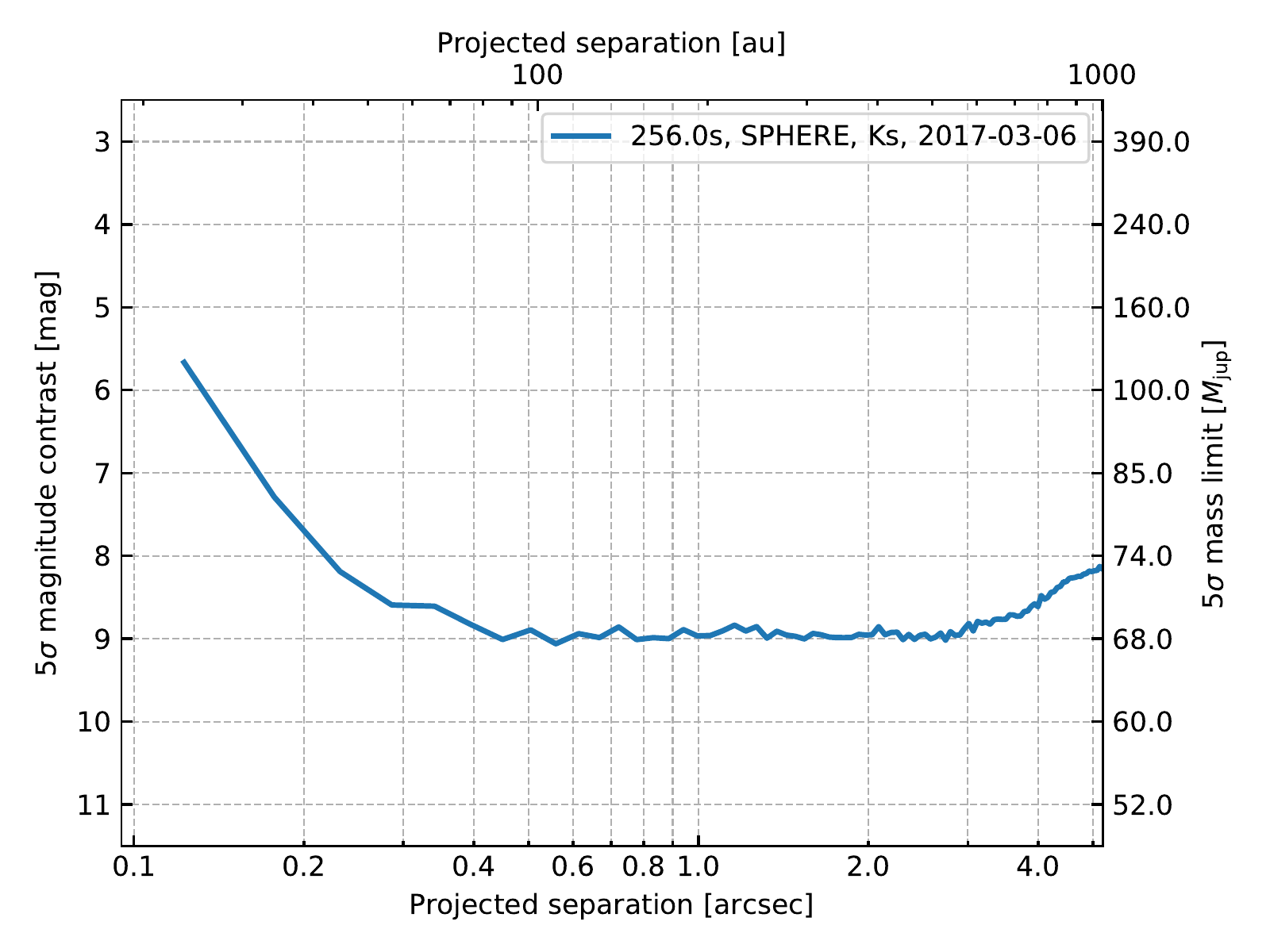}
\subcaption{K2-38}
\end{subfigure}
\begin{subfigure}[b]{0.3\textwidth}
\includegraphics[width=\textwidth]{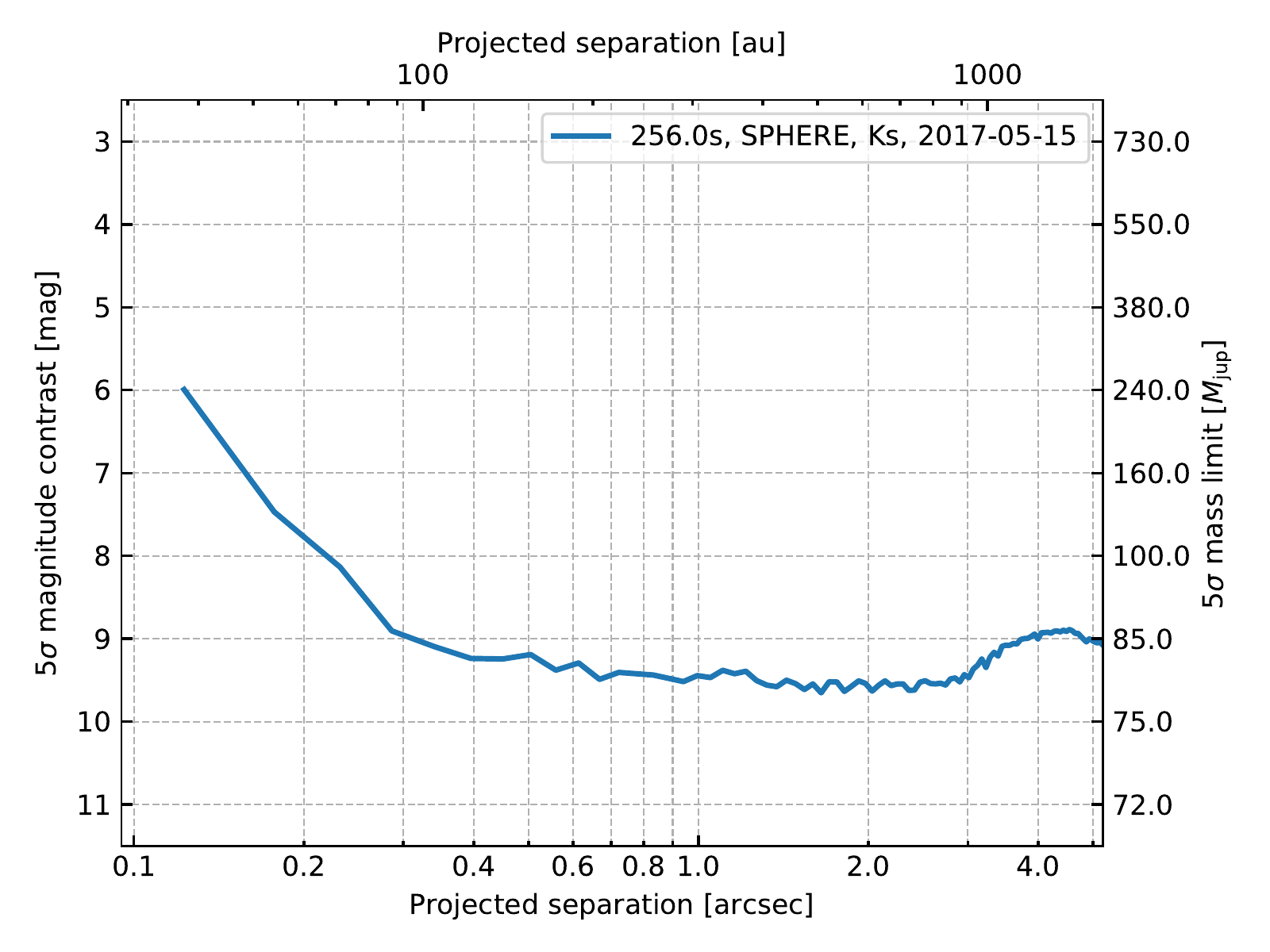}
\subcaption{K2-39}
\end{subfigure}

\begin{subfigure}[b]{0.3\textwidth}
\includegraphics[width=\textwidth]{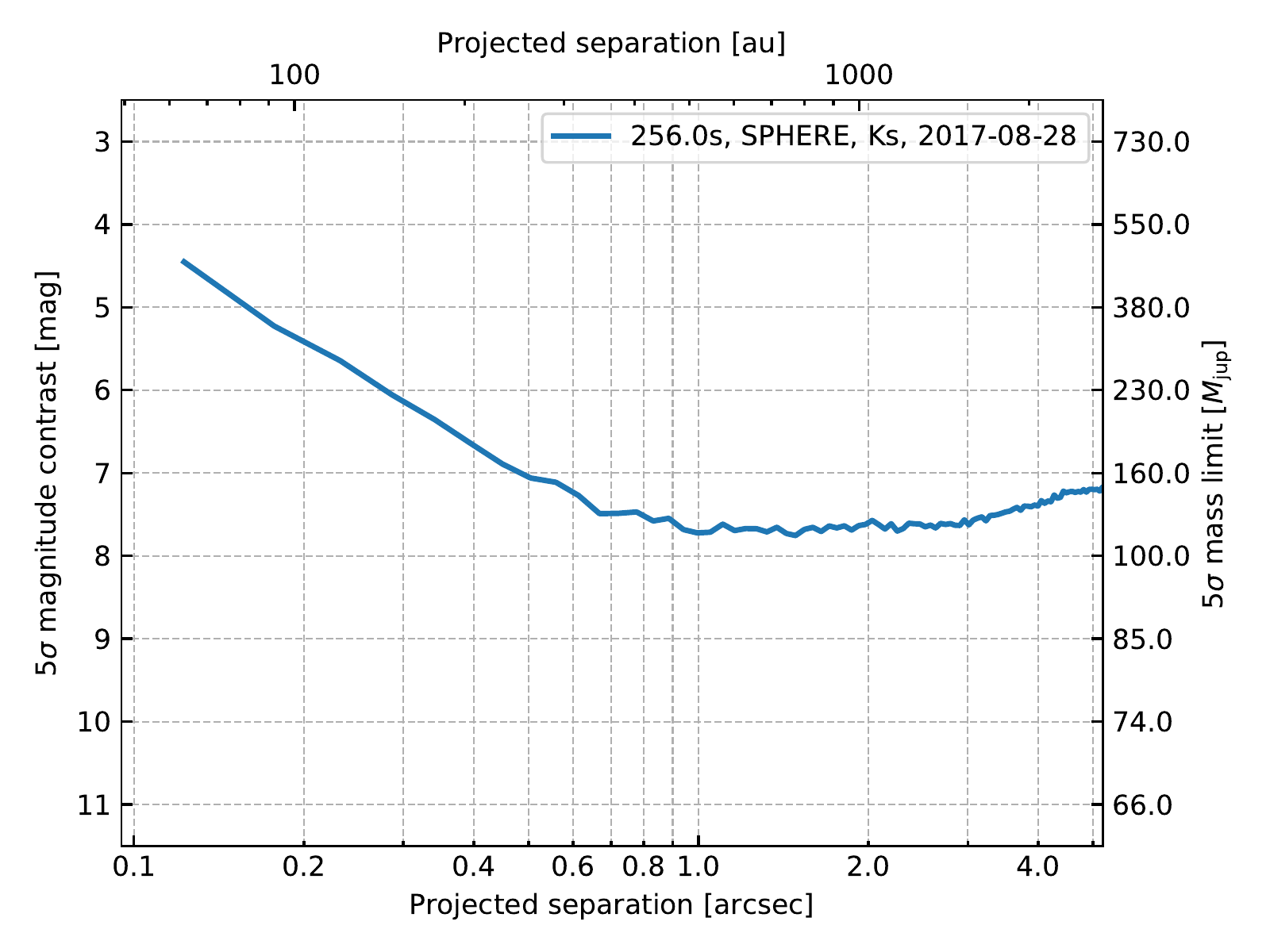}
\subcaption{K2-99}
\end{subfigure}
\begin{subfigure}[b]{0.3\textwidth}
\includegraphics[width=\textwidth]{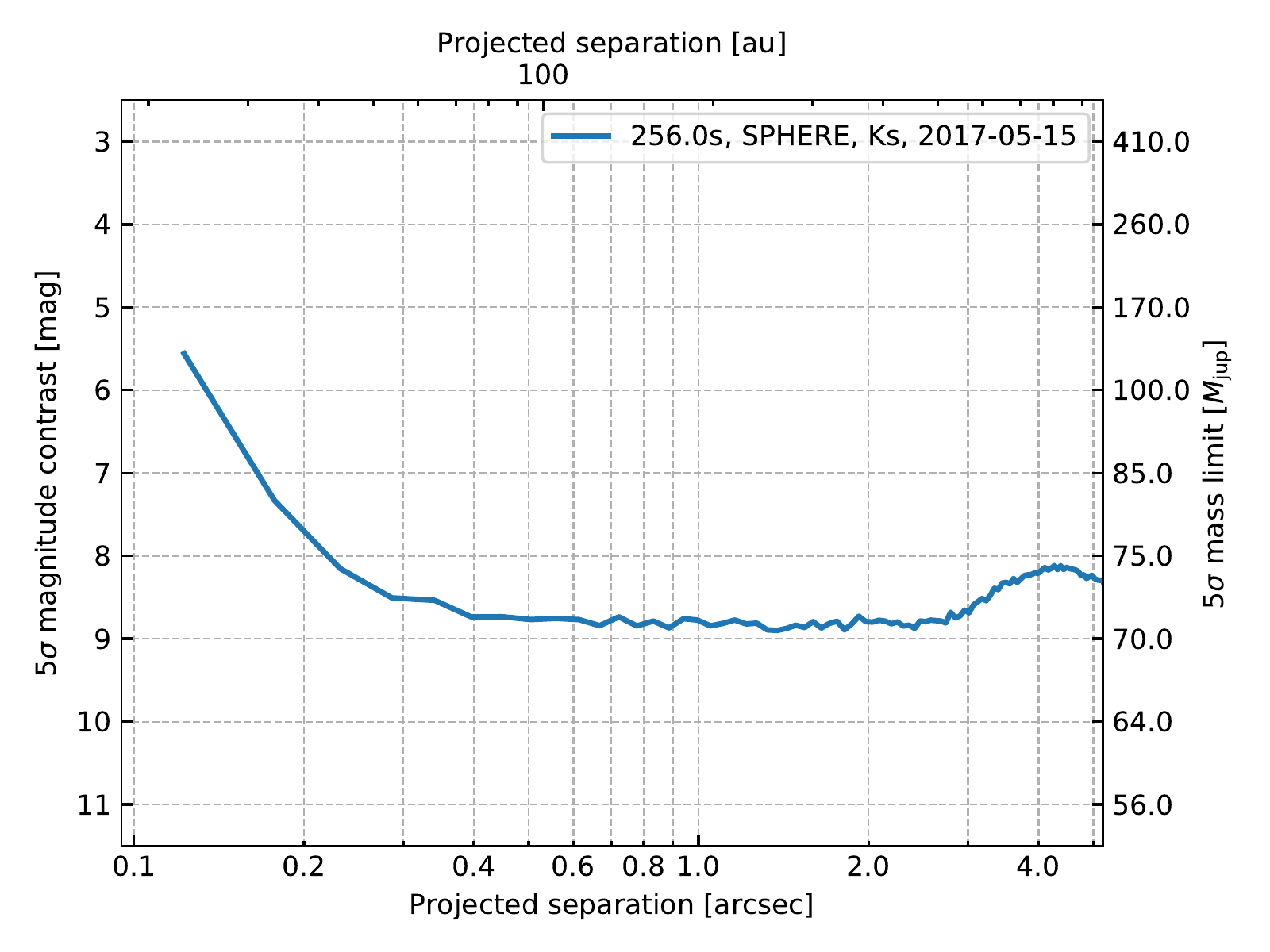}
\subcaption{KELT-10}
\end{subfigure}
\begin{subfigure}[b]{0.3\textwidth}
\includegraphics[width=\textwidth]{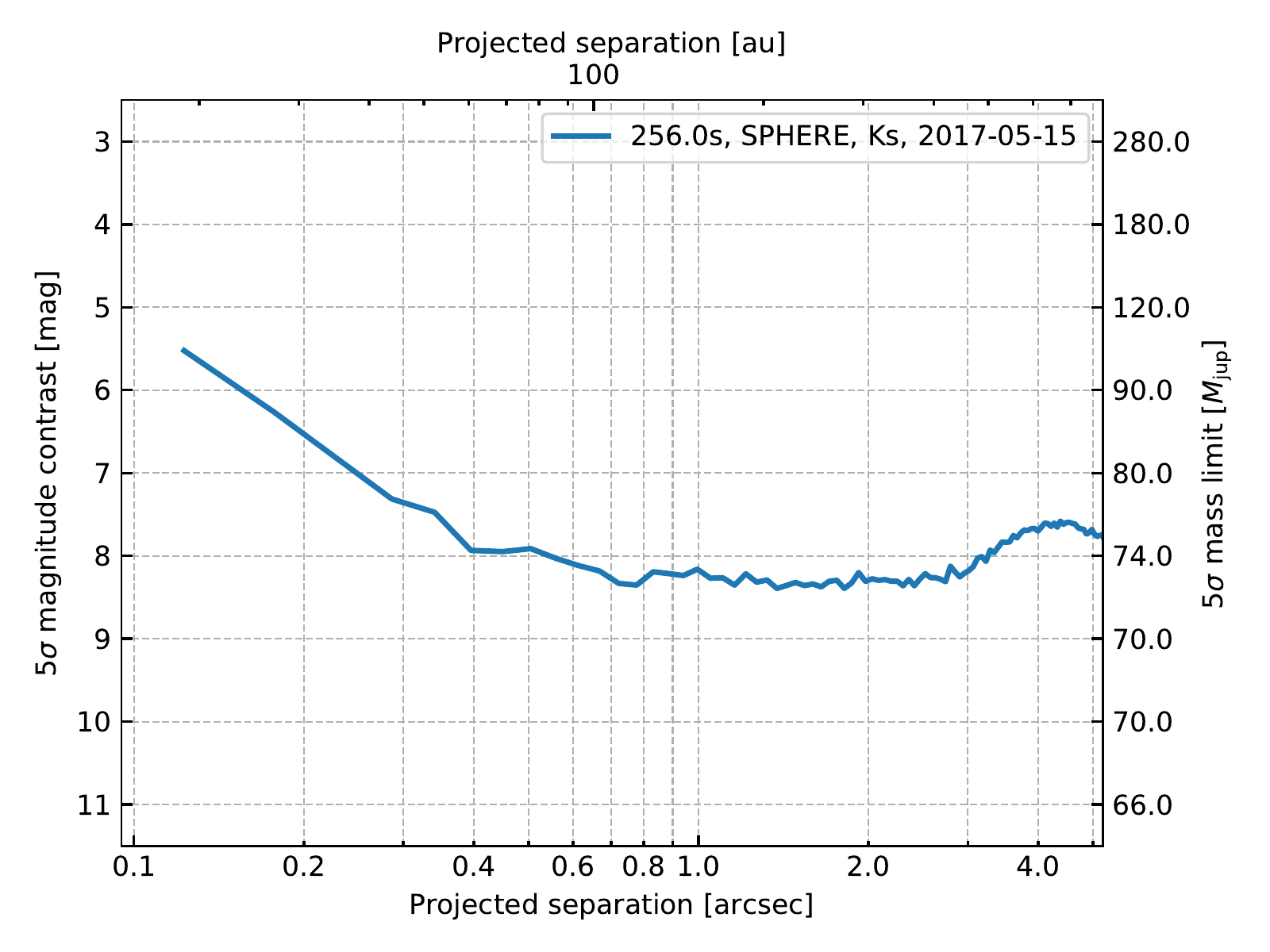}
\subcaption{WASP-2}

\end{subfigure}
\begin{subfigure}[b]{0.3\textwidth}
\includegraphics[width=\textwidth]{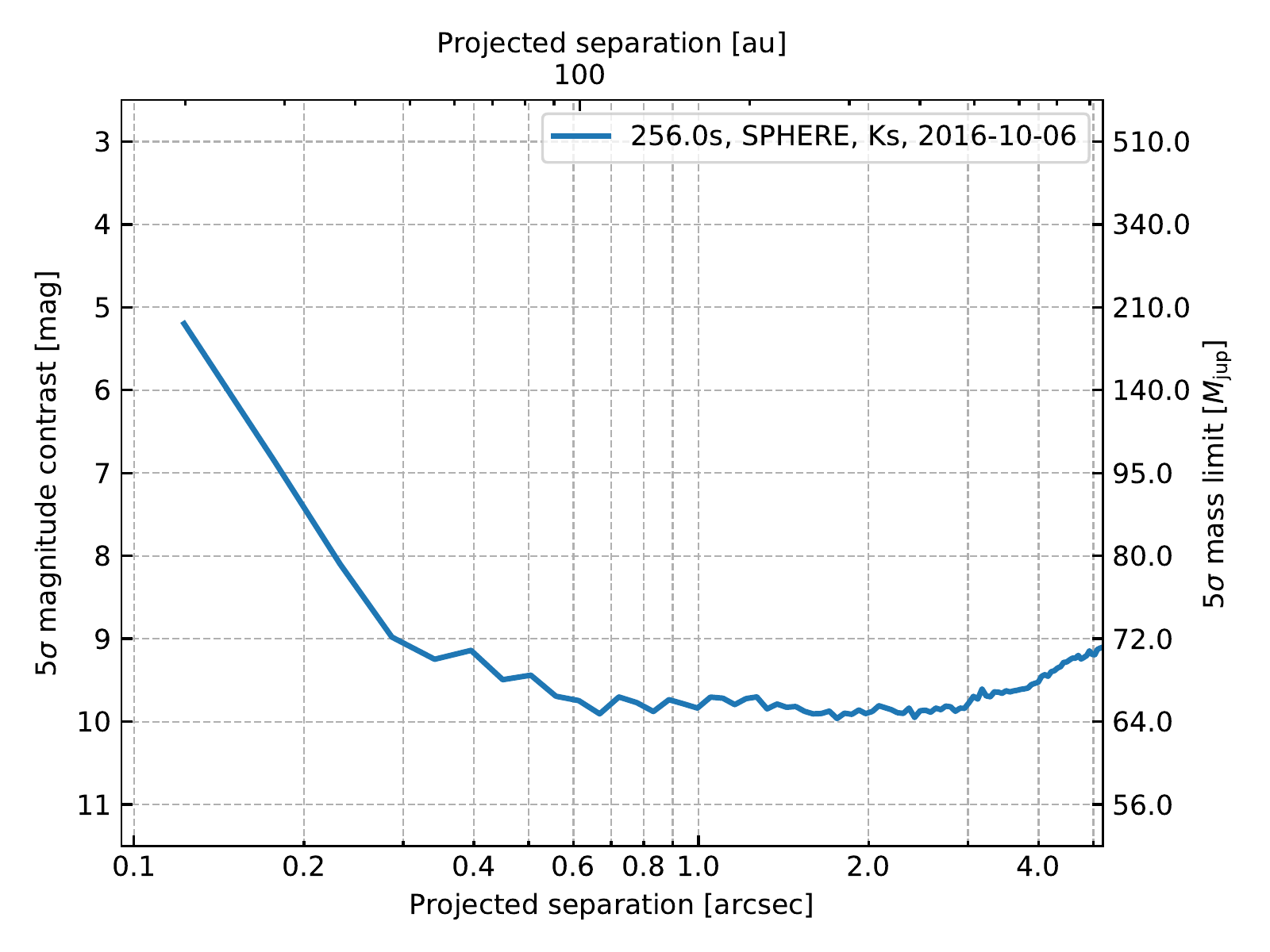}
\subcaption{WASP-7}
\end{subfigure}
\begin{subfigure}[b]{0.3\textwidth}
\includegraphics[width=\textwidth]{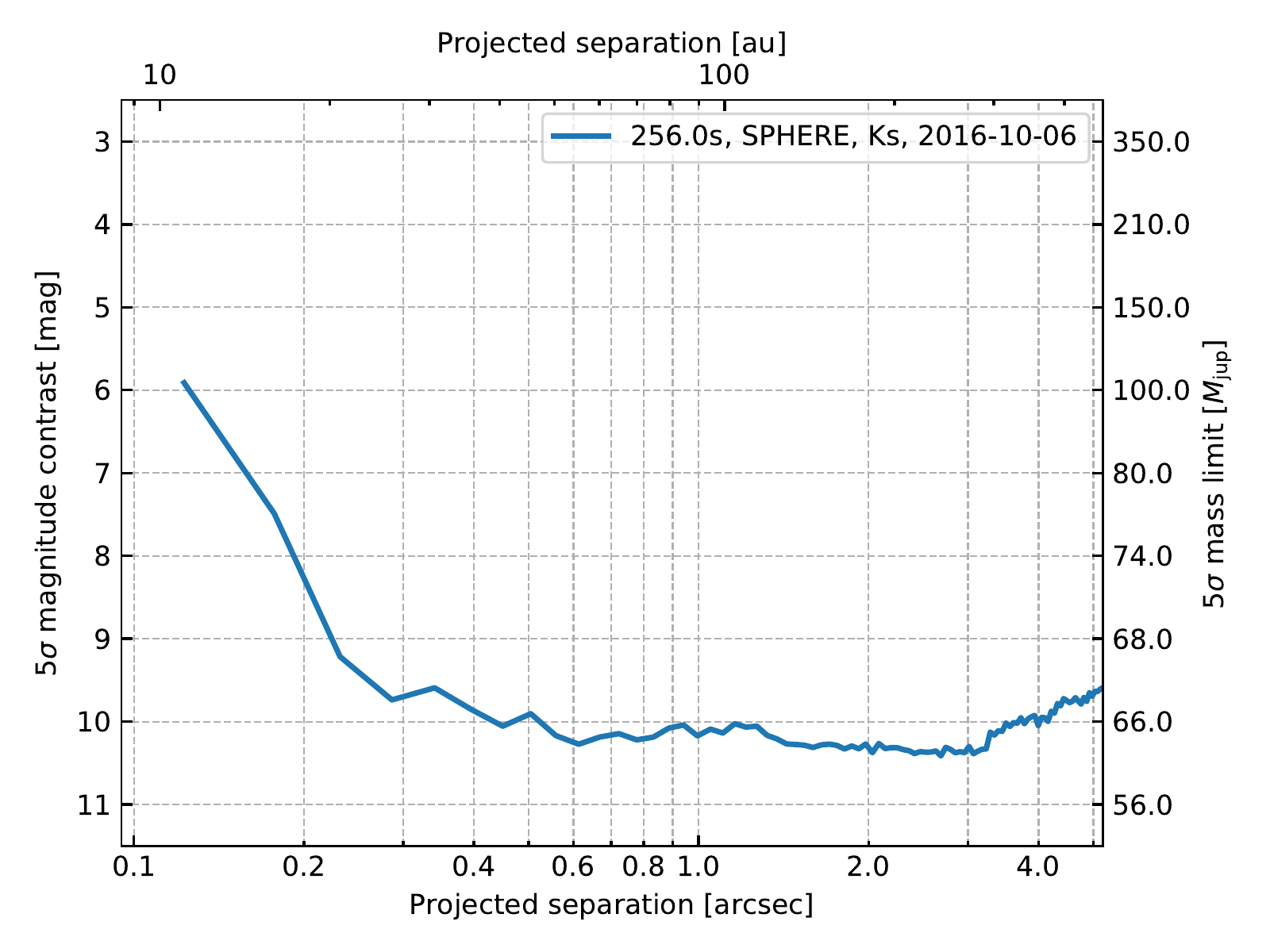}
\subcaption{WASP-8}
\end{subfigure}
\begin{subfigure}[b]{0.3\textwidth}
\includegraphics[width=\textwidth]{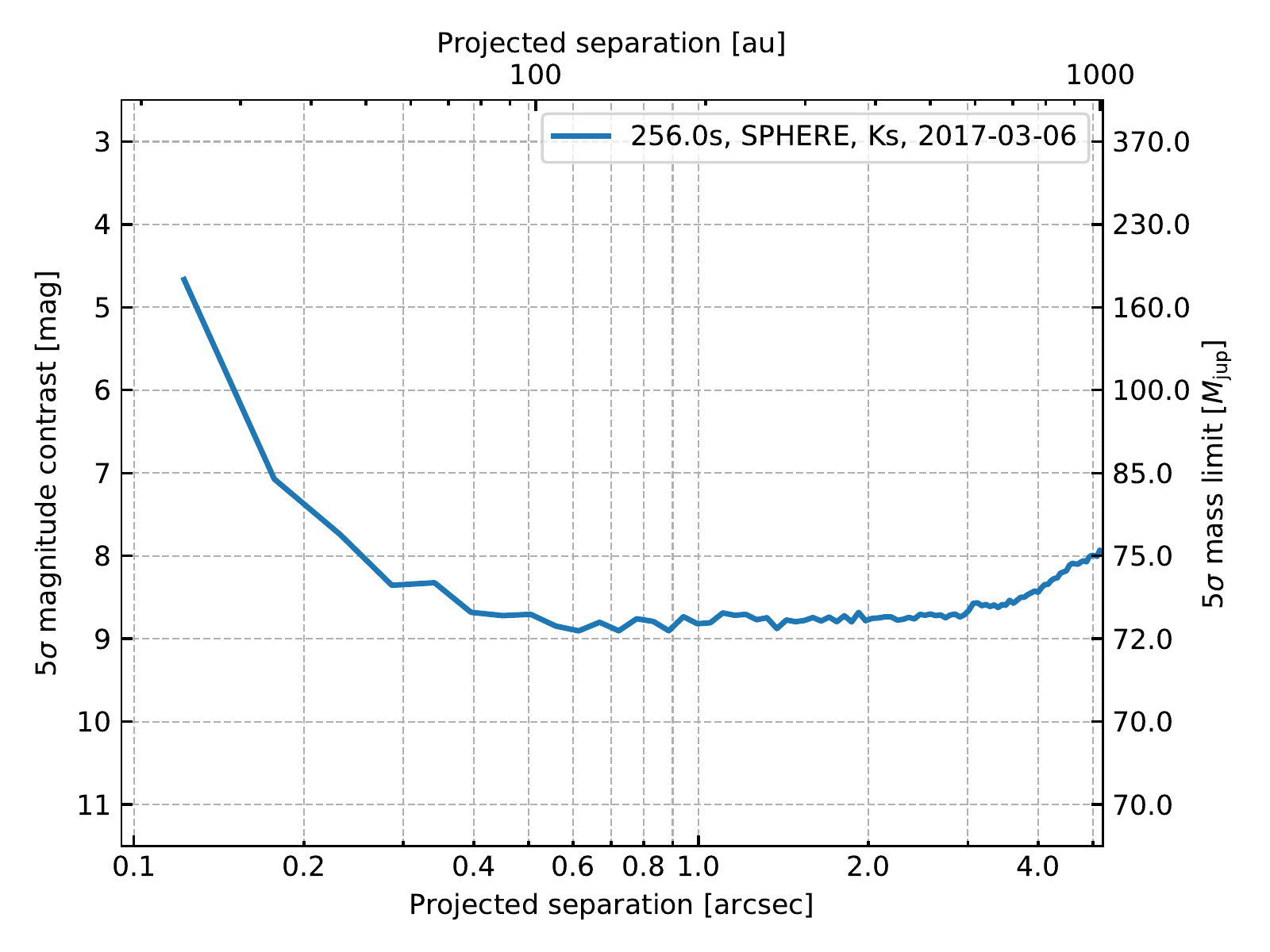}
\subcaption{WASP-16}
\end{subfigure}

\begin{subfigure}[b]{0.3\textwidth}
\includegraphics[width=\textwidth]{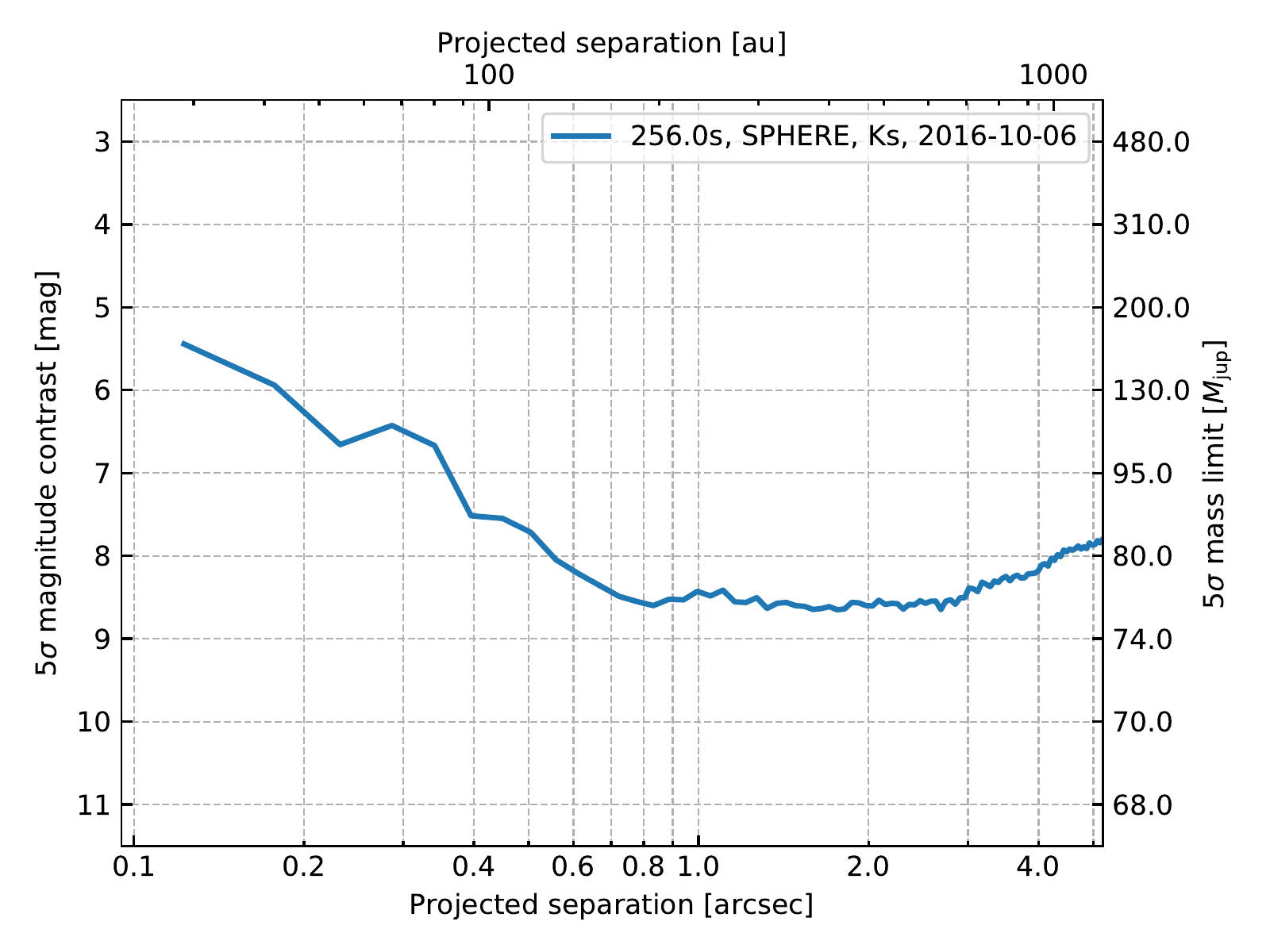}
\subcaption{WASP-20}
\end{subfigure}
\begin{subfigure}[b]{0.3\textwidth}
\includegraphics[width=\textwidth]{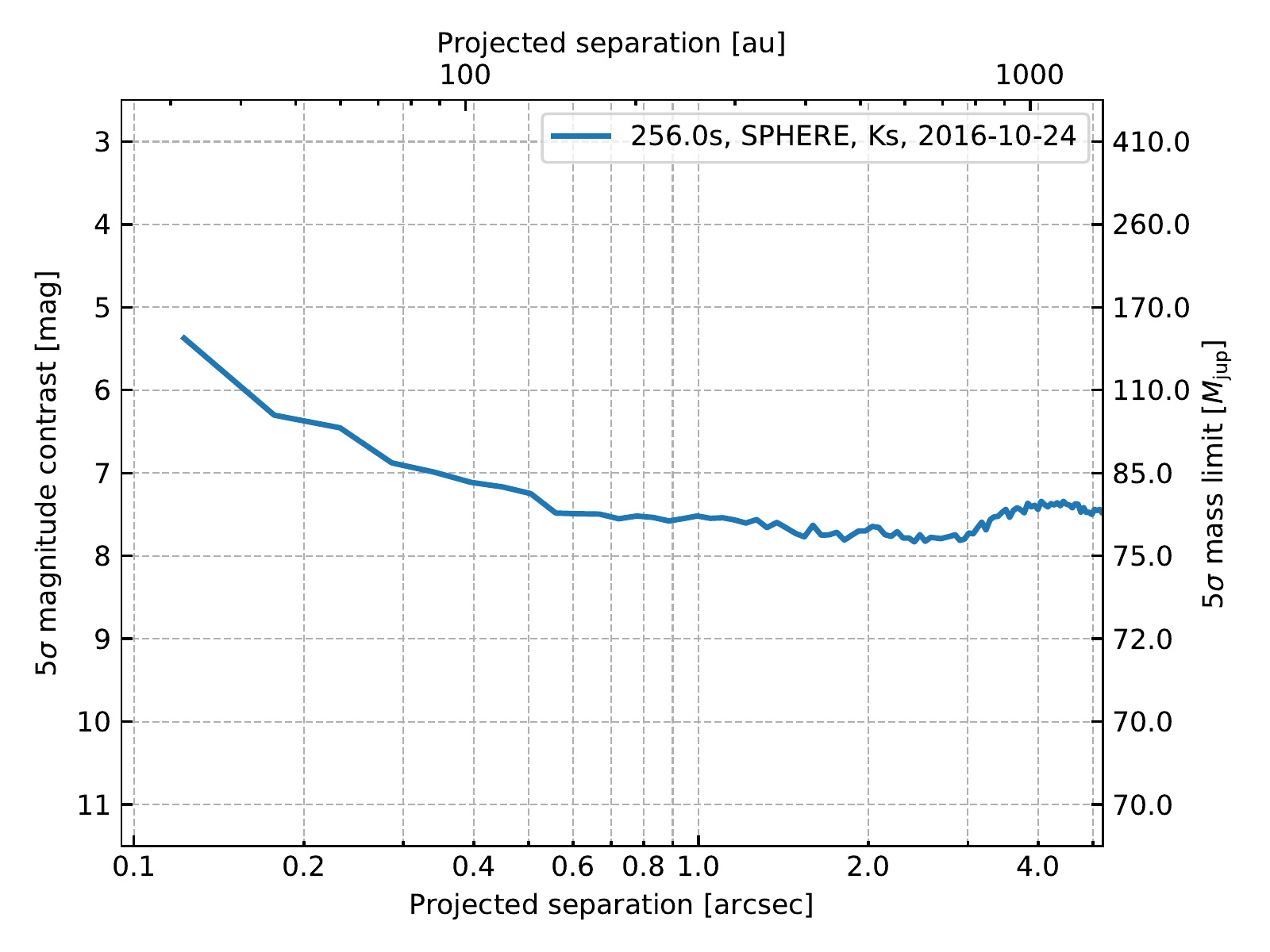}
\subcaption{WASP-21}
\end{subfigure}
\begin{subfigure}[b]{0.3\textwidth}
\includegraphics[width=\textwidth]{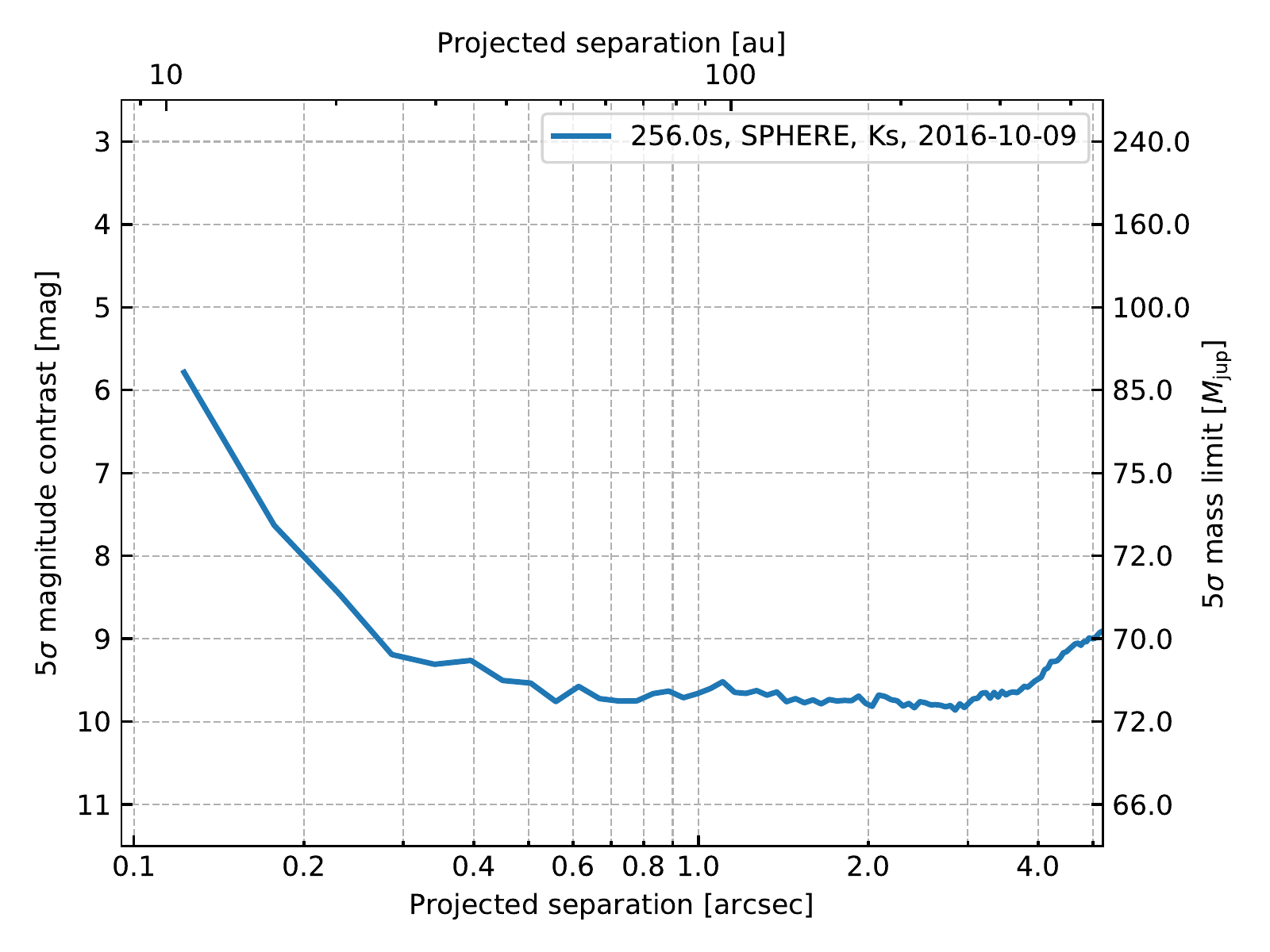}
\subcaption{WASP-29}
\end{subfigure}

\caption{
Detection limits of individual targets I.
We convert projected angular separations in projected physical separations by using the distances presented in Table~\ref{tbl:star_properties}.
The mass limits arise from comparison to AMES-Cond, AMES-Dusty, and BT-Settl models as described in Sect.~\ref{subsec:characterization_of_ccs}.
}
\label{fig:detection_limits_individual_1}
\end{figure*}

\begin{figure*}
\centering

\begin{subfigure}[b]{0.3\textwidth}
\includegraphics[width=\textwidth]{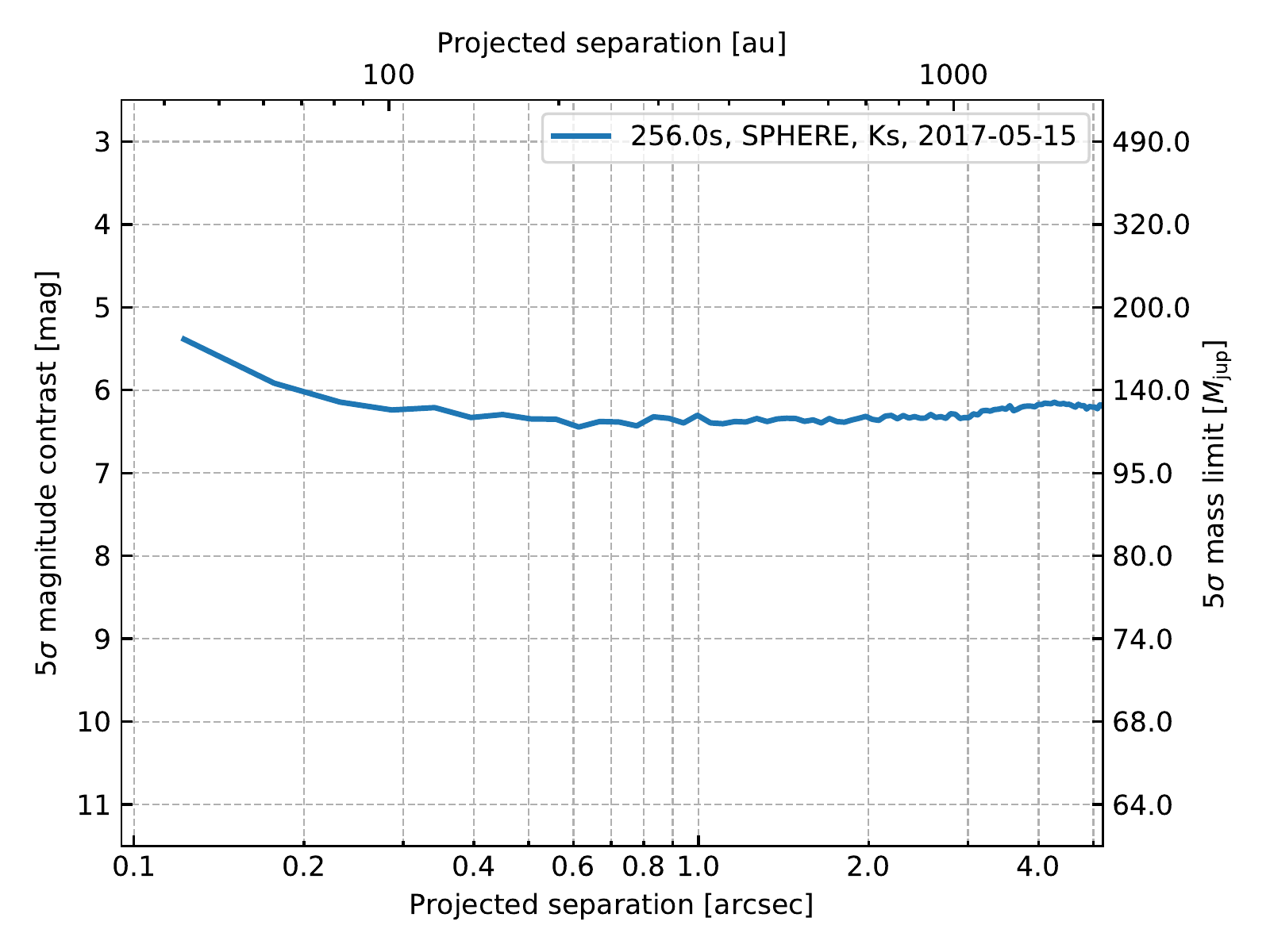}
\subcaption{WASP-30}
\end{subfigure}
\begin{subfigure}[b]{0.3\textwidth}
\includegraphics[width=\textwidth]{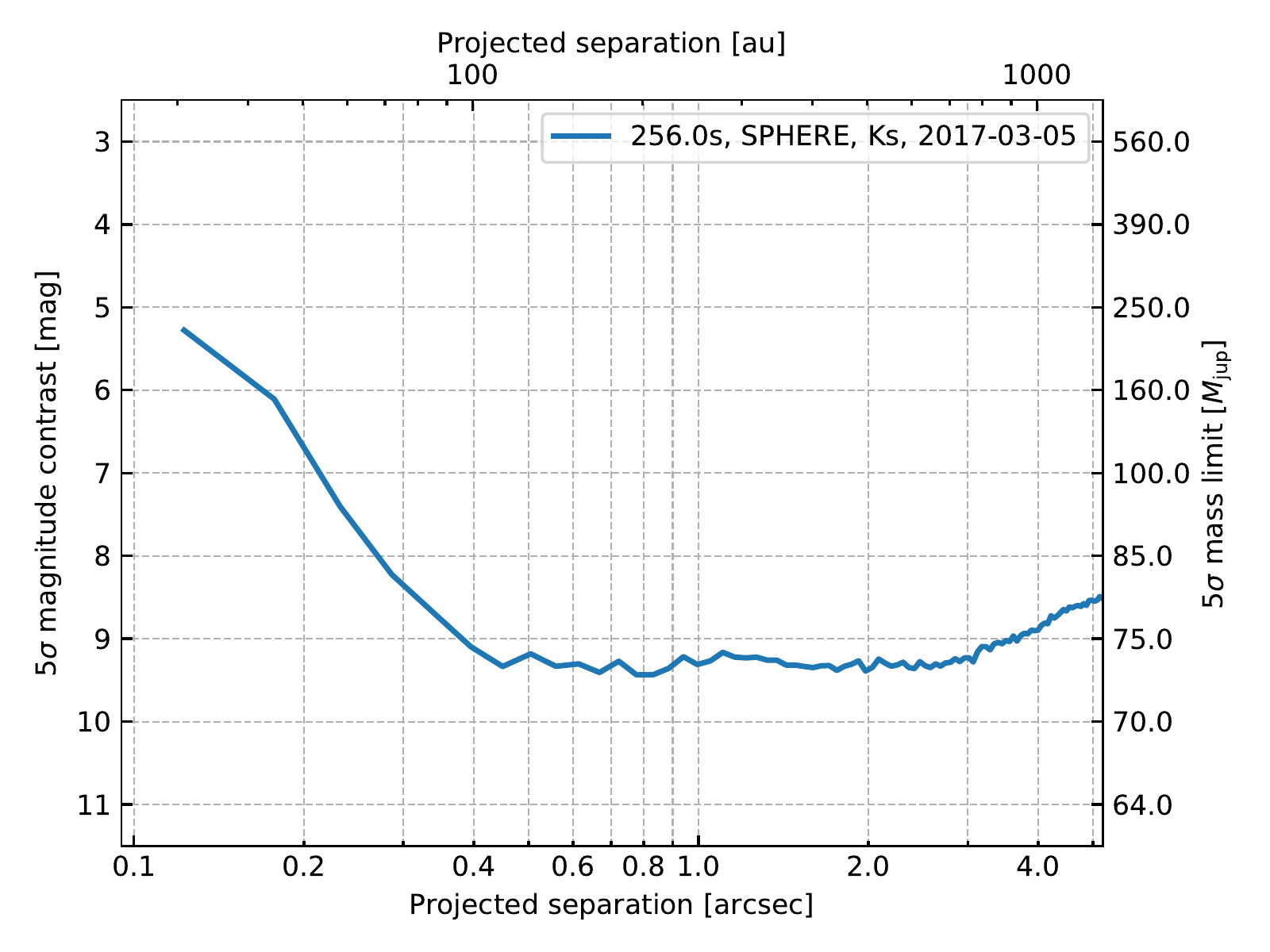}
\subcaption{WASP-54}
\end{subfigure}
\begin{subfigure}[b]{0.3\textwidth}
\includegraphics[width=\textwidth]{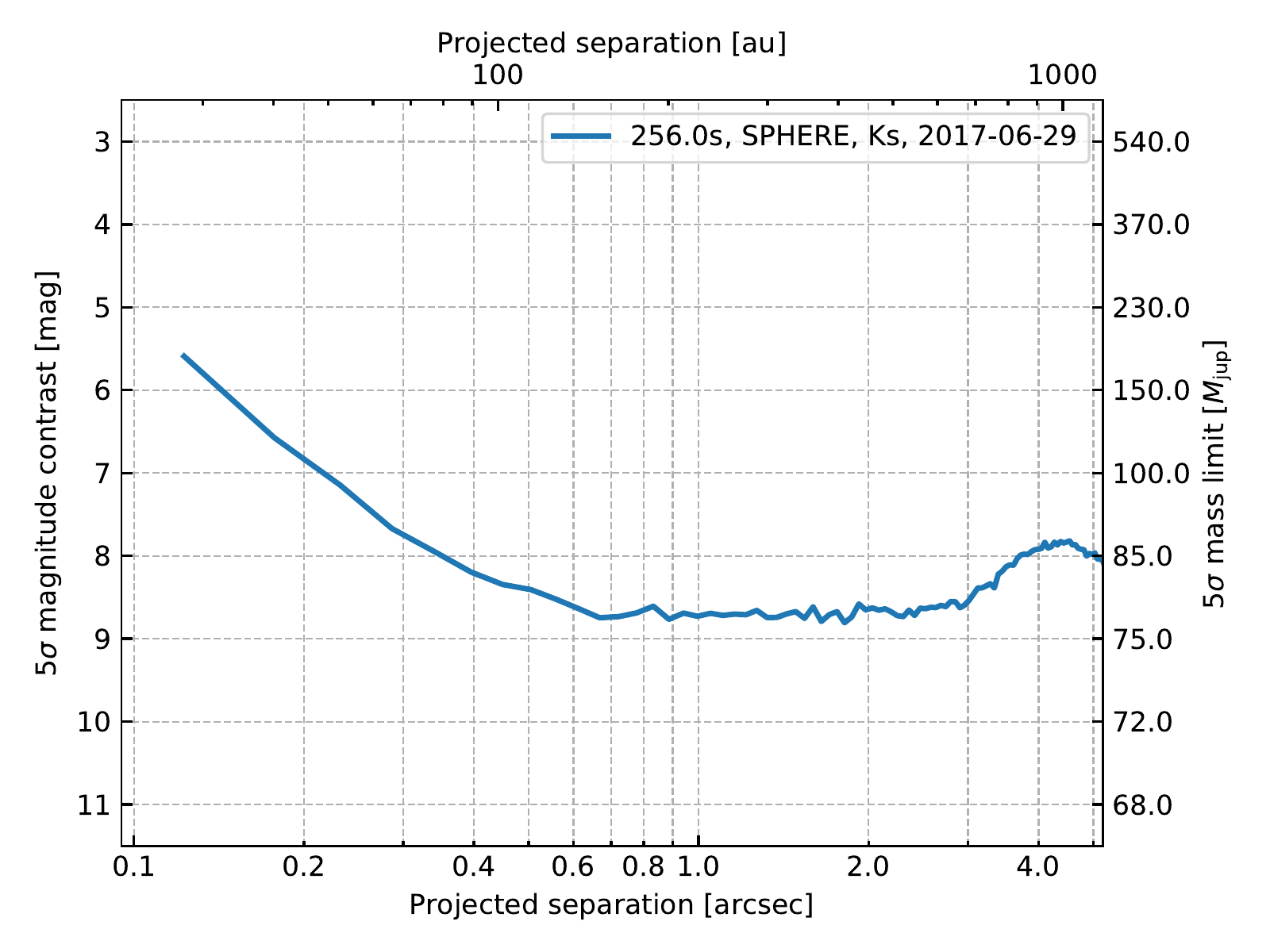}
\subcaption{WASP-68}
\end{subfigure}

\begin{subfigure}[b]{0.3\textwidth}
\includegraphics[width=\textwidth]{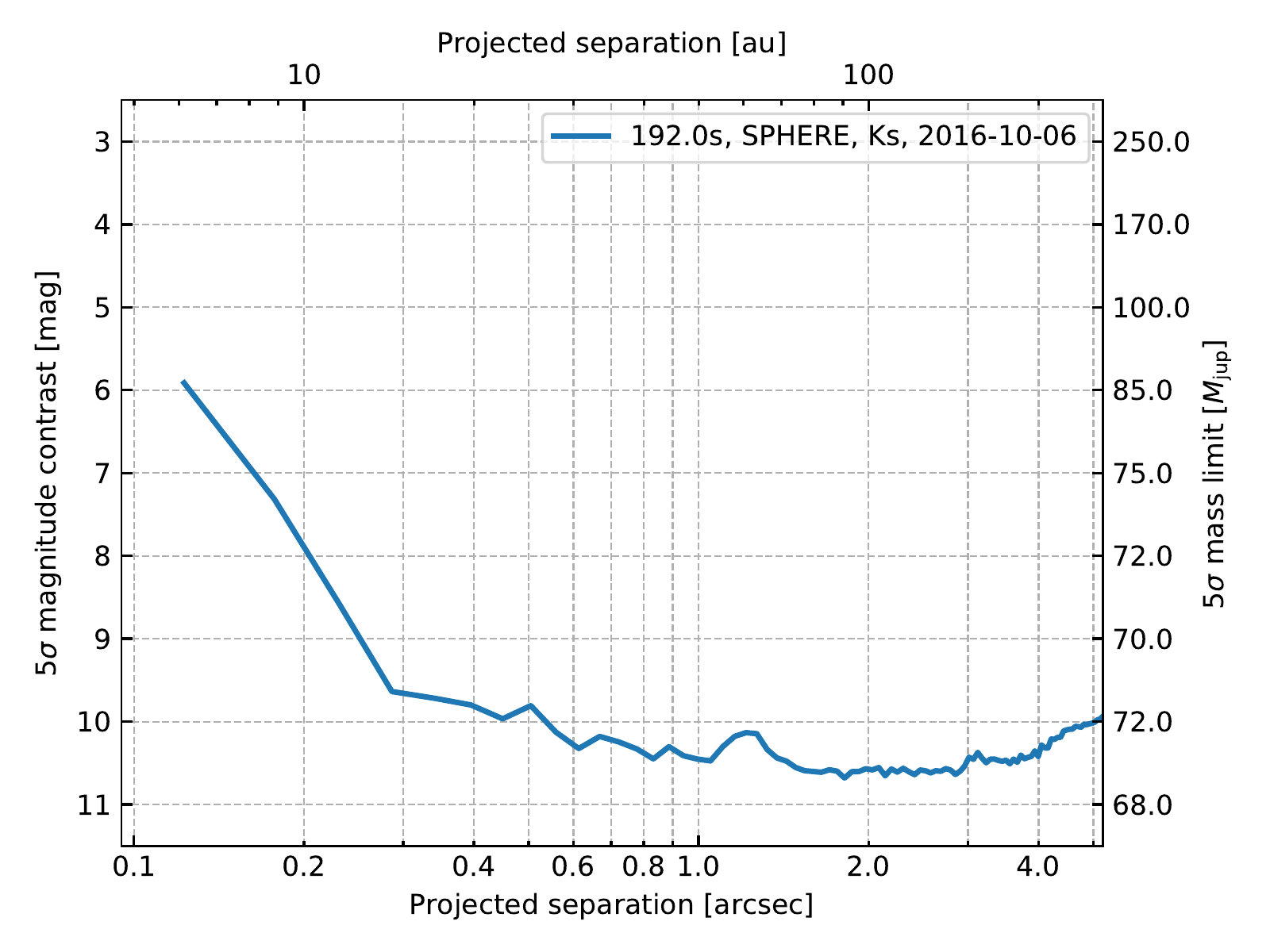}
\subcaption{WASP-69}
\end{subfigure}
\begin{subfigure}[b]{0.3\textwidth}
\includegraphics[width=\textwidth]{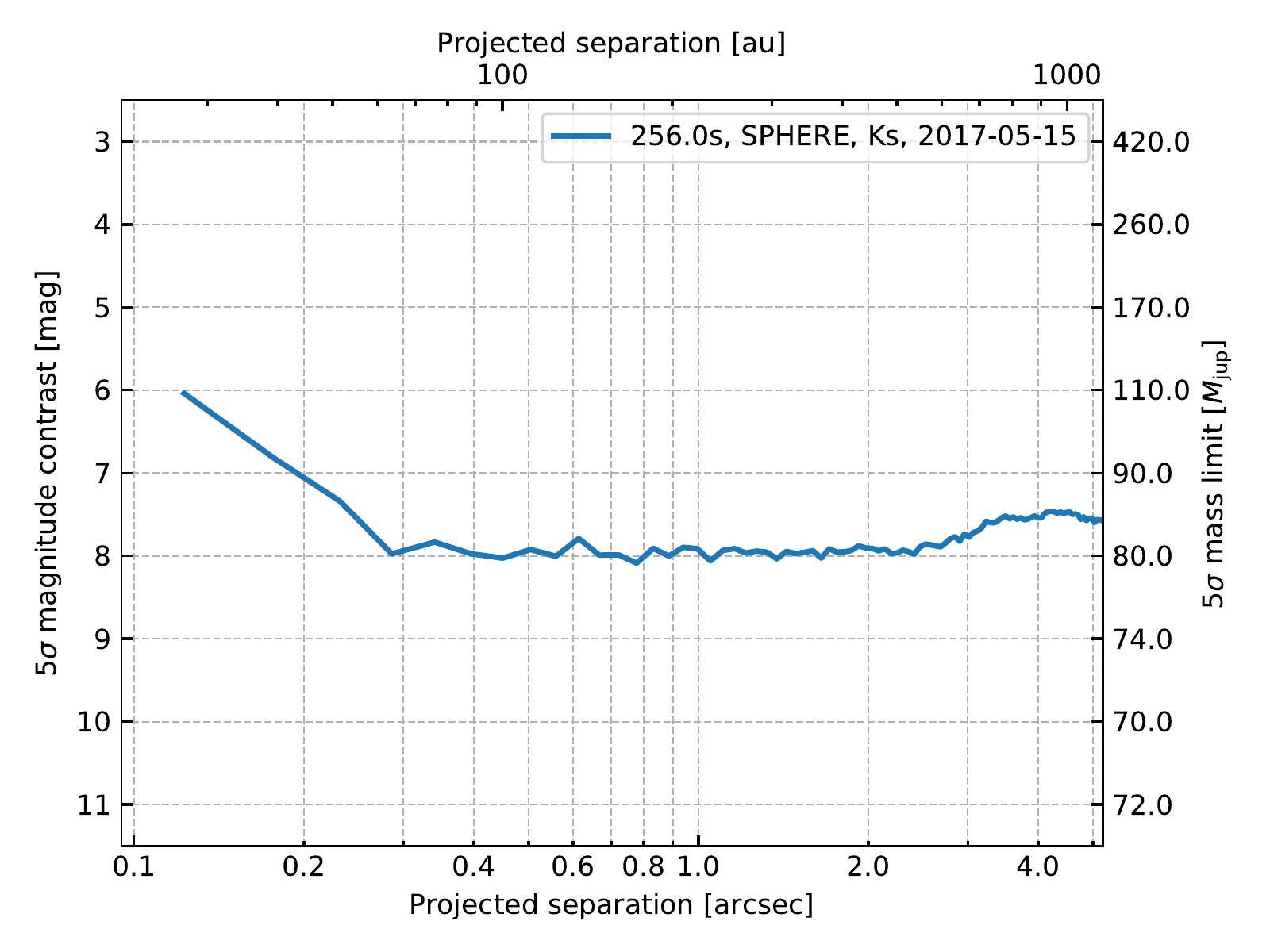}
\subcaption{WASP-70}
\end{subfigure}
\begin{subfigure}[b]{0.3\textwidth}
\includegraphics[width=\textwidth]{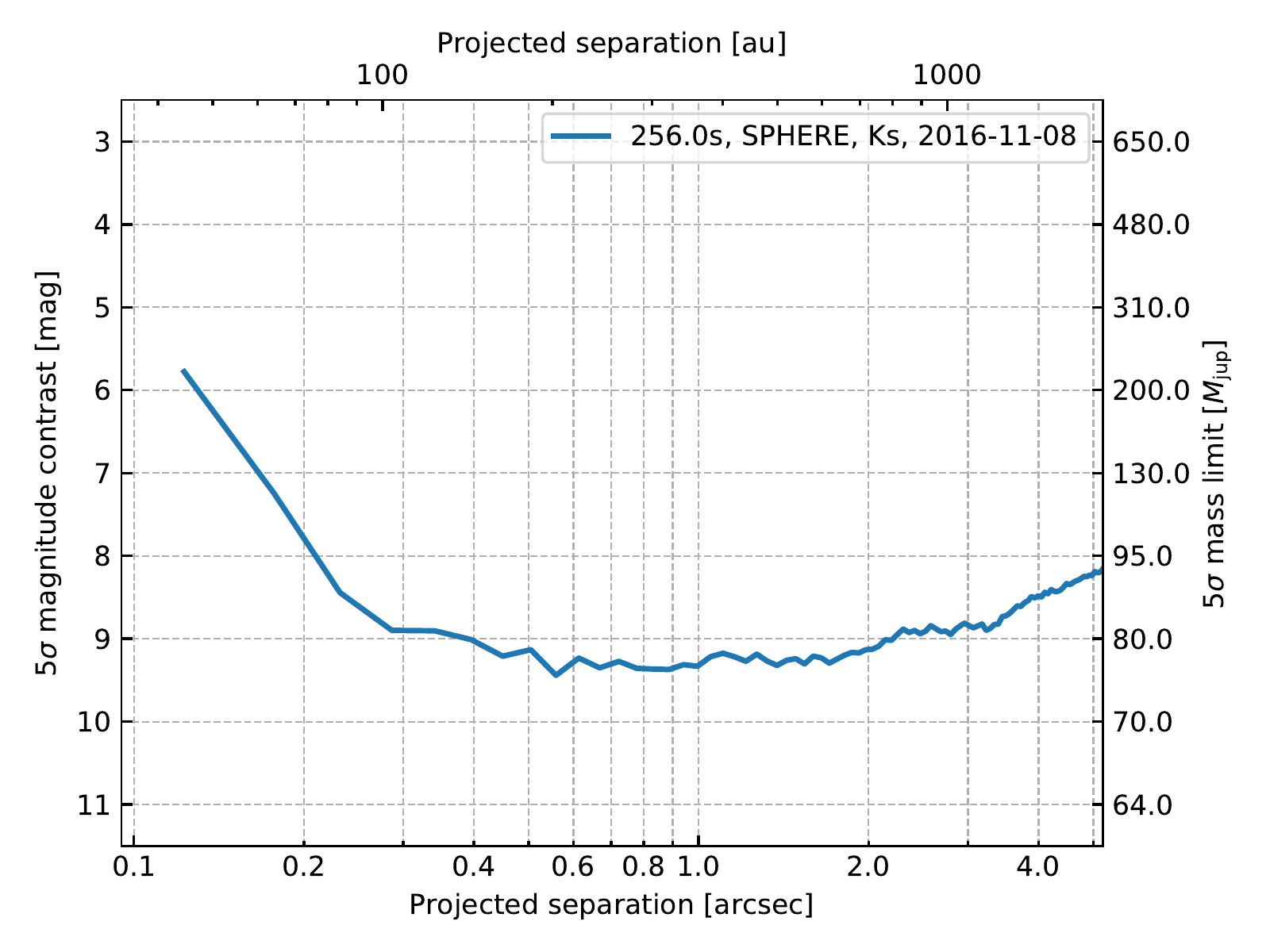}
\subcaption{WASP-71}
\end{subfigure}

\begin{subfigure}[b]{0.3\textwidth}
\includegraphics[width=\textwidth]{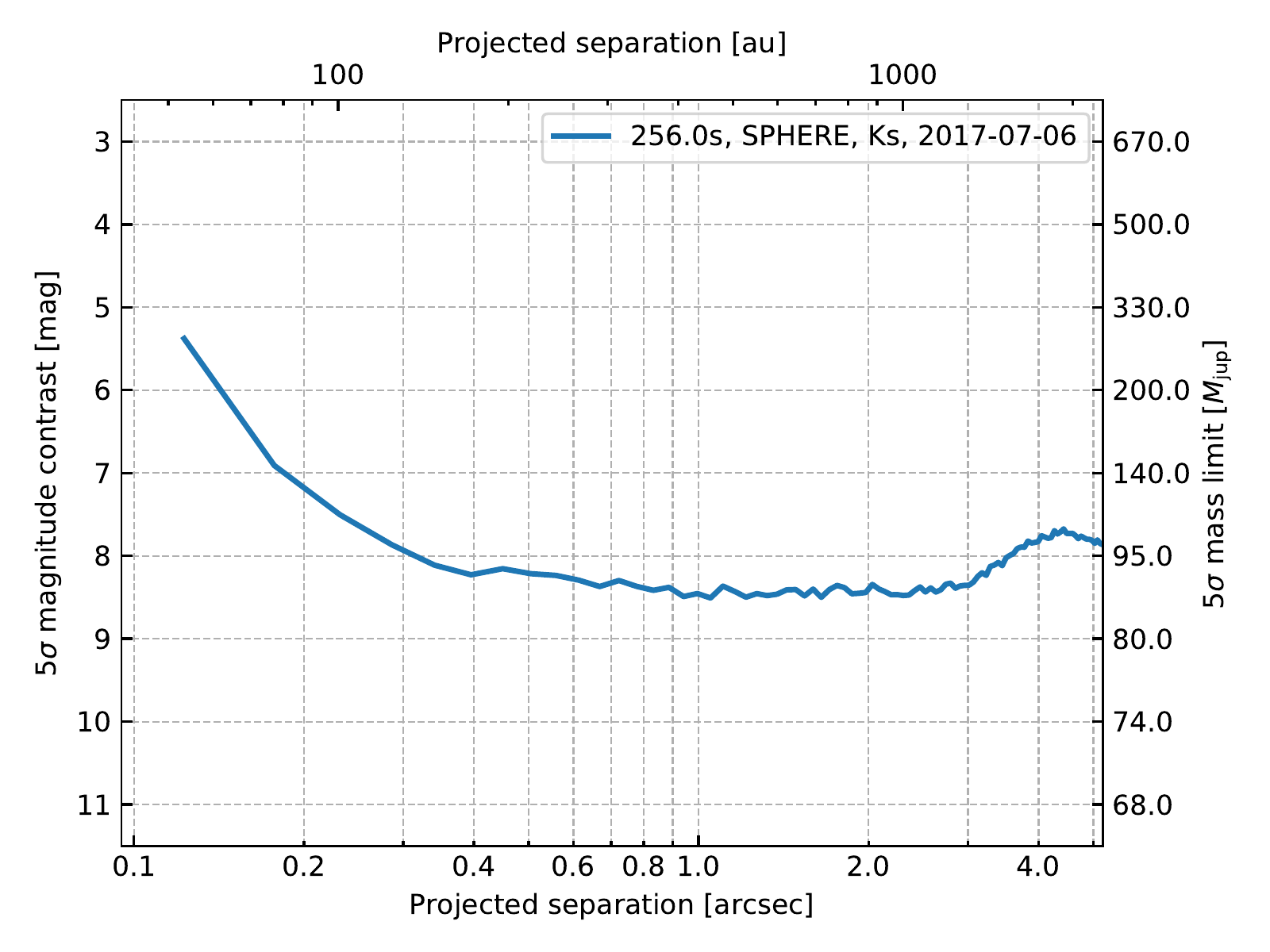}
\subcaption{WASP-72}
\end{subfigure}
\begin{subfigure}[b]{0.3\textwidth}
\includegraphics[width=\textwidth]{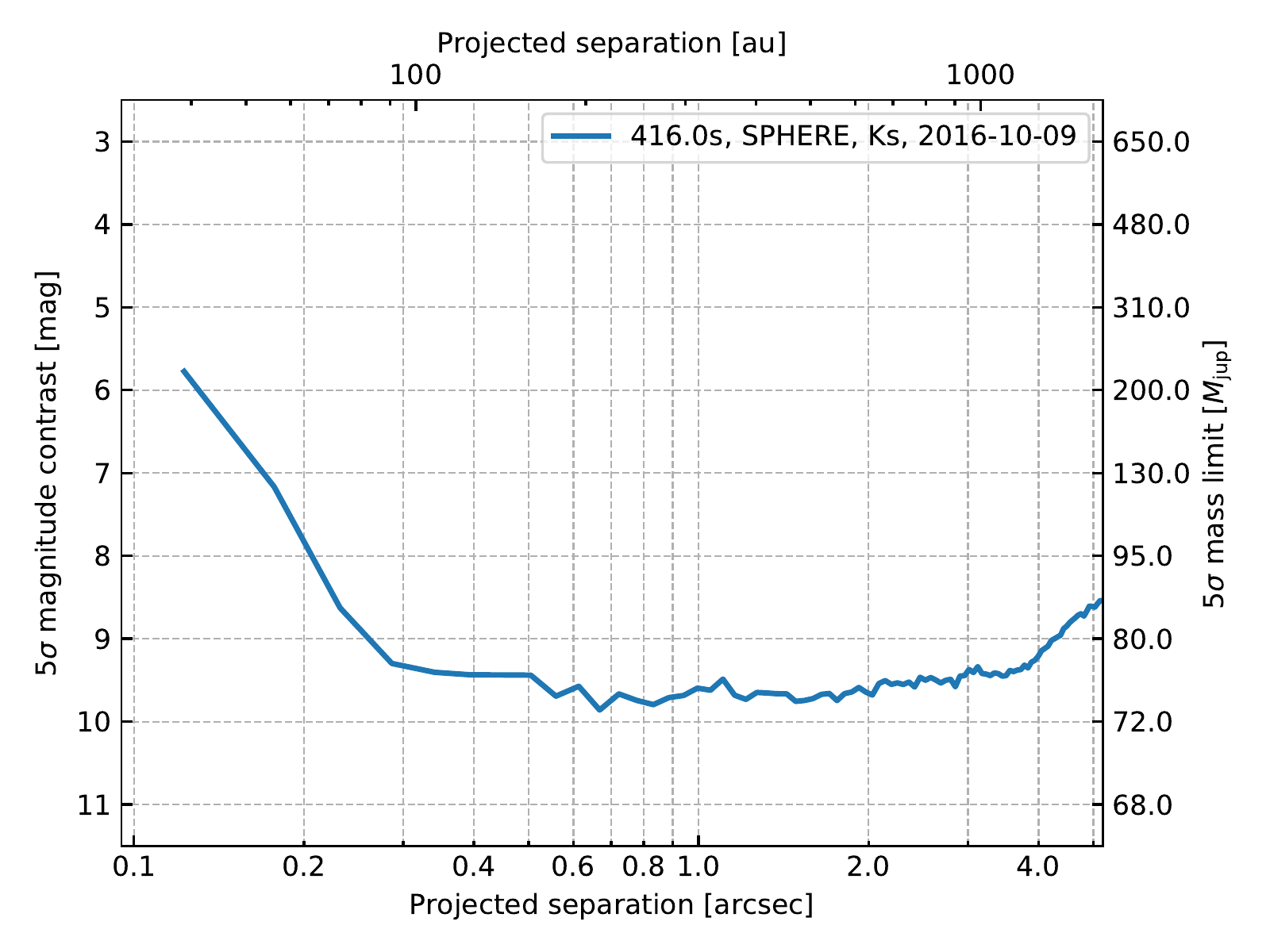}
\subcaption{WASP-73}
\end{subfigure}
\begin{subfigure}[b]{0.3\textwidth}
\includegraphics[width=\textwidth]{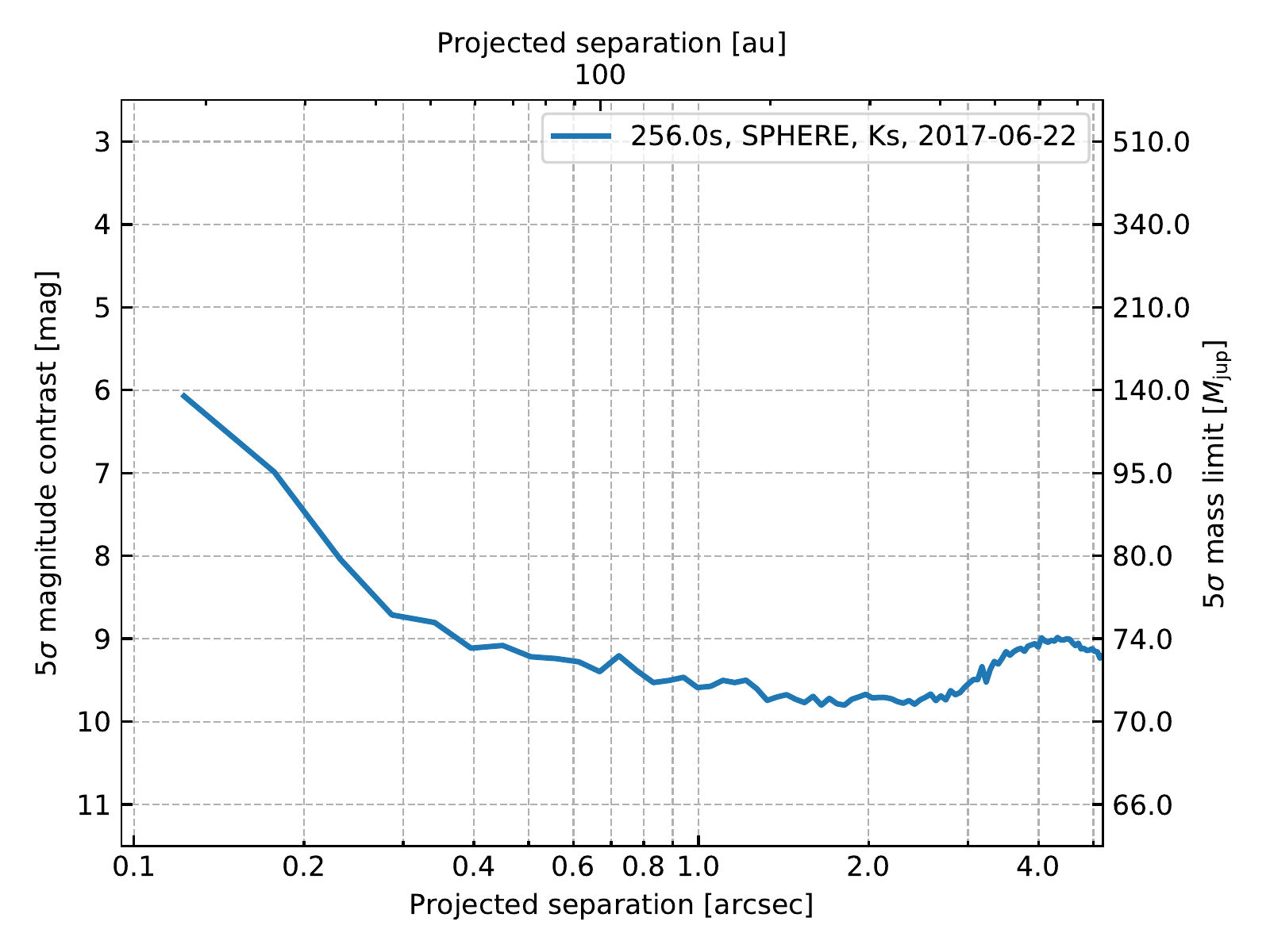}
\subcaption{WASP-74}
\end{subfigure}

\begin{subfigure}[b]{0.3\textwidth}
\includegraphics[width=\textwidth]{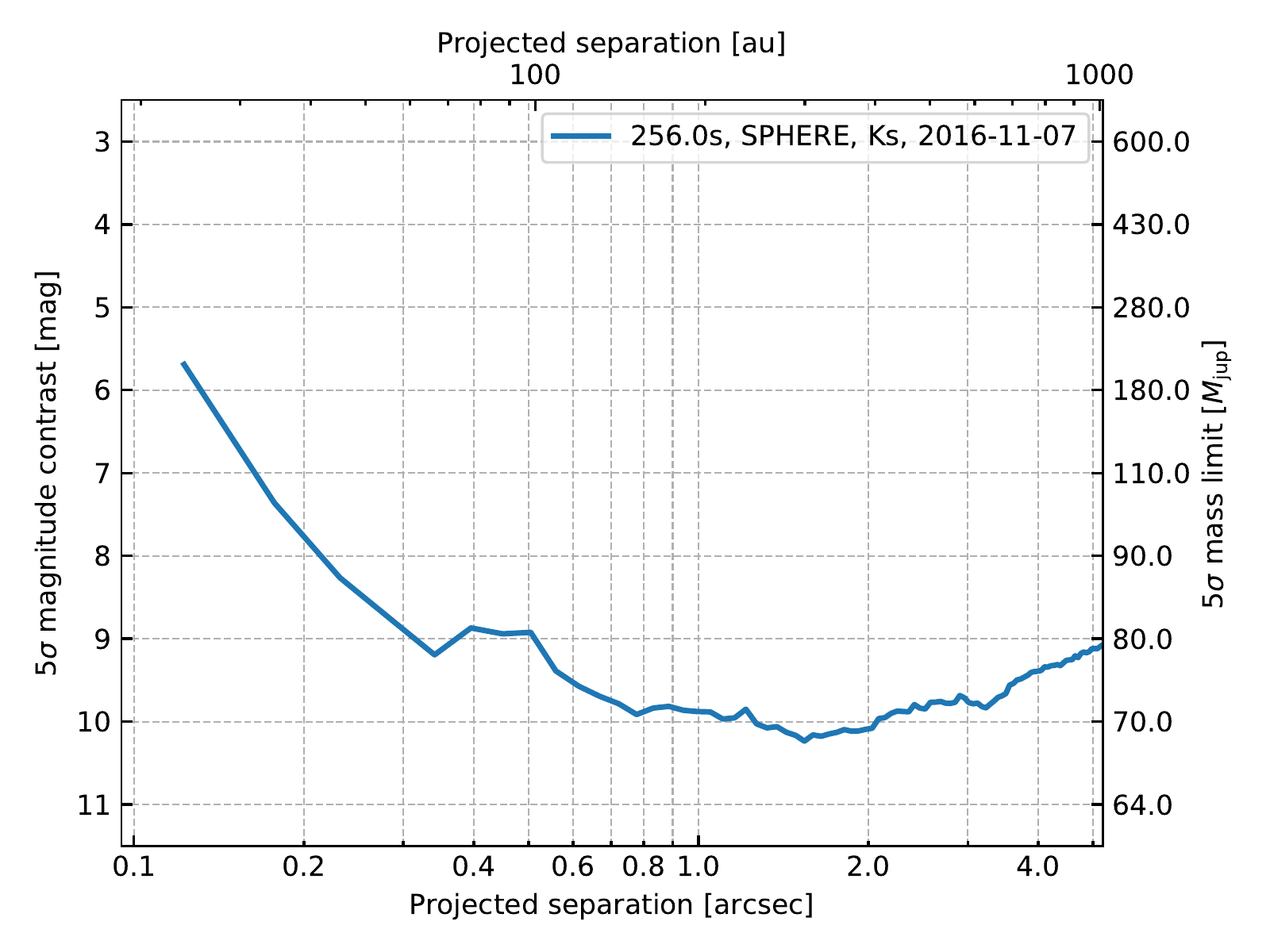}
\subcaption{WASP-76}
\end{subfigure}
\begin{subfigure}[b]{0.3\textwidth}
\includegraphics[width=\textwidth]{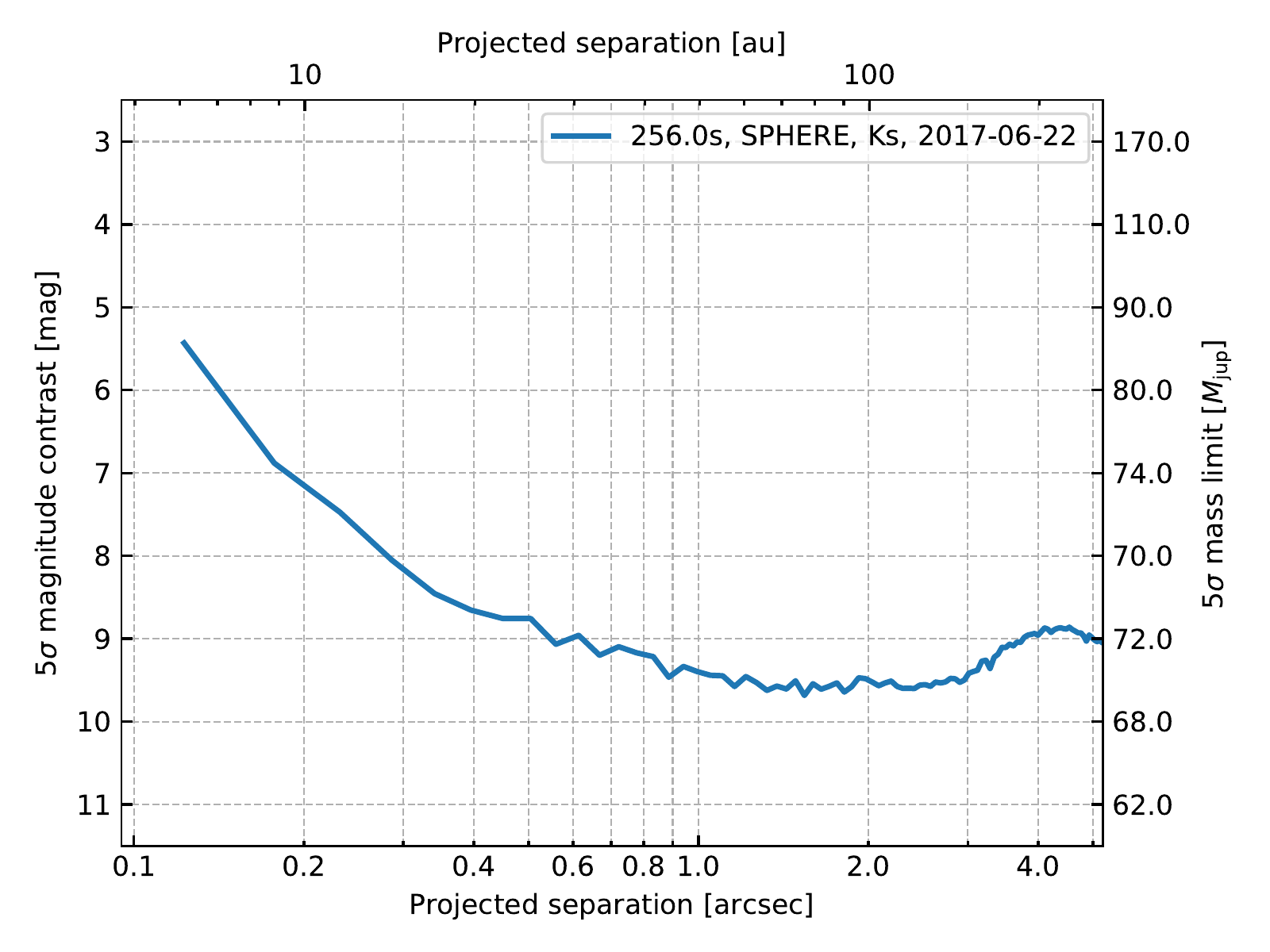}
\subcaption{WASP-80}
\end{subfigure}
\begin{subfigure}[b]{0.3\textwidth}
\includegraphics[width=\textwidth]{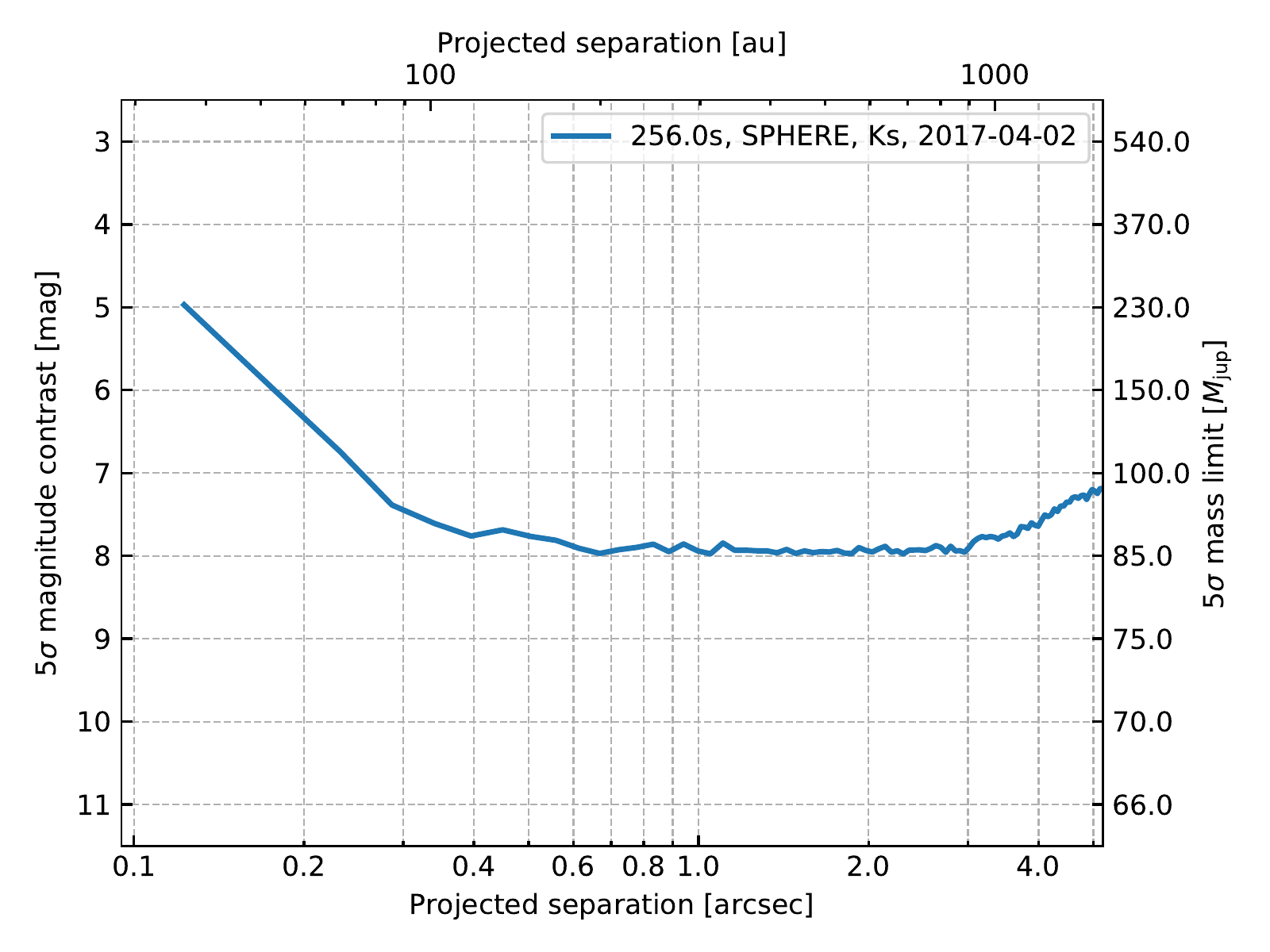}
\subcaption{WASP-87}
\end{subfigure}

\begin{subfigure}[b]{0.3\textwidth}
\includegraphics[width=\textwidth]{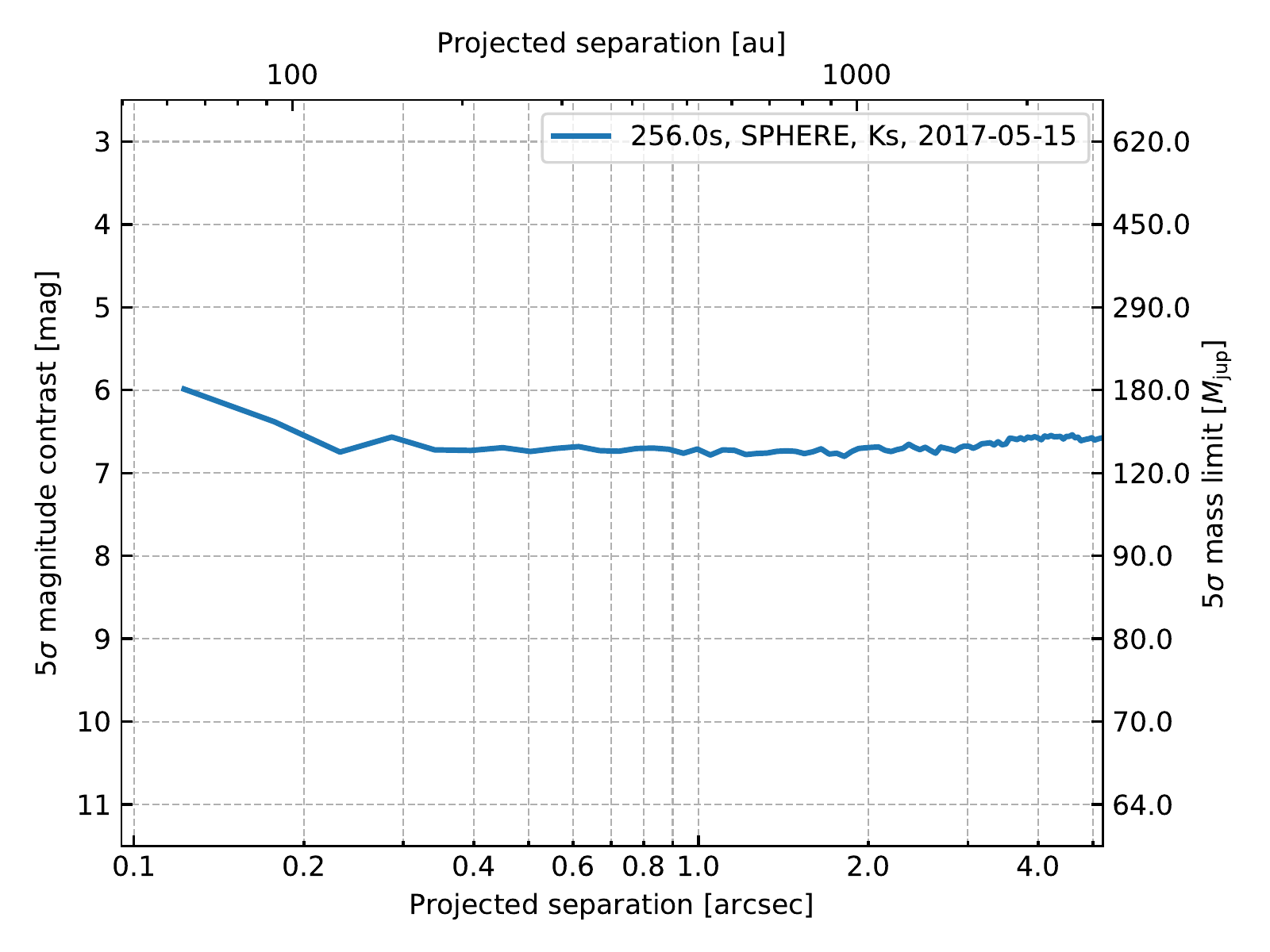}
\subcaption{WASP-88}
\end{subfigure}
\begin{subfigure}[b]{0.3\textwidth}
\includegraphics[width=\textwidth]{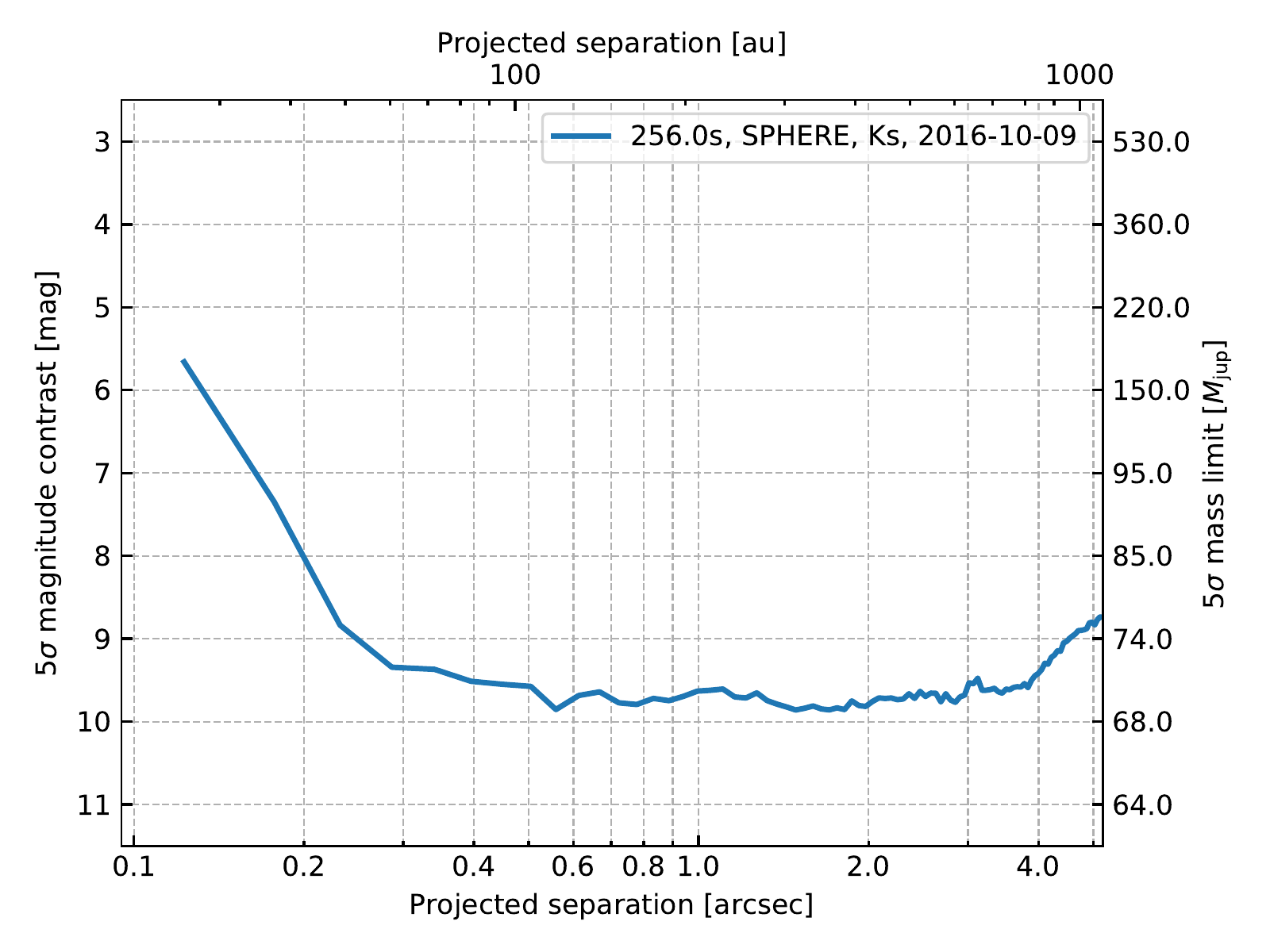}
\subcaption{WASP-94}
\end{subfigure}
\begin{subfigure}[b]{0.3\textwidth}
\includegraphics[width=\textwidth]{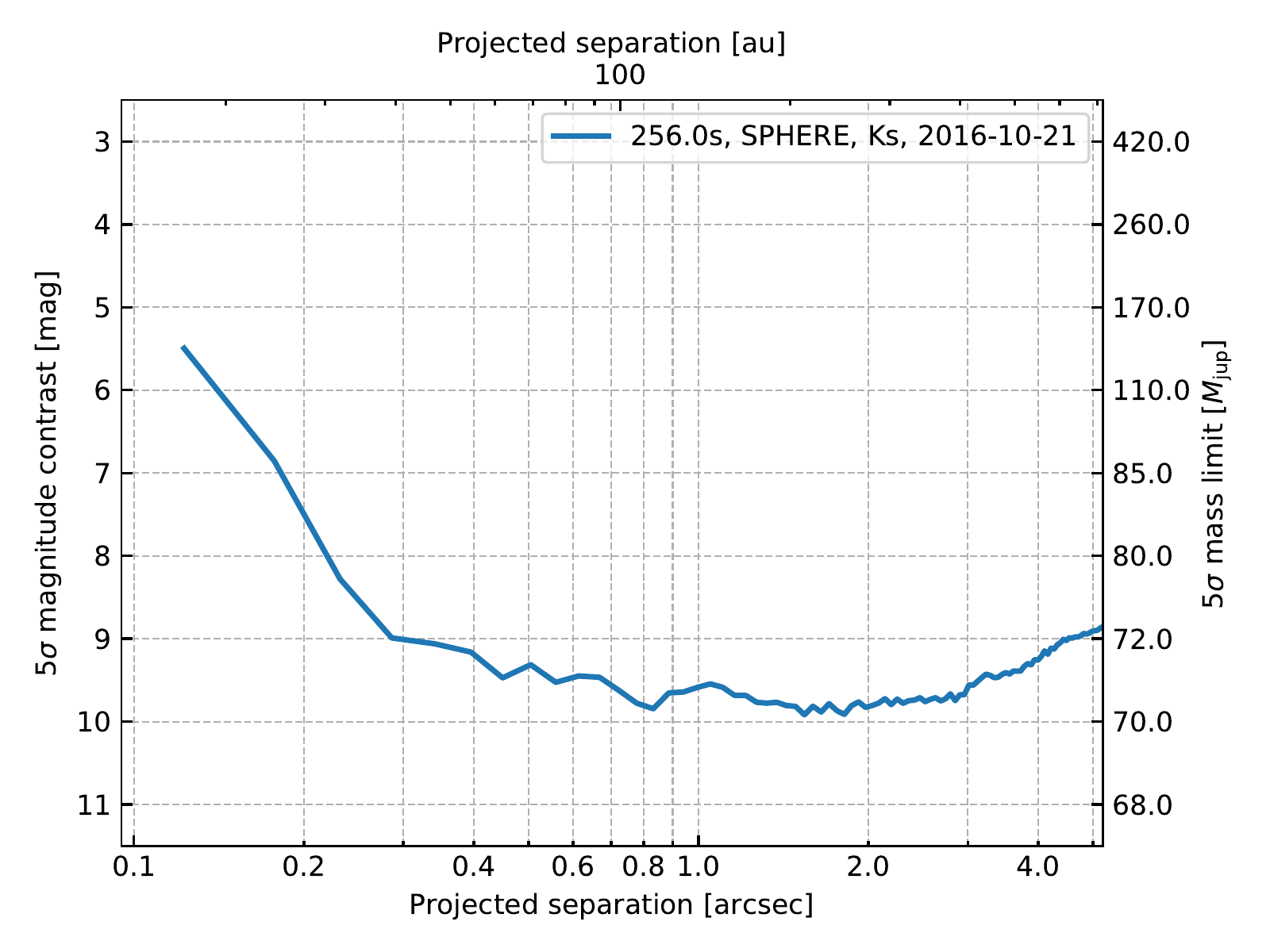}
\subcaption{WASP-95}
\end{subfigure}

\caption{
Detection limits of individual targets II.
We convert projected angular separations in projected physical separations by using the distances presented in Table~\ref{tbl:star_properties}.
The mass limits arise from comparison to AMES-Cond, AMES-Dusty, and BT-Settl models as described in Sect.~\ref{subsec:characterization_of_ccs}.
}
\label{fig:detection_limits_individual_2}
\end{figure*}

\begin{figure*}
\centering

\begin{subfigure}[b]{0.3\textwidth}
\includegraphics[width=\textwidth]{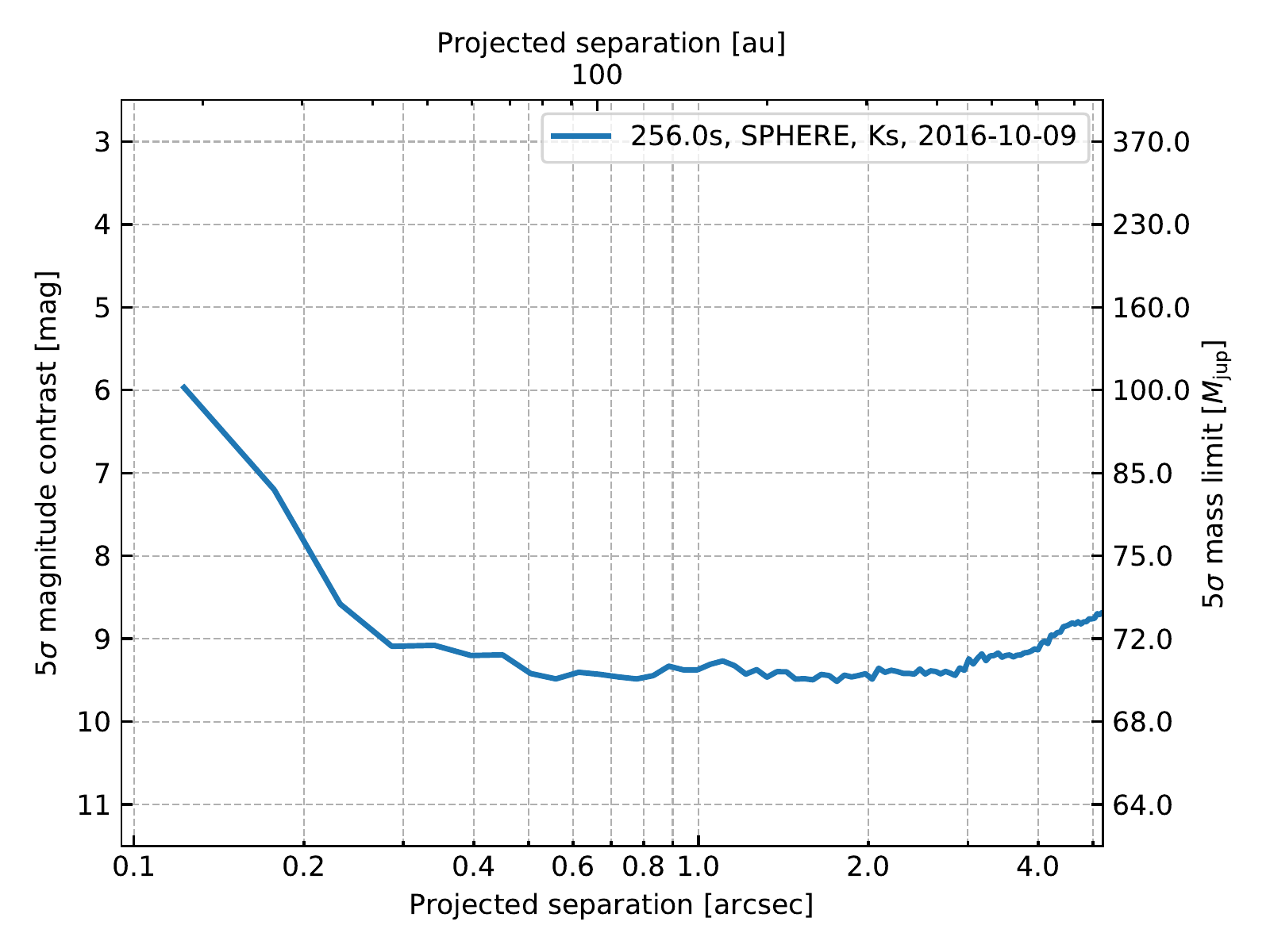}
\subcaption{WASP-97}
\end{subfigure}
\begin{subfigure}[b]{0.3\textwidth}
\includegraphics[width=\textwidth]{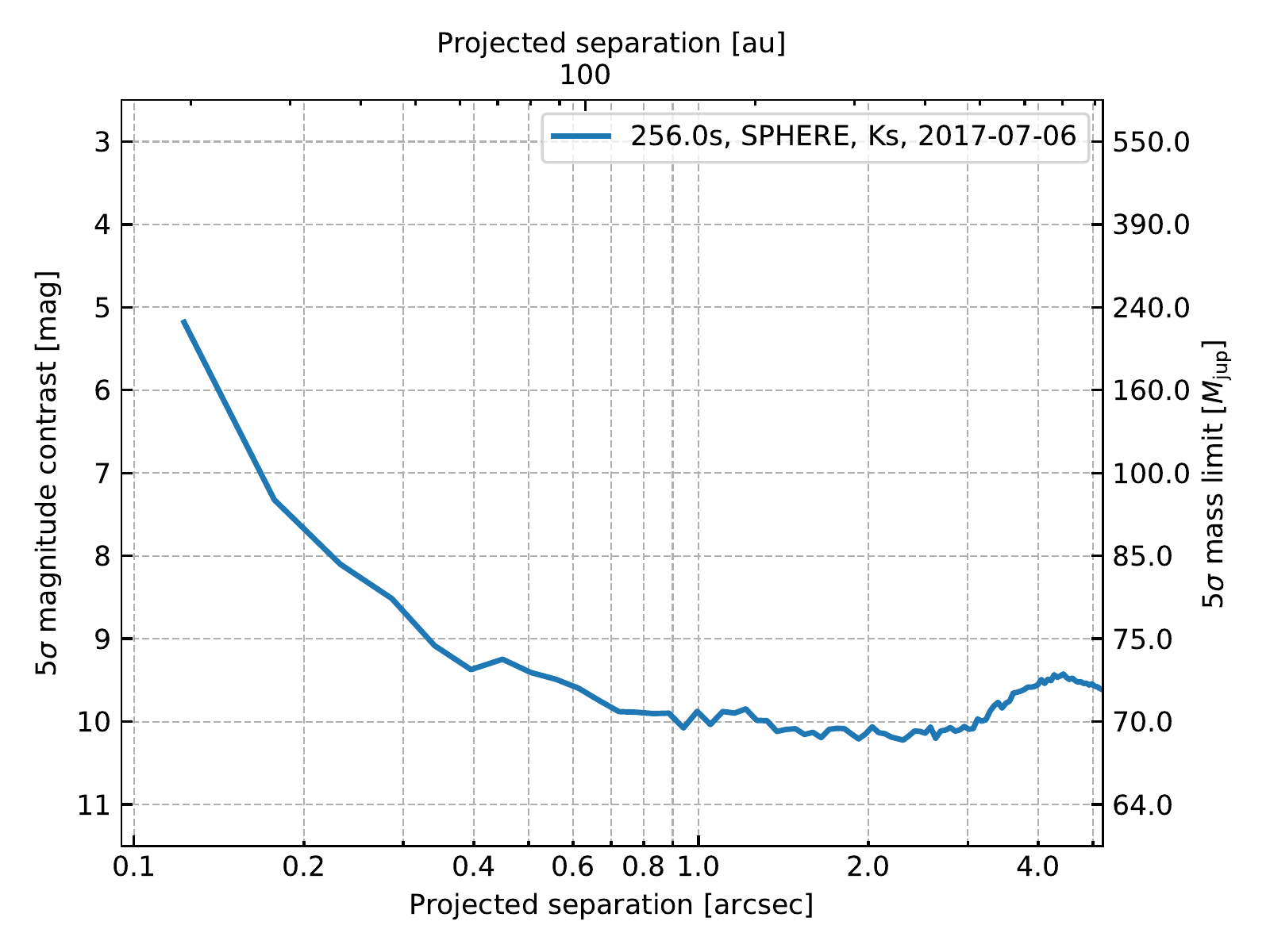}
\subcaption{WASP-99}
\end{subfigure}
\begin{subfigure}[b]{0.3\textwidth}
\includegraphics[width=\textwidth]{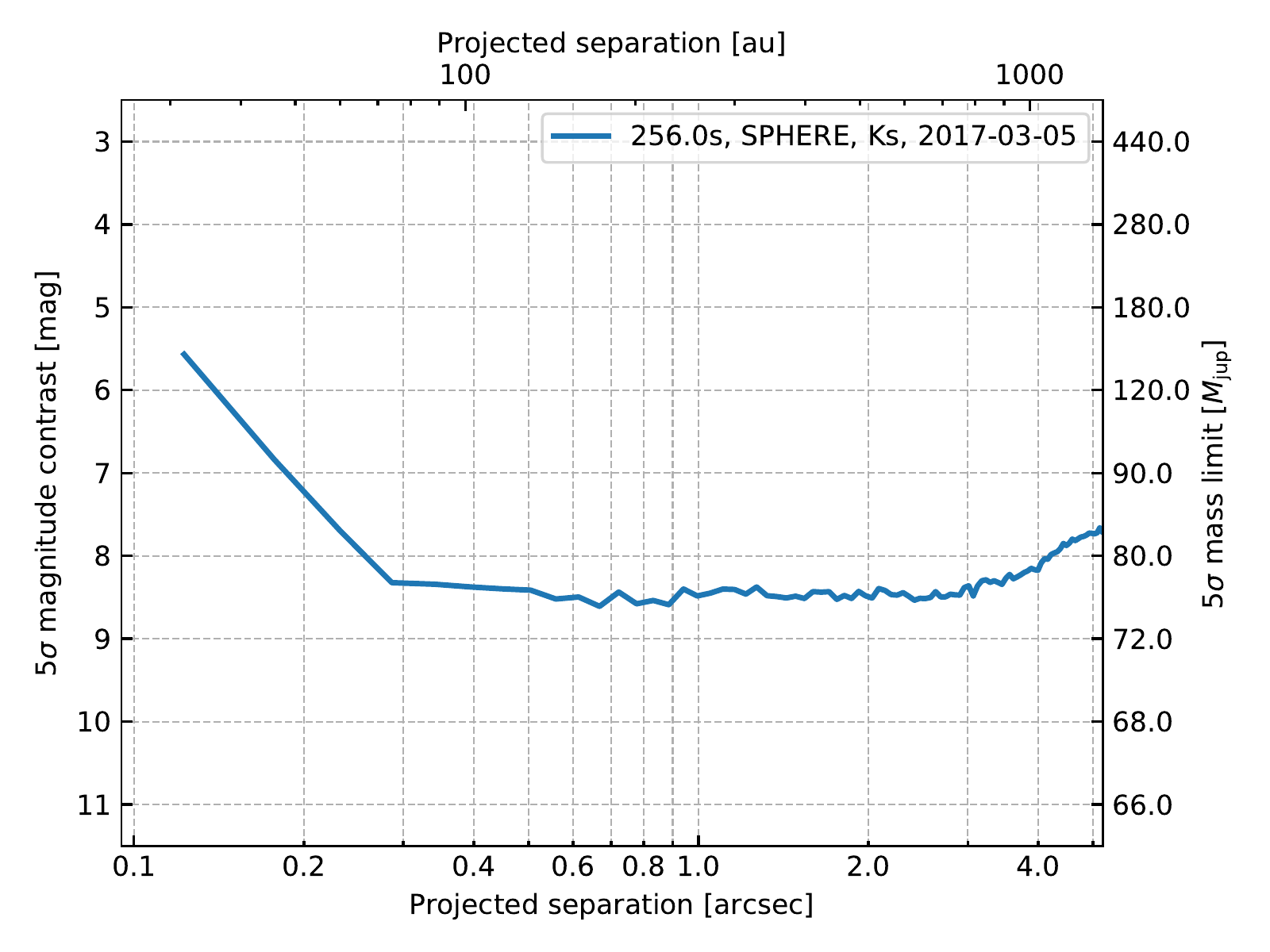}
\subcaption{WASP-108}
\end{subfigure}

\begin{subfigure}[b]{0.3\textwidth}
\includegraphics[width=\textwidth]{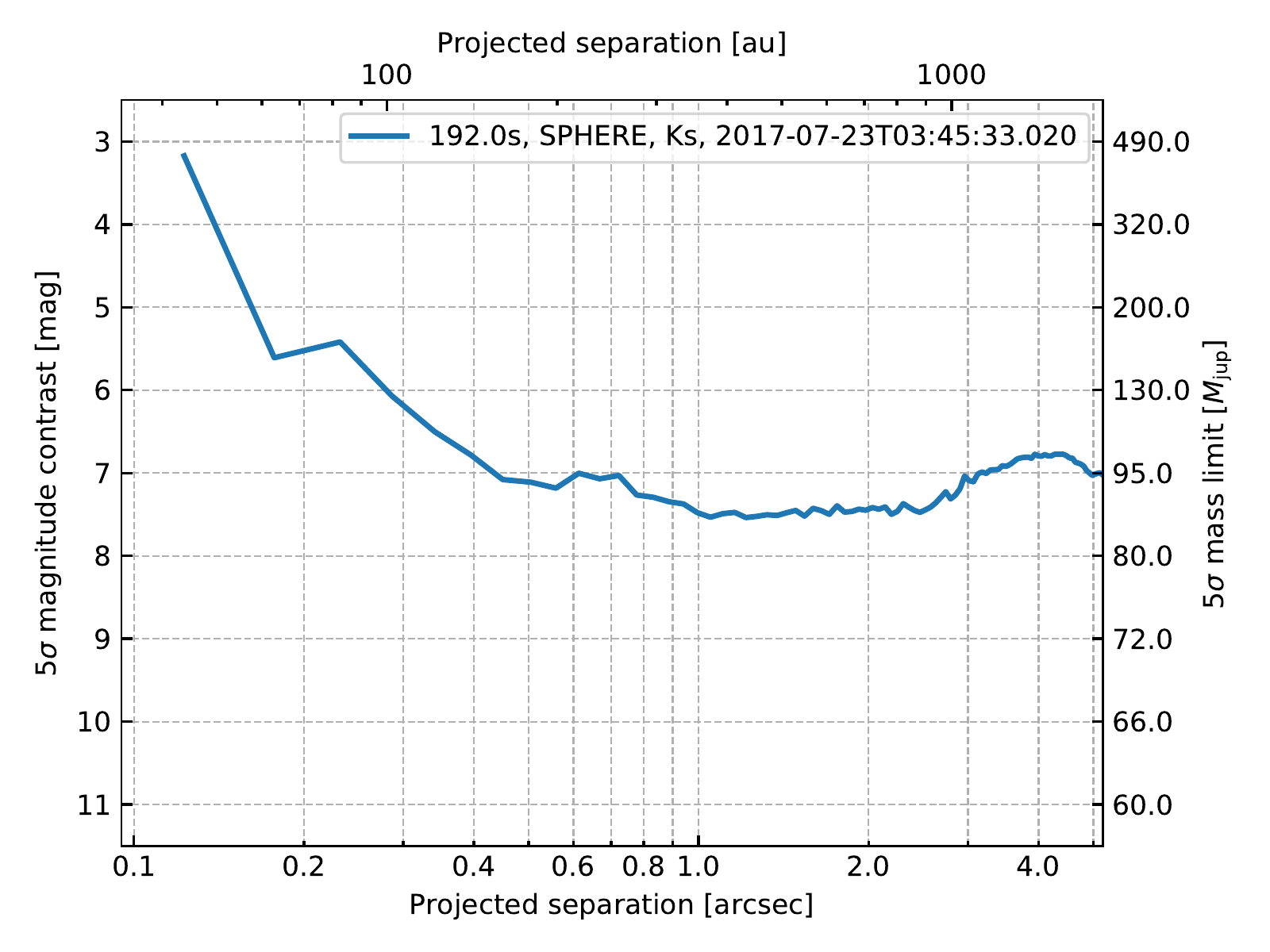}
\subcaption{WASP-109}
\end{subfigure}
\begin{subfigure}[b]{0.3\textwidth}
\includegraphics[width=\textwidth]{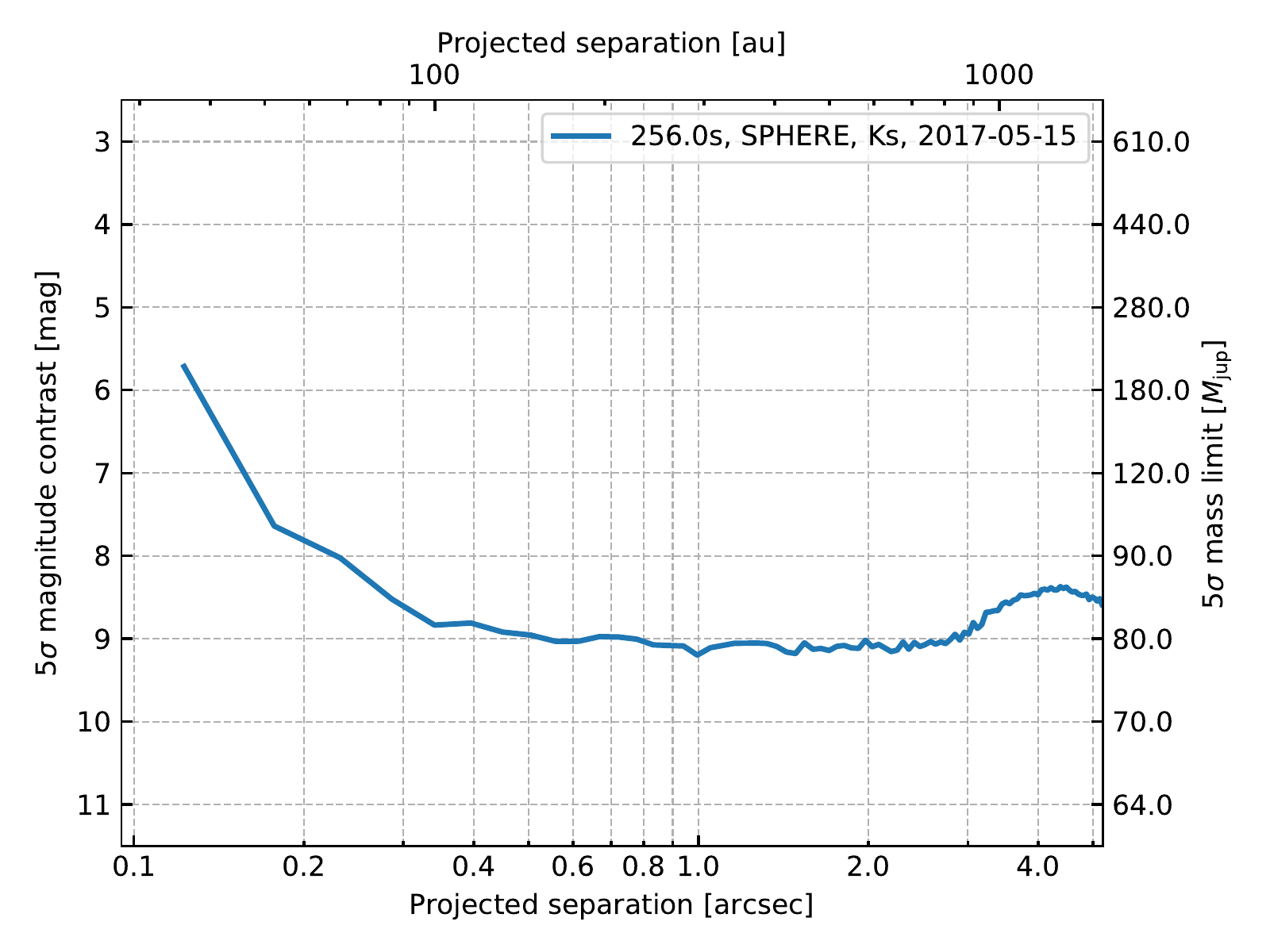}
\subcaption{WASP-111}
\end{subfigure}
\begin{subfigure}[b]{0.3\textwidth}
\includegraphics[width=\textwidth]{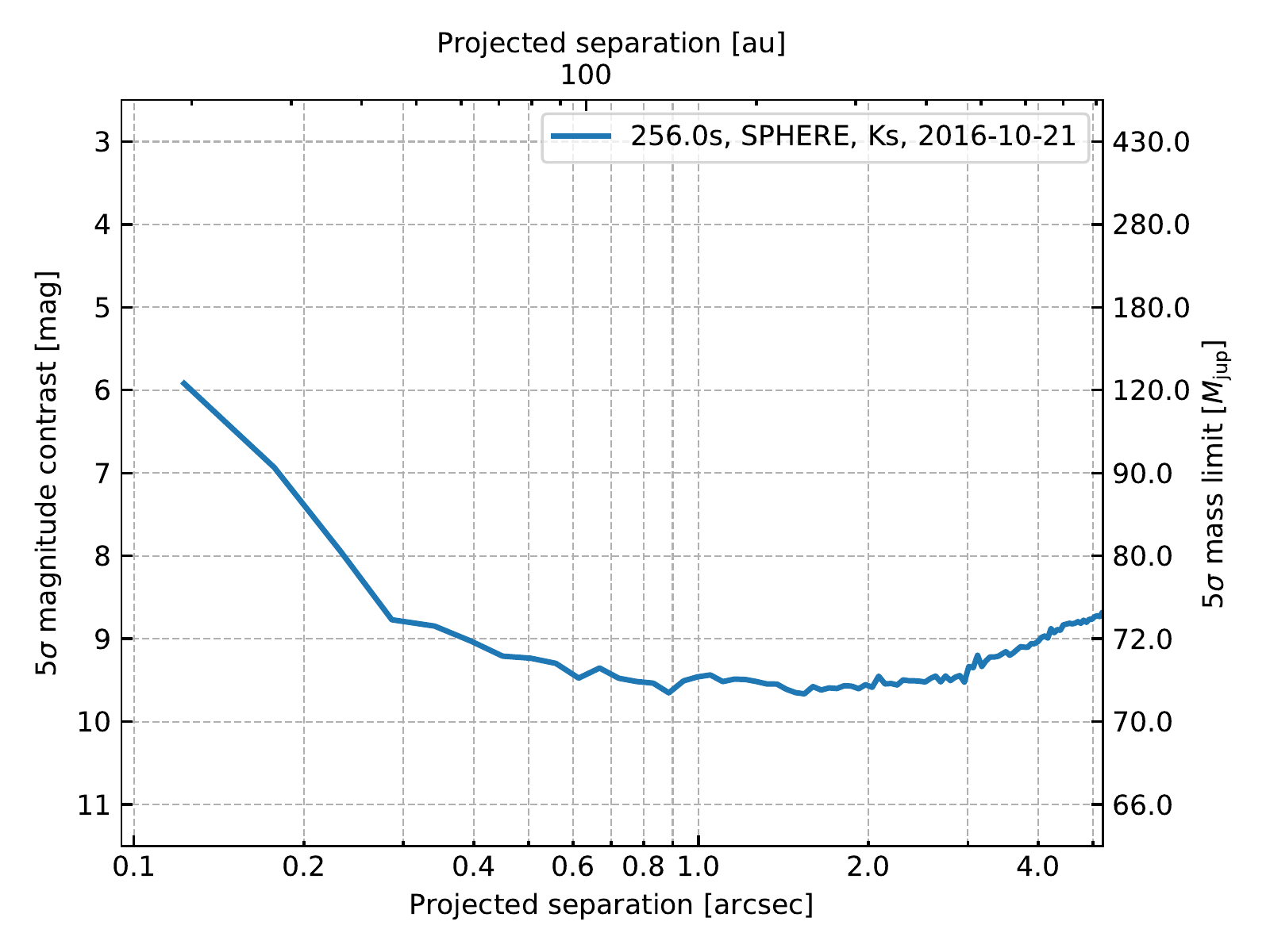}
\subcaption{WASP-117}
\end{subfigure}

\begin{subfigure}[b]{0.3\textwidth}
\includegraphics[width=\textwidth]{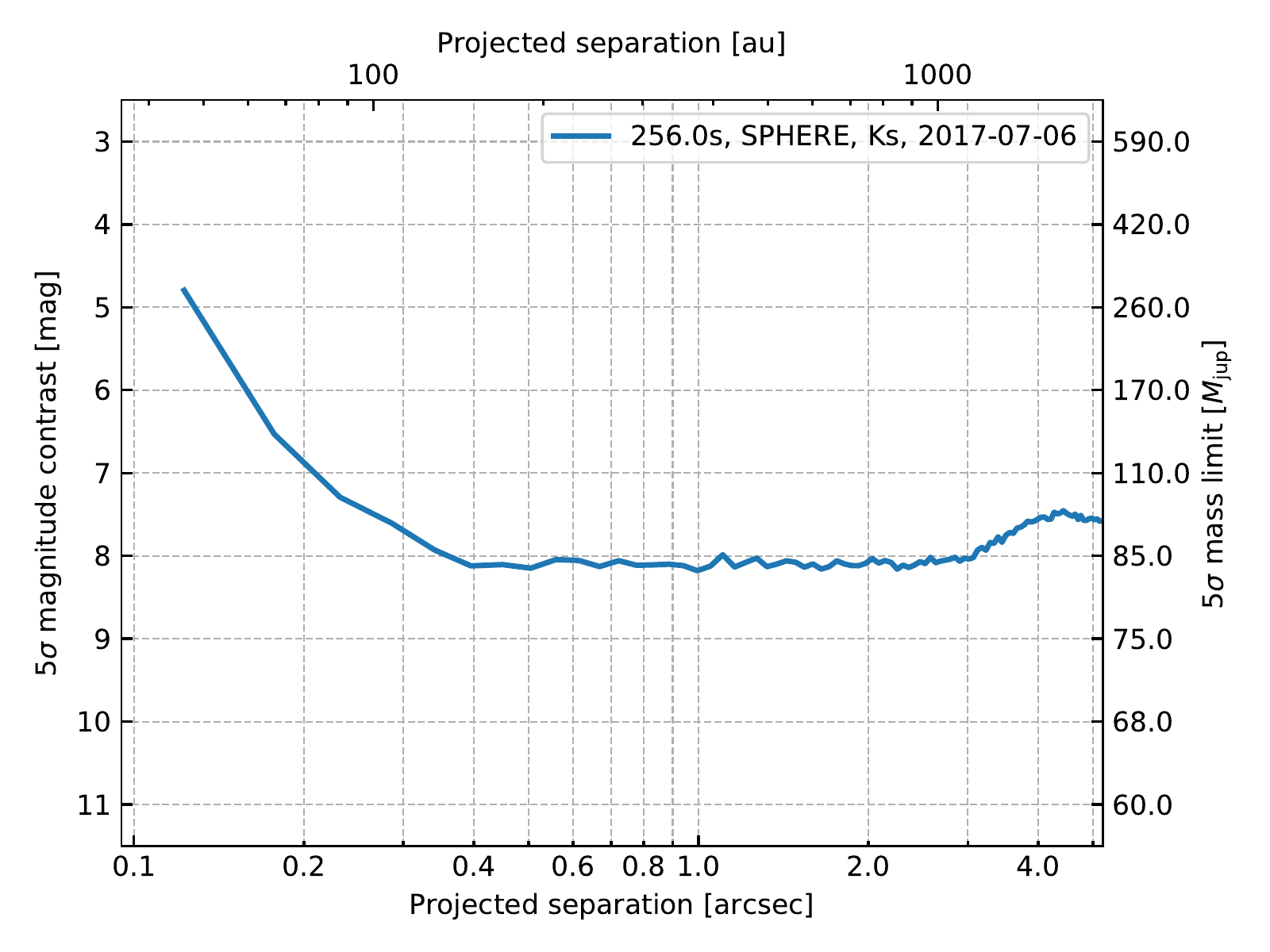}
\subcaption{WASP-118}
\end{subfigure}
\begin{subfigure}[b]{0.3\textwidth}
\includegraphics[width=\textwidth]{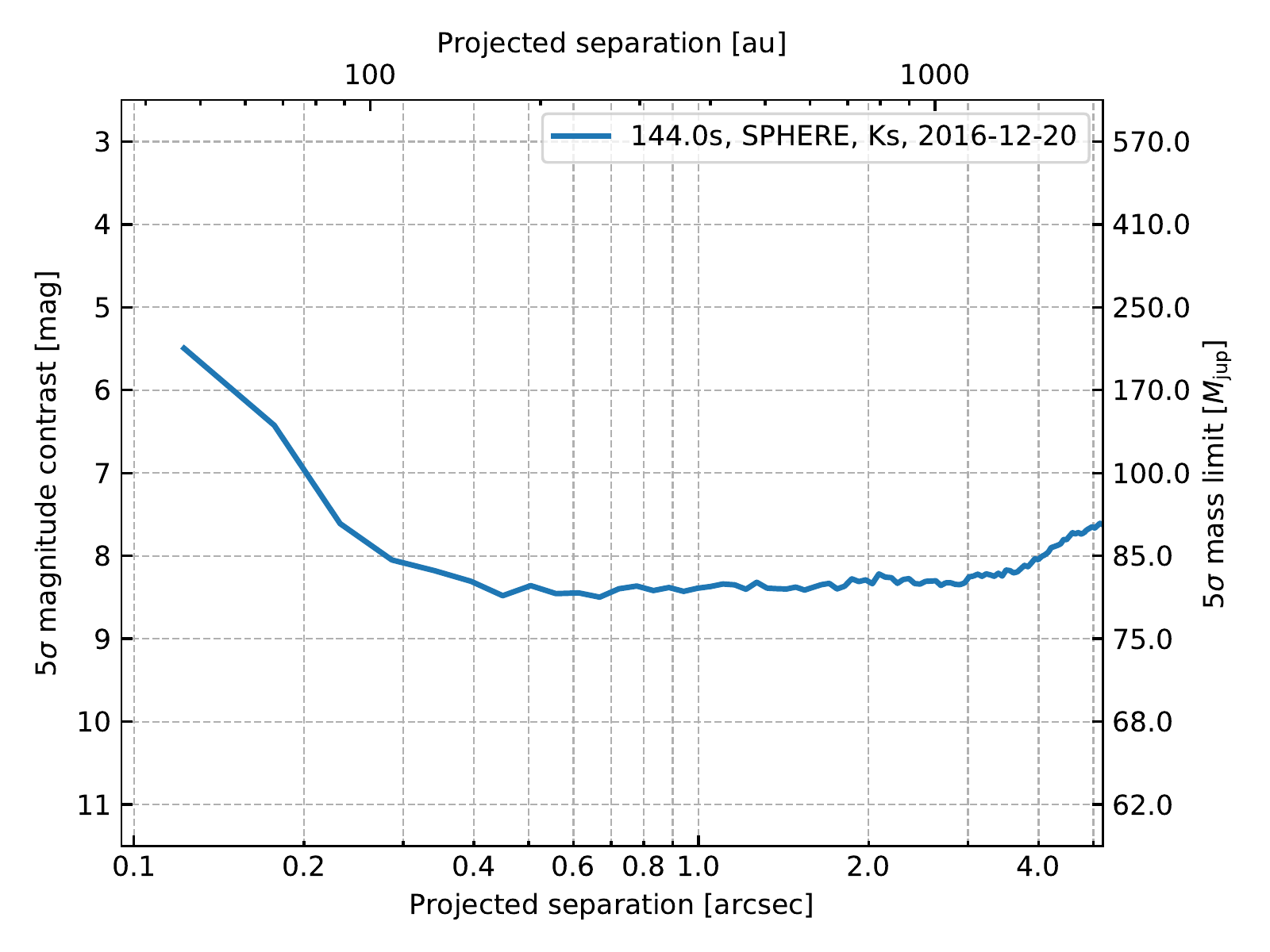}
\subcaption{WASP-120}
\end{subfigure}
\begin{subfigure}[b]{0.3\textwidth}
\includegraphics[width=\textwidth]{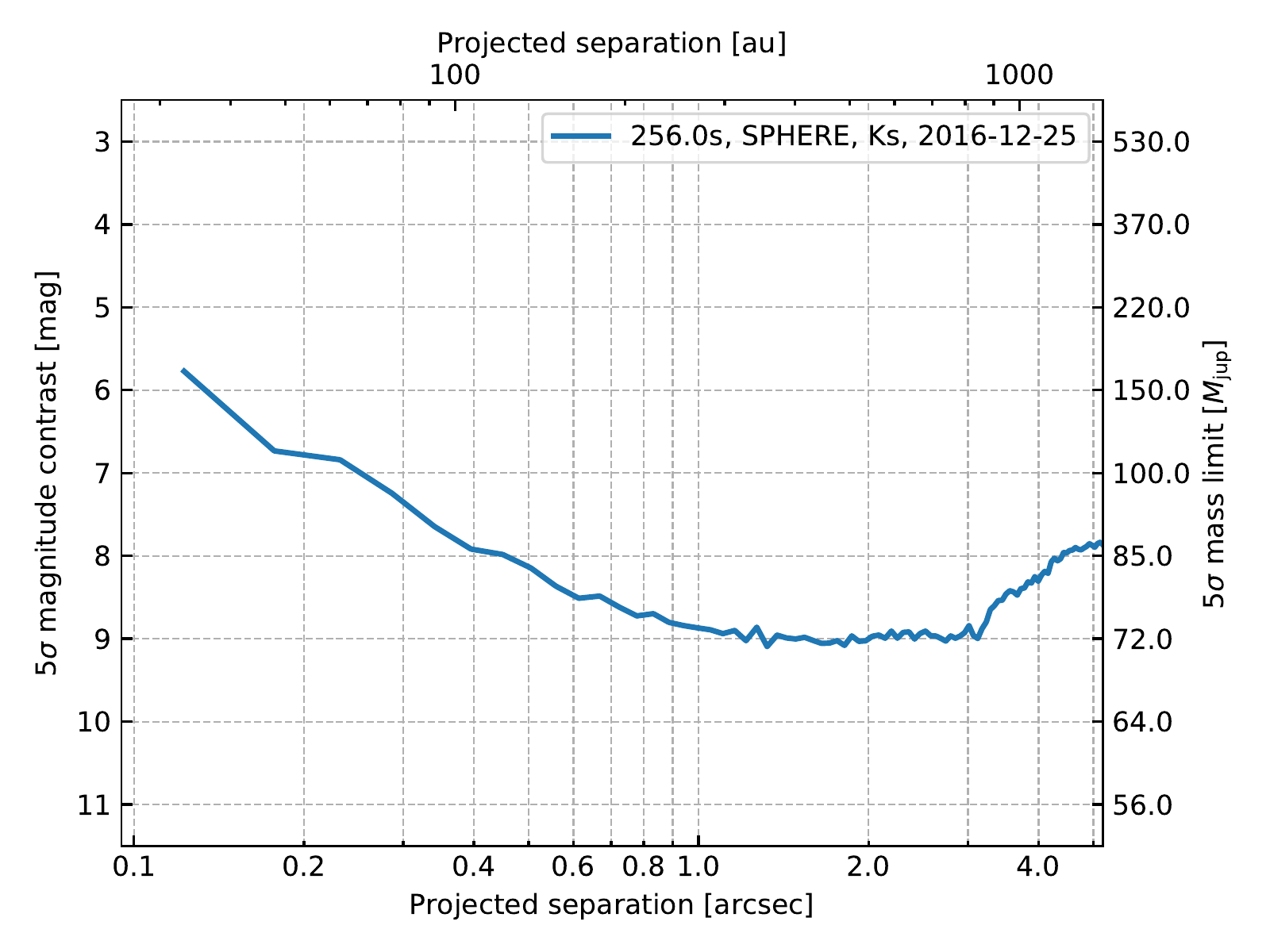}
\subcaption{WASP-121}
\end{subfigure}

\begin{subfigure}[b]{0.3\textwidth}
\includegraphics[width=\textwidth]{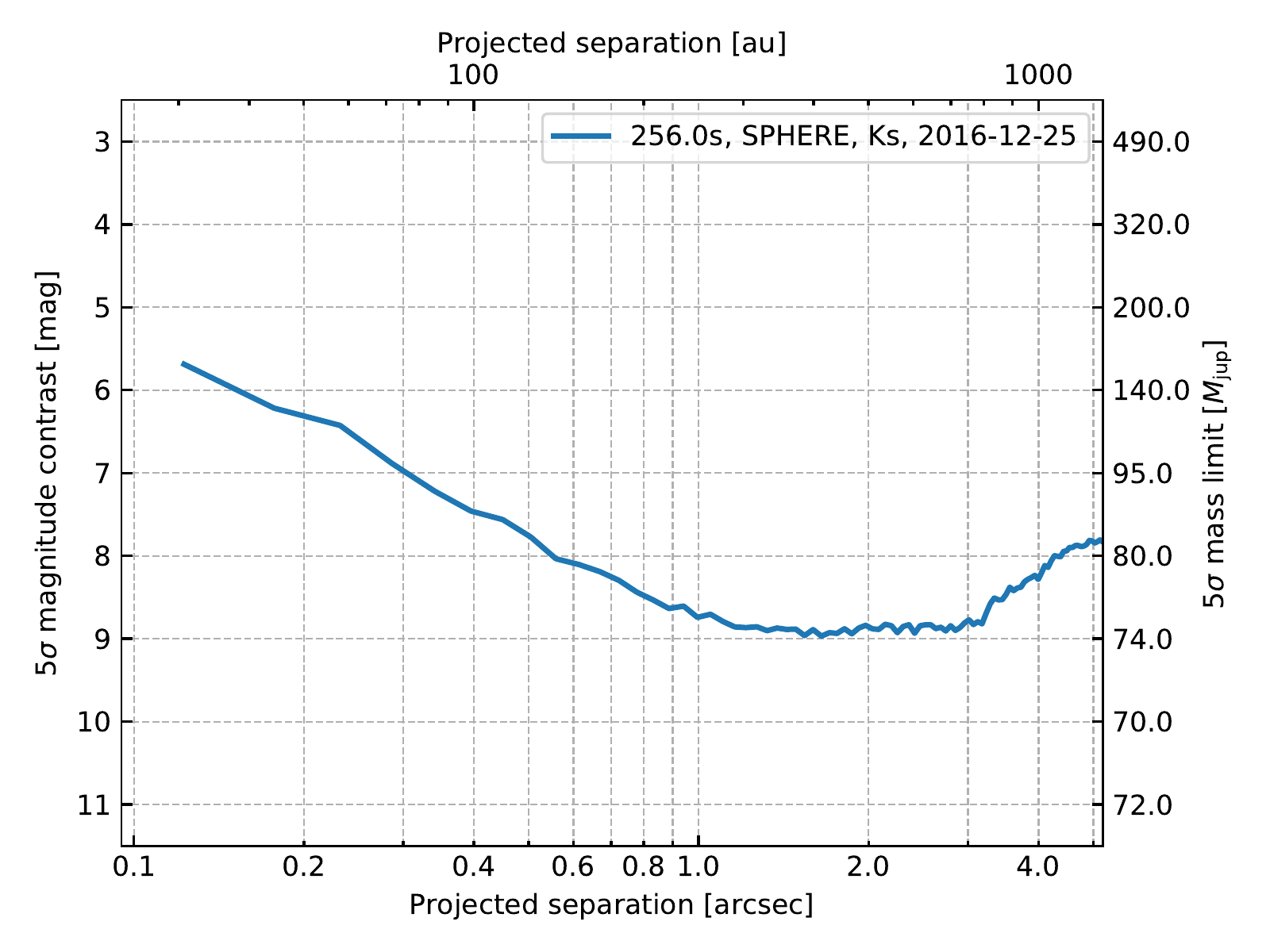}
\subcaption{WASP-122}
\end{subfigure}
\begin{subfigure}[b]{0.3\textwidth}
\includegraphics[width=\textwidth]{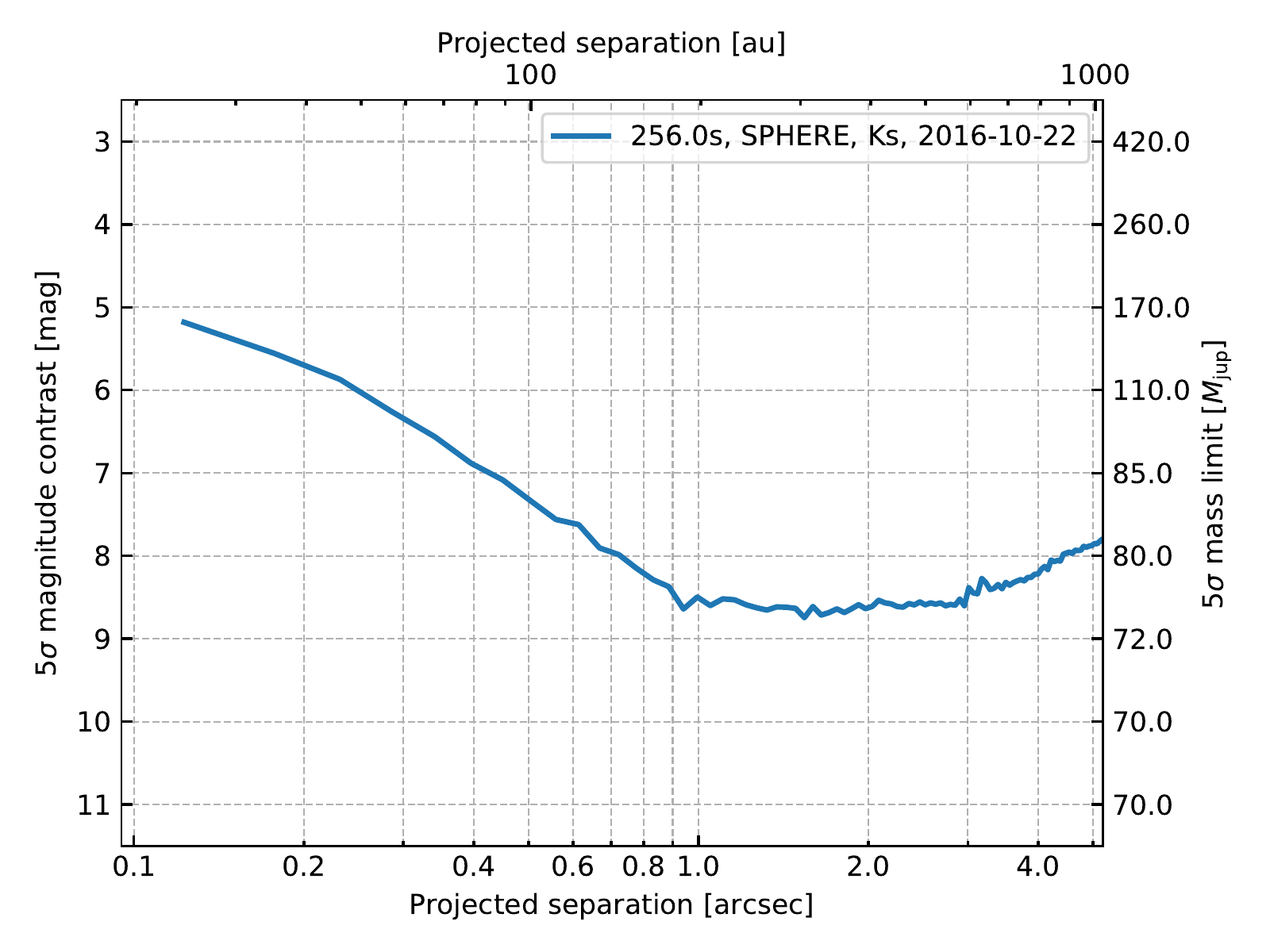}
\subcaption{WASP-123}
\end{subfigure}
\begin{subfigure}[b]{0.3\textwidth}
\includegraphics[width=\textwidth]{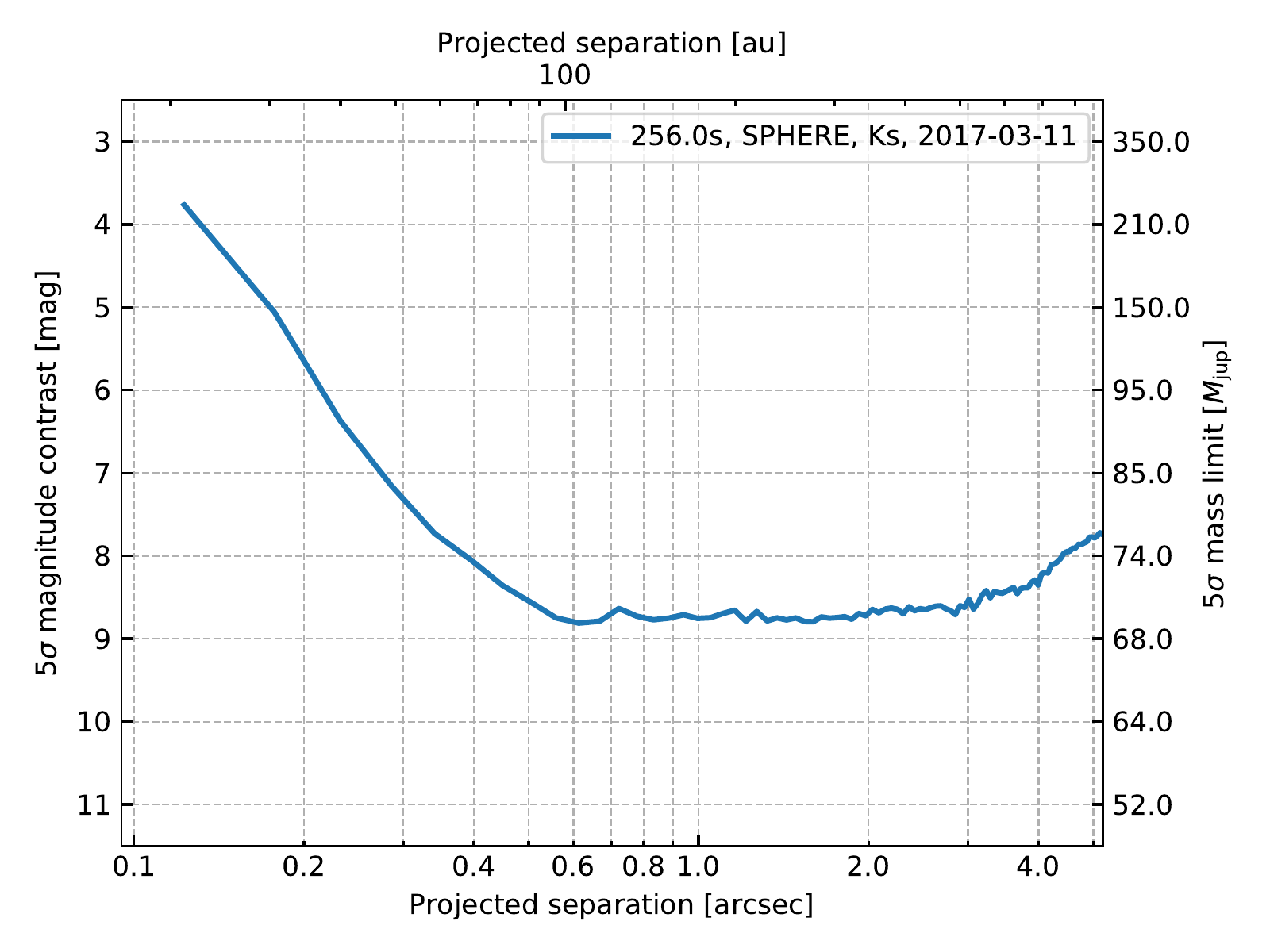}
\subcaption{WASP-130}
\end{subfigure}

\begin{subfigure}[b]{0.3\textwidth}
\includegraphics[width=\textwidth]{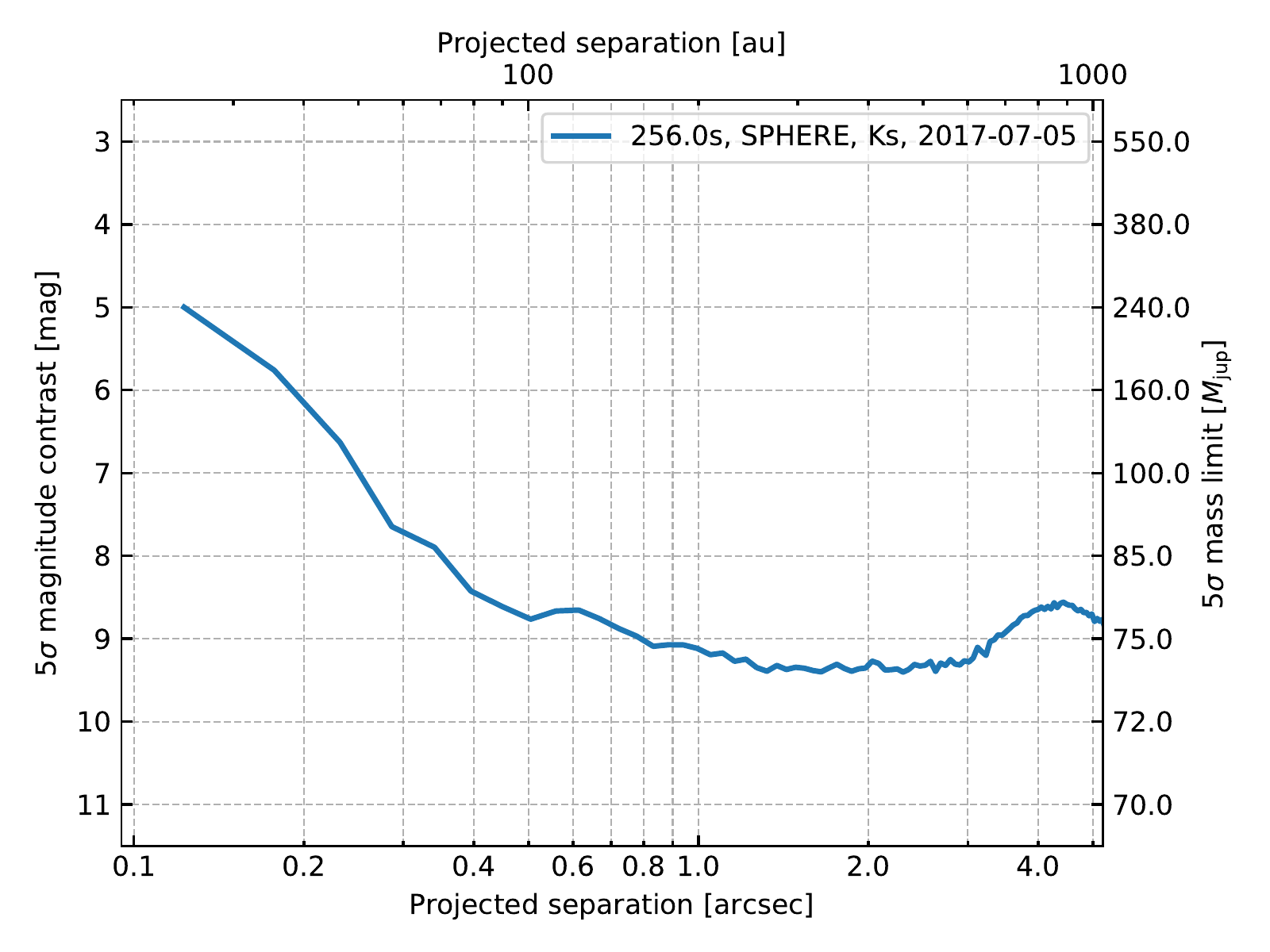}
\subcaption{WASP-131}
\end{subfigure}
\begin{subfigure}[b]{0.3\textwidth}
\includegraphics[width=\textwidth]{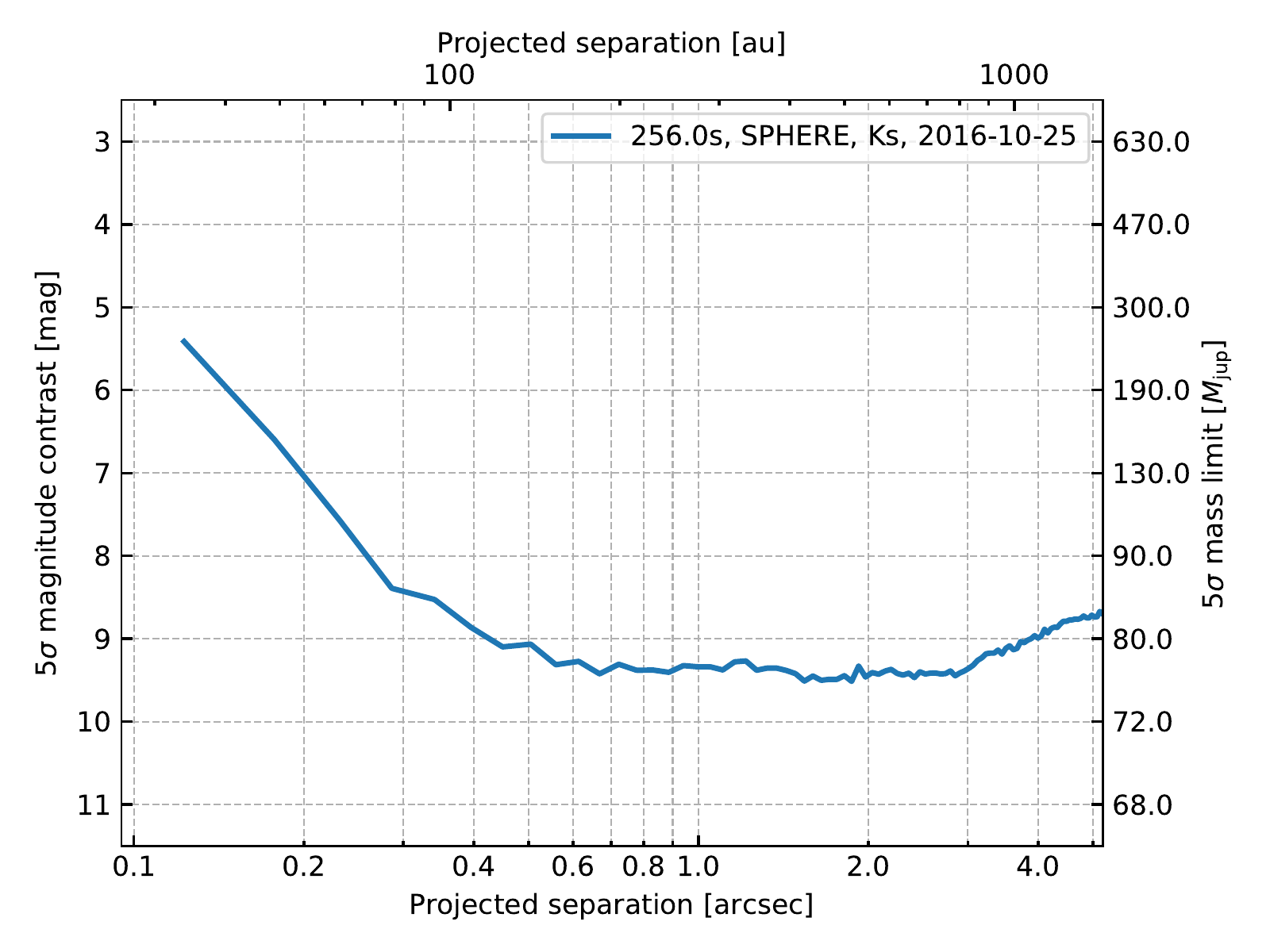}
\subcaption{WASP-136}
\end{subfigure}
\begin{subfigure}[b]{0.3\textwidth}
\includegraphics[width=\textwidth]{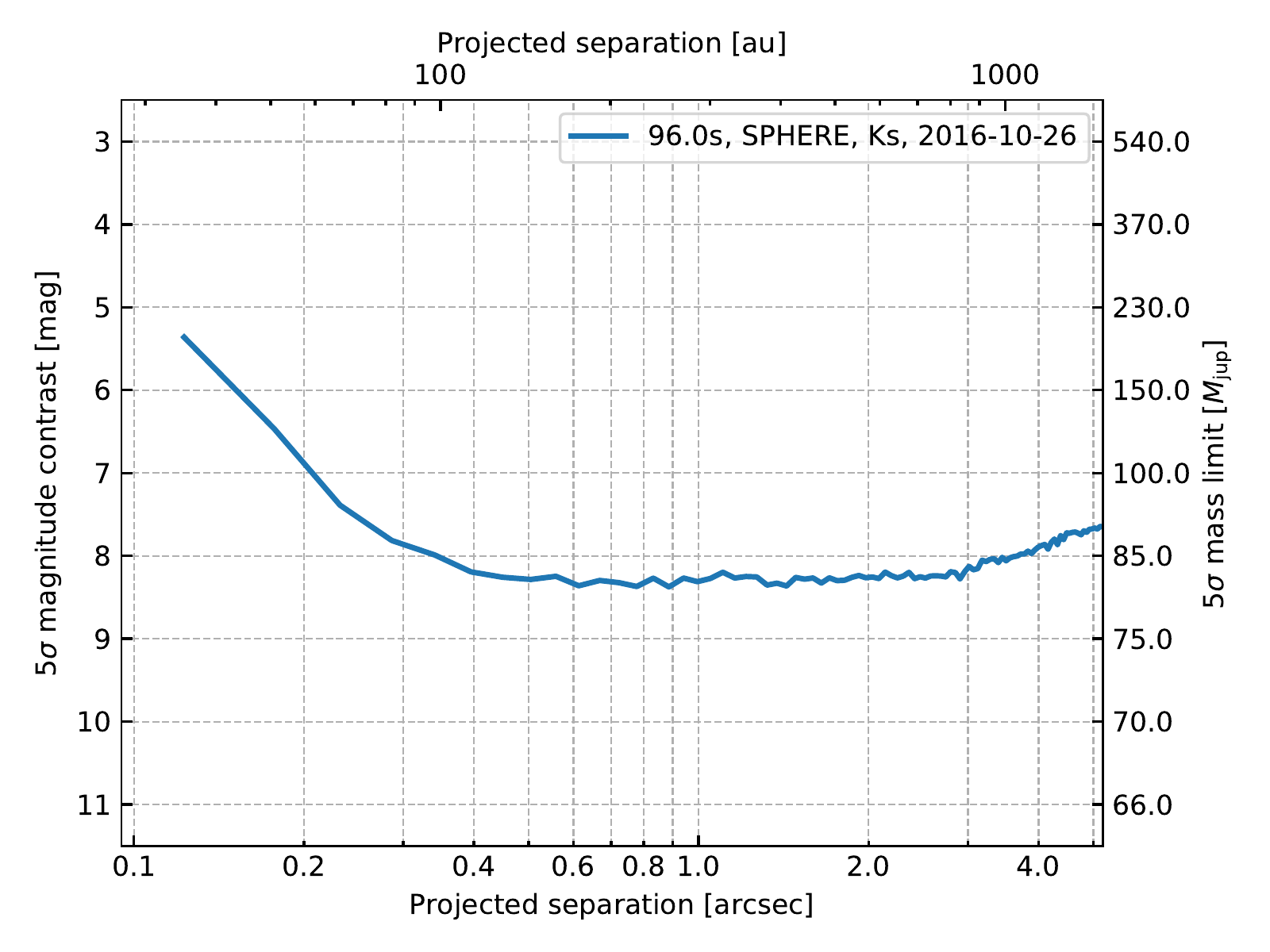}
\subcaption{WASP-137}
\end{subfigure}

\caption{
Detection limits of individual targets III.
We convert projected angular separations in projected physical separations by using the distances presented in Table~\ref{tbl:star_properties}.
The mass limits arise from comparison to AMES-Cond, AMES-Dusty, and BT-Settl models as described in Sect.~\ref{subsec:characterization_of_ccs}.
}
\label{fig:detection_limits_individual_3}
\end{figure*}

The detection limits for each individual target are presented in Fig.~\ref{fig:detection_limits_individual_1}, \ref{fig:detection_limits_individual_2}, and \ref{fig:detection_limits_individual_3}.
We used AMES-Cond, AMES-Dusty and BT-Settl models \citep{Allard2001,Baraffe2003} as illustrated in Sect.~\ref{subsec:characterization_of_ccs} to convert magnitude contrast to detectable Jupiter masses.
The data used for creating these plots will be published online in the Strasbourg astronomical Data Center (CDS).

\end{appendix}

\end{document}